\documentclass[11pt,preprint]{aastex}

\newcommand{\Ni}{$^{56}$Ni}
\newcommand{\Co}{$^{56}$Co}
\newcommand{\Fe}{$^{56}$Fe}
\newcommand{\Ms}{$M_{\odot}$}
\newcommand      \grays       {$\gamma$-rays}

\shorttitle{THE CHEMISTRY OF POP. III SN EJECTA. II.}
\shortauthors{Cherchneff \& Dwek}

\begin{document}


\title{THE CHEMISTRY OF POPULATION III SUPERNOVA EJECTA: II - THE NUCLEATION OF MOLECULAR CLUSTERS AS A DIAGNOSTIC FOR DUST IN THE EARLY UNIVERSE}


\author{Isabelle Cherchneff \altaffilmark{1} \& Eli Dwek\altaffilmark{2}}

\altaffiltext{1}{Departement Physik, Universit{\"a}t Basel, CH-4056 Basel, Switzerland; isabelle.cherchneff@unibas.ch}
\altaffiltext{2}{Observational Cosmology Laboratory, Code 665, NASA Goddard Space Flight Center, Greenbelt, MD 20771, USA; eli.dwek@nasa.gov}


\begin{abstract}
We study the formation of molecular precursors to dust in the ejecta of Population~III supernovae using a chemical kinetic approach to follow the evolution of small dust cluster abundances from day 100 to day 1000 after explosion. Our work focuses on zero-metallicity 20~\Ms~and 170 \Ms progenitors, and we consider fully-macroscopically mixed and unmixed ejecta. The dust precursors comprise molecular chains, rings and small clusters of chemical composition relevant to the initial elemental composition of the ejecta under study. The nucleation stage for small silica, metal oxides and sulphides, pure metal, and carbon clusters is described with a new chemical reaction network highly relevant to the kinetic description of dust formation in hot circumstellar environments. We consider the effect of the pressure dependence of critical nucleation rates, and test the impact of microscopically-mixed He$^+$ on carbon dust formation. Two cases of metal depletion on silica clusters (full and no depletion) are considered to derive upper limits to the amounts of dust produced in SN ejecta at 1000 days, while the chemical composition of clusters gives a prescription for the type of dust formed in Pop. III supernovae. 

We show that the cluster mass produced in the fully-mixed ejecta of a 170\Ms\ progenitor is $\sim$ 25 \Ms\  whereas its 20~\Ms\ counterpart forms $\sim$ 0.16~\Ms\ of clusters. The unmixed ejecta of a 170~\Ms\ progenitor supernova synthesizes $\sim 5.6$~\Ms\ of small clusters, while its 20~\Ms\ counterpart produces $\sim 0.103$~\Ms. Our results point to smaller amounts of dust formed in the ejecta of Pop.~III supernovae by  a factor $\sim$ 5 compared to values derived by previous studies, and to different dust chemical composition. Such deviations result from some erroneous assumptions made, the inappropriate use of classical nucleation theory to model dust formation, and the omission of the synthethis of molecules in supernova ejecta. We also find that the unmixed ejecta of massive Pop.~III supernovae chiefly form silica and/or silicates, and pure silicon grains whereas their lower mass counterparts form a dust mixture dominated by silica and/or silicates, pure silicon and iron sulphides. Amorphous carbon can only condense in ejecta where the carbon-rich zone is deprived of He$^+$ via the nucleation of carbon chains and rings characteristic of the synthesis of fullerenes. The first dust enrichment to the primordial gas in the early universe from Pop.~III massive supernova comprises primarily pure silicon, silica and silicates. If carbon dust is present at redshift  $ z>$ 6, alternative dust sources must be considered.
\end{abstract}


\keywords{astrochemistry --- supernovae: general --- early universe --- molecular processes --- dust}



\section{Introduction}

For the last two decades, many efforts have been developed to try and understand the role of supernovae (hereafter, SNe) as potential dust contributors to the Universe. On the observational front, the first evidence for dust formation in a SN event was observed during the explosion of the Type IIp supernova SN SN1987A. Its extensive observational coverage at mid-infrared (IR) wavelengths revealed the presence of the fundamental and overtone transitions of a few molecules, specifically CO and SiO, as early as $\sim$ 120 days post-explosion, and the formation of dust grains after day 400 \citep{meik89,mos89,roch91,dan91,wood93}. The exact composition of these condensates is yet not known exactly but recent modeling of the IR excess measured by Spitzer using a clumpy ejecta proposes a mixture of amorphous carbon and silicate dust  \citep{erco07}. Several other galactic and extra-galactic SN explosions have been monitored since then, and the detection of CO, SiO and dust in their ejecta has proved positive \citep{ko05, ko06, ko09, sug06}. However, wether a SN explosion produces large amounts of dust or not in its ejecta is still a matter of debate. Indeed, the values of the dust mass synthesized derived from observational IR data are never larger than $\sim 10^{-2}$ \Ms. Large amounts of dust ($\ge 10^8$ \Ms) at high redshift (z $ >$ 6) have been conjectured to explain the sub-millimeter observations of distant quasars and metal measurements in Damped-Lyman-Alpha Systems \citep{pei91, pet94, ber03}. In view of the small amount of dust detected in local SN ejecta, these findings pose problem for a dust SN origin in the early universe and the possible synthesis of solids in the explosive ends of massive Population III stars (hereafter Pop. III, \citep{dwek07}). 

The formation of dust in SN events has also been tackled theoretically and several studies have been conducted for local and high redshifted SNe. Two approaches are usually used for modeling purposes: (1) the thermodynamical equilibrium (TE) assumption, and (2) the classical nucleation theory, or CNT. The first approach is assumed in models aiming at deriving  dust and mixing properties in local SN ejecta from the study of isotopic anomalies in meteorites \citep{lod96,tra99}. The assumption of TE implies chemical equilibrium. The derivation of condensation temperatures and chemical compositions in a pressure-temperature equilibrium phase diagram for the various solids under study may be adequate to describe the formation of molecules in stellar photospheres where high pressures prevail \citep{tsu73, cher92, hel96} but is clearly inappropriate for lower density circumstellar environments, in particular the dynamical flows of SN ejecta. As we will see below,  the chemistry is neither at equilibrium nor at steady state over the timespan under which dust condenses in SN ejecta, while local thermal equilibrium may apply. The second approach based on CNT has been extensively used by various groups for low and high redshift SNe \citep{koz89, clay99, tod01,noz03,sch04,bian07}. CNT was first developed to explain the formation of water droplets in the Earth atmosphere under equilibrium conditions  \citep{fed66} and used in initial attempts to model dust formation in the interstellar medium \citep{dra79}. However,  \citet{don85} first objected to the appropriateness of CNT to describe dust synthesis in space, i.e.,  when the formation of solids from the gas phase occurs in low-density systems far from equilibrium, e.g., in circumstellar environments such as stellar winds of evolved stars or SN ejecta. Indeed, CNT involves the calculation of critical cluster sizes for nucleation and bulk material concepts like specific surface energies used in the derivation of the free energy barrier for nucleation. However, the derived critical clusters are usually on the atomic scale, making clear that CNT bulk property concepts are arguable for atoms and small molecules. The existence of a steady-state critical cluster distribution is questionable if the molecular phase from which those clusters form is not at steady state. Finally, even in high-pressure experiments where equilibrium conditions hold, the nucleation steps involved and the nucleation products are different from those predicted by CNT. These limitations have been recognized for some time by the various scientific communities modeling the synthesis of ceramics or soot in flames in the laboratory.  In such experiments, the formation of dust proceeds through a two-step mechanism involving the nucleation of small molecular clusters and the growth of those clusters from surface deposition and coagulation  \citep{mcm96, pra98, woo98}. All those processes are kinetic in essence and usually occur far from equilibrium. 

A stochastic, kinetically-driven approach must thus be used to describe dust formation in space, in which gas-phase molecules transform into small dust clusters via chemical kinetic processes with no prior assumption on the nature of solids that should condense from the initial gas mixture under study. A major drawback is the partial lack of information on the different steps involved in nucleation and their related chemical reaction rates, and on the intermediate species formed during coagulation. However, many recent studies of ceramic synthesis in flames highlight the role of key species and provide new insights on these matters. A first attempt to model the formation of dust in SN ejecta at high redshift using a chemical kinetic approach based on a careful description of the formation of molecules and the nucleation of small solid clusters was carried out by \citet{cher08}. In a more complete study with revised chemical processes,  \citet[hereafter CD09]{cher09} show that Pop. III SNe are efficient at forming key molecules such as O$_2$, CO, SiS and SO in substantial amounts (up to $\sim 35$~\% of the progenitor mass). Those molecules have a crucial impact on the dust synthesis as a) they deplete the gas phase from potential heavy elements that could otherwise be included in grains and b) they act as coolants to the local gas through emission in their ro-vibrational transitions. For example, O$_2$ and CO are the prevalent molecular coolants through their rotational transitions during the collapse phase of molecular clouds  \citep{gol78}. CD09 also show that the formation of chemical species in the ejecta is neither at equilibrium, nor at steady state, a common assumption used in all existing studies. 

In the present paper, we study the formation of small molecular clusters involved in the nucleation step of the synthesis of dust in the explosive ends of Pop III stars.  Those clusters are of paramount importance in the determination of the dust chemical composition as well as the final dust mass formed as they are the bottleneck to condensation processes. We consider small clusters involved in the condensation of several solid compounds commonly found in evolved, circumstellar environments and whose chemical compositions are in agreement with the elemental composition pertaining to SNe. They include pure metallic grains, metal oxides, metal sulphides, silica and silicates, silicon carbides and amorphous carbon dust. We study their formation for the two mixing cases described in CD09, i.e., fully microscopically-mixed and unmixed ejecta, for a zero-metallicity, very massive progenitor exploding as a pair-instability SN (or PISN) and a low-mass progenitor exploding as a core-collapse SN (CCSN). In \S2, we fully describe the chemistry of cluster formation from the gas phase. \S3 briefly presents the SN models assumed in this study whereas results are presented in \S4 along with a critical assessment of existing studies, and an analysis of key parameters to the dust formation processes. Finally, a summary of the results and their interpretation in the context of the early universe are presented in \S5. 

\section{THE CHEMISTRY OF DUST NUCLEATION}

Dust formation in circumstellar environments does not occur under equilibrium or steady state conditions due to the complexity of the dynamics and the physics involved in such media. As in the laboratory, dust condensation and growth occur through stochastic, kinetically-controlled processes. For example, in Asymptotic Giant Branch stars, the chemistry characterizing the gas layers located above the stellar photosphere and where dust forms is not at equilibrium despite the large gas densities and temperatures encountered, owing to the action of periodic shocks induced by the stellar pulsations \citep{cher06}. In supernova ejecta, the passage of the blast wave through the helium core and the deposition of radioactive energy in the ejecta trigger an active chemistry far from equilibrium or steady state, as shown in CD09. As such, the formation of dust resembles that occurring in pyrolysis, flame, or vaporization experiments, and consist of two phases: (1) the nucleation of small molecular clusters from the gas phase determined by chemical kinetic processes; (2) the growth of those clusters to dust grains from stochastic coagulation and surface addition processes. 

Depending on the initial composition of the gas, the clusters are of different chemical nature. In an oxygen-rich environment, metal oxides and various silicates are expected to form whereas a large carbon content fosters the formation of metal carbides and various forms of solid carbons (polycyclic aromatic hydrocarbons, i.e., PAHs, fullerenes, carbon chains and rings etc.). Clusters of pure iron, silicon, and magnesium also form in vaporization experiments as well as metal sulphides, when atomic sulphur is present.  We will describe below the various clusters considered in the present study and their related formation chemistry. Those clusters are the building blocks of the various 'astronomical' types of dust observed in circumstellar environments and are all synthesized in flame, condensation, and vaporization experiments in the laboratory. The nucleation stage described in this study involves the formation of 1D- molecular chains, 2D- rings and 3D- small cages up to a maximum of 10 atoms. This cluster size  upper limit is somehow arbitrary and is motivated by the impact of the cluster geometry on the growth process, i.e., a cluster with a 3D structure like a cage or a twisted ring can be seen as a small seed nucleus for coagulation with a larger collision cross section than those of chains or rings. The clusters and the dust that condenses out of them are listed in Table {1}.  

All chemical pathways leading to the formation of chains include neutral-neutral processes like bimolecular and radiative association reactions whereas destruction is described  by thermal fragmentation, reaction with Compton electrons, and neutral-neutral processes (i.e., oxidation reactions for carbon chains). In the special case of rings, their structure precludes a large chemical  reactivity and we assume that the polymerization of the gas-phase monomer proceeds via neutral-neutral processes while the destruction of polymers results from thermal fragmentation {\it only}. Hence, reactions of atomic oxygen and He$^+$ ions with rings are ignored. All processes are discussed in detail below for the various clusters and a list of all chemical reactions involved in the nucleation of clusters and their respective rates is given in Table 2. The full chemical network considered in this study includes Table 2 and Table 5 of CD09, the latter describing the chemistry involved in the synthesis of molecules in SN ejecta. As all those chemical processes apply to hot ($\sim 2000$~K) and dense ($\sim 10^9$~mol. cm$^{-3}$) gas, these two tables represent an extremely useful compilation of reactions and rates for modeling the dust chemistry of other circumstellar environments such as Type Ic supernovae, Novae, the shocked inner regions of AGB stellar envelopes, Supergiants and LBVs outflows, the wind of RCrB stars, or the colliding winds of Wolf-Rayet stars. 

\subsection{Oxygen-Rich Clusters}

Amorphous and crystalline silicates are the most abundant family of solids found in space. Olivine grains including forsterite (Mg$_2$SiO$_4$) and fayalite (Fe$_2$SiO$_4$) are observed in the wind of oxygen-rich low-mass stars ascending the AGB when pyroxene dust (enstatite MgSiO$_3$ and ferrostite FeSIO$_3$) are found around young stars  \citep{mol03}. For silicates, no gas phase monomer precursors exist from which solids can grow by direct coagulation. Despite the wealth of studies dedicated to the formation of silicates and ceramics in the laboratory, the identification of molecular precursors to silicate formation is still a difficult task. Flame aerosol technology experiments highlight the important role of heterogeneous seed clusters in the formation of silicate compounds \citep{mcm96,pra98,woo98}. In presence of an initial Fe/Mg/Si/O$_2$-rich vapor mixture in condensation experiments, \citep{rie99,rie02} show that the final products include eutectic Mg-SiO magnesiosilica or Fe-SiO ferrosilica condensates but {\it no} mixed ferromagnesiosilica grains. The formation of magnesio- and ferrosilica involves the coalescence of heterogenous small clusters of silica, magnesium oxides and iron oxides, and pure metallic clusters dependent of the initial ratio of metals to gas-phase SiO \citep{kai03,kam05}. We discuss below the possible reaction channels to the formation of small silica and metallic oxide clusters. Cluster structure and stability have been derived for silicon oxide clusters \citep{lu03}, metal oxides \citep{zie90, kho97, bha07,rei07}, and pure metallic clusters \citep{jel02}. The structures for (SiO)$_n$, (MgO)$_m$ and (Mg)$_m$ are schematically represented in Figure 1 for $n=1-5$ and $m=1-4$, respectively. 

\subsubsection{Silica}
In space, the 20 \micron\ band of SiO$_2$ was detected in Herbig Ae/Be stars with ISO \citep{bou01}, while the modeling of Cas A IR spectra from Spitzer suggests  SiO$_2$ as a potential dust candidate \citep{rho08}. In the laboratory, silica clusters form following two routes depending on the hydrogen content of the gas from which they nucleate. Where H is present, \citet{zac94} report the importance of the OH hydroxyl radical in the formation of silica clusters in H/O/Ar-rich flames fueled with silane (SiH$_4$), using Laser-Induce Fluorescence techniques to characterize the gas phase species entering in the conversion of gas to particle processes. Indeed, OH triggers the formation of gaseous SiO$_2$ through the reactions (see CD09 for more detail)
\begin{equation}
\label{sio1}
SiO + OH \longrightarrow SiO_2 + H,
\end{equation}
\begin{equation}
\label{sio2}
SiO + H_2O \longrightarrow SiO_2 + H_2,
 \end{equation}  
and consequent cluster growth occurs from the SiO$_2$ polymerization reaction
\begin{equation}
\label{sio3}
SiO_2 + (SiO_2)_n \longrightarrow (SiO_{2})_{n+1}.
\end{equation}
where $1 \le n \le 4$. 

When the gas is hydrogen-poor, the nucleation of silica occurs through direct polymerization of SiO to form small (SiO)$_n$ molecular clusters, precursors to solid silicon monoxide SiO. At temperatures $\ge$ 900 K, solid SiO is unstable and directly decomposes to silica SiO$_2$ and solid Si.  We thus consider that half of the total number of (SiO)$_5$ particles will be in the form of (SiO$_2$)$_5$, whereas the other half will be in solid (Si)$_5$. It is not clear wether this disproportionation effect gives rise to two separated condensate populations of SiO$_2$ and pure Si or to solids made of pure Si clusters embedded in an amorphous silica matrix \citep{mam01}. We neglect this aspect in the present study and consider that solid SiO gives rise to the formation of two types of solids, whatever their final form. SiO is first synthesized from the Si oxidation reaction (see CD09 for more detail)
\begin{equation}
\label{sio4}
Si + O_2 \longrightarrow SiO + O,
\end{equation}
by reaction with CO
\begin{equation}
\label{sio5}
Si + CO \longrightarrow SiO + C,
\end{equation}
and the radiative association reaction 
\begin{equation}
\label{sio6}
Si + O \longrightarrow SiO + h\nu,
\end{equation}
followed by the polymerization of SiO
\begin{equation}
\label{sio7}
SiO + (SiO)_n \longrightarrow (SiO)_{n+1}.
\end{equation}
with $1 \le n\le 4$. In pyrolysis experiments, reaction \ref{sio7} usually happens at almost collision rate and is at or near high-pressure limit at room temperature. However, as the temperature increases, the rate decreases due to the onset of energy transfer effects. The rate is also highly pressure-dependent and is estimated by \citet[hereafter ZT93]{zac93} for the high temperatures encountered in the combustion flames and applicable to SN ejecta. For our ejecta, we scale the rates derived by ZT93 for 1 atm pressure according to the total number of collisions per cm$^{-3}$ for the ejecta pressure under consideration. The resulting rates for the polymerization reactions of type reaction \ref{sio7} are a factor $\sim$ 100 smaller than those of ZT93 at 1 atm pressure. 
We model both the formation of (SiO$_2$)$_n$  and (SiO)$_m$ up to $n=4$ and $m=5$. The most stable form of (SiO$_2$)$_4$ has a rhombus chain structure with double oxygen bridged whereas the most stable structure of (SiO)$_4$ is a buckled eight-membered ring and that of (SiO)$_5$ has a double ring structure \citep{lu03}. 

The destruction process for those silica clusters is the thermal fragmentation reaction

\begin{equation}
\label{sio8}
(SiO)_{n} + M \longrightarrow (SiO)_{m} + (SIO)_{n-m} + M,
\end{equation}
where M represents the ambient gas, and $2 \le m <  5$. Theoretical estimation of dissociation energies for silica clusters by \citet{lu03} shows that the most common fragmentation products are gas-phase SiO, (SiO)$_2$ and (SiO)$_3$, as found in experiments on silica cluster material. This result implies that those species must be rather stable in the gas phase, and for the sake of simplicity, we assume a rate for reaction \ref{sio8} equals to the rate for the thermal fragmentation of SiO (see Table 9 in CD09).   

\subsubsection{Iron Oxide}

Iron oxide has been proposed as a potential contributor to the 23 \micron\ bands in young stars \citep{bou00}, in the binary AFGL 4106 system \citep{mol99}, and in Cas A \citep{rho08}. The reaction of iron with oxygen to generate iron oxides is at the basis of fundamental processes in environmental corrosion, biological oxygen transport or oxide film formation.  Of special interest to the present study is the gas phase iron monoxide formation reaction 
\begin{equation}
\label{feo1}
Fe + O_2 \longrightarrow FeO + O,
\end{equation}
whose rate was measured at high temperatures by \citet{fon73} and \citet{ak88}. The following reaction with carbon dioxide also occurs
\begin{equation}
\label{feo2}
Fe + CO_2 \longrightarrow FeO + CO,
\end{equation}
and its rate was measured by \citet{smi08}. Finally, we consider destruction of iron monoxide through thermal fragmentation and its reaction with atomic oxygen
\begin{equation}
\label{feo3}
FeO + O  \longrightarrow Fe + O_2,
\end{equation}
for which a rate was derived by \citet{sel03} in their study of iron oxide formation in the Earth upper mesosphere. The thermodynamically stable neutral clusters resulting from laser ablation of an iron rod bathed in an oxidizing gas are of the form (FeO)$_m$ for $m \leq 10$ \citep{shi04}. We thus define the nucleation mechanism to small (FeO)$_m$ clusters in the gas phase by successive addition of the FeO monomer 
\begin{equation}
\label{feo4}
FeO + (FeO)_n \longrightarrow (FeO)_{n+1},
\end{equation}
with $1 \le n \le 3$. The rate for this reaction is chosen similar to that of reaction \ref{sio7}. Destruction of small iron oxide clusters proceeds through thermal fragmentation such as reaction \ref{sio8}. Iron oxide clusters are characterized by the following ground state structures: linear for FeO, rhombus for (FeO)$_2$, triangular for (FeO)$_3$ and tetrahedral for (FeO)$_4$ \citep{wan96,rei07}. 

\subsubsection{Magnesium oxide}

Magnesium oxide MgO (periclase) was proposed as a component of the dust around the low-redshift quasar PG 2112+059 for its contribution to the 20 \micron\ spectrum region \citep{mar07} and was suggested as potential seeds to silicate condensation in O-rich AGBs \citep{bha07}. The nucleation of  magnesium oxides involves the formation of $MgO$ molecules in the first place through reaction
\begin{equation}
\label{mgo1}
Mg + O_2 \longrightarrow MgO + O,
\end{equation}
and 
\begin{equation}
\label{mgo2}
Mg + CO_2 \longrightarrow MgO + CO,
\end{equation}
and MgO destruction by thermal fragmentation and reaction with atomic oxygen 
\begin{equation}
\label{mgo3}
MgO + O \longrightarrow Mg + O_2.
\end{equation}
The rate for reaction \ref{mgo1} was estimated by \citet{kas82} whereas those of reactions \ref{mgo2} and \ref{mgo3} were assumed equal to the equivalent reactions involving atomic Fe.  We follow the formation of small magnesium oxide clusters $(MgO)_n$ with $n=1-4$. As seen from Figure 1, the $(MgO)_2$ and $(MgO)_3$ clusters have a rhombus and triangular structures, respectively, whereas the lowest energy structure of $(MgO)_4$ is cubic \citep{kho97, bha07}. We therefore describe MgO cluster formation by the direct polymerization of MgO according to reaction
\begin{equation}
\label{mgo4}
MgO + (MgO)_n \longrightarrow (MgO)_{n+1},
\end{equation}
where a rate similar to that of reaction \ref{sio7} has been assumed. Small periclase clusters are destroyed according to the thermal fragmentation reactions similar to reaction \ref{sio8}. 

\subsubsection{Alumina}

Alumina Al$_2$O$_3$ has been detected through its 11 \micron\ band in evolved, oxygen-rich AGB stars \citep{ste90,cam02} and was found in pre-solar meteorites with an isotopic composition pointing to its O-rich AGB stellar origin \citep{nit97}. Models of the mid-IR excess of the low-redshift quasar PG 2112+059  \citep{mar07}, the Blue Luminous Variable Eta Carinae \citep{che05}, and the SNR Cas A \citep{rho08} also use alumina as a potential dust component. In the laborarory, laser-induced aluminum evaporation in an helium bath seeded with oxygen produces small aluminum oxide clusters \citep{de96}. Small cluster molecules like Al$_2$O$_x$ ($x=3-5$) show a rhombus structure, including Al$_2$O$_2$ rings. The chemical pathway to those small clusters involves the formation of the molecule AlO, and its further reaction with oxygen and itself \citep{ca03}. In our SN ejecta, we only consider the formation of AlO as aluminum has an initial low mass abundance  compared to other metals reacting with oxygen (see Tables 2 and 3 of CD09). Furthermore, the formation routes to alumina are still very speculative and we prefer deriving AlO mass yields as a first indication of the alumina content of the ejecta. AlO synthesis occurs through the following reactions
\begin{equation}
\label{alo1}
Al + O_2 \longrightarrow AlO + O,
\end{equation}
\begin{equation}
\label{alo2}
Al+ O + M \longrightarrow AlO + M,
\end{equation}
\begin{equation}
\label{alo3}
Al + CO_2 \longrightarrow AlO + CO,
\end{equation}
and finally, in the presence of hydrogen,
\begin{equation}
\label{alo4}
Al+ H_2O \longrightarrow AlO + OH.
\end{equation}
Reaction rates for reactions \ref{alo1}, \ref{alo2}, \ref{alo3}, and \ref{alo4} were derived by \citet{gar92}, \citet{ca03}, and \citet{mcc93}. Destruction process for gas-phase AlO molecules are thermal fragmentation by the ambient gas, collisions with Compton electron,  and reaction with He$^+$.

\subsection{Metal sulphides}
The metal sulphides MgS and FeS have been proposed as carriers for several IR bands observed in evolved, carbon-rich, circumstellar environments. Specifically, MgS is most likely responsible for the broad 30 \micron~emission band observed in carbon stars, proto-planetary and planetary nebulae \citep{beg94,vol02}, while FeS has been proposed as the carrier of the 23 \micron~feature observed in two planetary nebulae with ISO \citep{ho03}. On Earth, the iron sulphide clusters (FeS)$_2$ and (FeS)$_4$ act as evolutionary ancient prosthetic groups in sustaining fundamental life processes such as biological electron transport in proteins \citep{jo05}.

As no reaction between metals and sulphur is documented, we use the S and O isovalence to assess reaction rates for metal sulphide formation. Following reaction \ref{mgo1}, MgS formation starts with reactions
\begin{equation}
\label{mgs1}
Mg+ S_2 \longrightarrow MgS + S,
\end{equation}
and 
\begin{equation}
\label{mgs2}
Mg+ SO \longrightarrow MgS + O.
\end{equation}
to both of which we assign the rate of reaction \ref{mgo1}. Destruction of MgS occurs via thermal fragmentation and the reverse processes of reactions \ref{mgs1} and \ref{mgs2} with assigned rates similar to those for reaction \ref{mgo3}. Finally the clustering of MgS proceeds via the reaction
\begin{equation}
\label{mgs3}
MgS + (MgS)_n \longrightarrow (MgS)_{n+1},
\end{equation}
where a rate similar to that of reaction \ref{sio7} has been assumed. The ground state structures of MgS clusters are as follows: a rhombus structure for (MgS)$_2$, a planar triangular structure for (MgS)$_3$, and a distorted cubic structure for (MgS)$_4$. 

Similar reactions and rates are considered for the growth of troilite FeS clusters. Formation processes are 
\begin{equation}
\label{fes1}
Fe+ S_2 \longrightarrow FeS + S,
\end{equation}
and 
\begin{equation}
\label{fes2}
Fe+ SO \longrightarrow FeS + O,
\end{equation}
followed by
\begin{equation}
\label{fes3}
FeS + (FeS)_n \longrightarrow (FeS)_{n+1}.
\end{equation}
Destruction of FeS occurs via thermal fragmentation and the reverse processes of reactions \ref{fes1} and \ref{fes2}. Clusters of FeS have the following structures: rhombic for (FeS)$_2$, planar and triangular for (FeS)$_3$, and distorted cubic for (FeS)$_4$. 

We do not consider the formation of solid SiS in the present study, although gaseous SiS being isovalent to SiO, similar nucleation trends to (SiO)$_n$ synthesis could be expected. Solid SiS is known to disproportionate into solid SiS$_2$ and solid Si upon annealing at temperatures greater than 900 K \citep{by73}. SiS$_2$ solid was proposed to explain the 21 \micron\ emission band observed in several carbon-rich Proto-Planetary Nebulae \citep{go93,be96}, although the identification is still controversial \citep{zha09}. However, the large abundances of gaseous SiS predicted to form in SNe ejecta by CD09 could spawn the synthesis of solid silicon disulphide in SNe. 

\subsection{Pure Metal Clusters}

There exists no observational evidence for the existence of pure metal clusters in evolved circumstellar environments. However, iron clusters have been proposed as a component of dust in O-rich AGB stars to explain the required near-IR opacity \citep{ke02}. They were also included in dust models aiming at reproducing the IR excess of Cas A observed by Spitzer \citep{rho08}. Pure iron, magnesium and silicon clusters are easily formed in the laboratory. In particular, Fe nanoparticles gas-phase synthesis from iron pentacarbonyl Fe(CO)$_5$ decomposition has been extensively studied as iron particles are important catalysts in the formation of carbon nanotubes  \citep{we07}. The Fe small cluster nucleation proceeds according to the following steps
\begin{equation}
\label{f1}
Fe+ (Fe)_n+ M \longrightarrow (Fe)_{n+1} + M,
\end{equation}
where $n=1,3$. The coagulation of small clusters is also considered 
\begin{eqnarray}
\label{f2}
Fe_2+ Fe_2 &\longrightarrow & Fe_4 \\ \nonumber
& \longrightarrow & Fe_3 + Fe,
 \end{eqnarray}       
when destruction occurs through the thermal fragmentation reaction 
\begin{equation}
\label{f3}
Fe_n+ M \longrightarrow Fe_{n-1} + M.
\end{equation}
The rates for reactions \ref{f1}, \ref{f2}, and \ref{f3} have all been estimated by \citep{gie03}. Similar processes and rates have been assumed for the nucleation of pure Si and Mg clusters. 

 \subsection{Carbon-Rich Clusters}

Along with silicates, carbon-rich clusters like amorphous carbon (AC) and silicon carbide (SiC) are ubiquitous condensates in space. In the present study, we consider the nucleation of both AC and SiC clusters in primordial SN ejecta. 

\subsubsection{Amorphous Carbon}

The main providers of AC grains in galaxies are carbon-rich AGB stars in the late stages of their evolution, but other formation loci include RCrB stars, the colliding-wind region of carbon-rich Wolf-Rayet binary systems, and SN ejecta. Whenever the carbon content of a gaseous medium is larger than that of oxygen, amorphous carbon dust may condense. The nucleation to solid clusters involves different carbon routes dependent of the temperature and the hydrogen content of the gas \citep{cherc09}. When hydrogen is present and for the low temperatures encountered in premixed acetylene flames ($T_{gas} \sim 1500$ K),  unsaturated hydrocarbons form, among which acetylene C$_2$H$_2$ and its isomeric radical vinylidene (HHCC), the ethynyl radical C$_2$H, the propargyl radical C$_3$H$_3$, and vinylacetylene $C_4H_4$. The formation of the planar, aromatic ring benzene $C_6H_6$ results from successive additions of acetylene and its isomer, or the direct reaction of two propargyl radicals  \citep{mi02}. Its further growth to Polycyclic Aromatic Hydrocarbons (PAHs) occurs through successive addition of acetylene on aromatic radical sites  \citep{fre89,cher292,ri02}. 

When the environment is hot ($T_{gas} \sim 4000$ K) and hydrogen-free, the nucleation of carbon clusters involves the synthesis of pure carbon chains of sp hybridization up to C$_9$. For C$_{10}$, the most stable structure is ringed and the closure of the chain to large monocyclic rings occurs. The coalescence of such rings in bicyclic and tricyclic rings as precursors to fullerene formation is observed in pulsed laser vaporization experiments \citep{hel93,hun93} and the condensate is in the form of fullerene-like soot \citep{ja09}. 

For the purpose of this study, we explore the kinetics of the formation of carbon chains up to $C_9$ and the closure to the monocyclic ring $C_{10}$. As our ejecta are considered hydrogen-free as hydrogen is not microscopically mixed with other constituents, we disregard the formation routes to aromatics involving hydrocarbons. Our kinetic network for chain formation includes the radiative association reactions of the type \citep{clay99,clay01,den06}.
\begin{equation}
\label{c1}
C + C_n \longrightarrow C_{n+1} + h\nu,
\end{equation}

with $n=1-9$. For $n=1$, \citet{an97} derived a low associative rate ($k_{\ref{c1}}= 1\times 10^{-17}$ cm$^3$ s$^{-1}$) whereas the rate for larger chain synthesis is assumed much faster ($k_{\ref{c1}}= 1\times 10^{-10}$ cm$^3$ s$^{-1}$) as larger chains are able to quickly stabilize their formation complex owing to their numerous vibrational degrees of freedom \citep{clay99}. An additional growth process for chains is their mutual collision and binding by edge carbon atoms. According to \citet{wei95}, such mechanisms have energy barriers less than 0.5 eV.  The thermodynamics of the addition reactions are estimated using thermodynamic data for small carbon chains calculated by \citet{clay01} and we find that these additions are all exothermic processes. Therefore, we include the additional growth pathways to chain synthesis

\begin{equation}
\label{c2}
C_{n-m} + C_m \longrightarrow C_{n},
\end{equation}
with $m=2-7$. The rate for reaction \ref{c2} is assumed to be collisional with an estimated value of $1\times 10^{-10}$ cm$^3$ s$^{-1}$.

The dominant destruction channels of carbon clusters included in the model are firstly, the thermal fragmentation of the chains through collision with the ambient gas
\begin{equation}
\label{c3}
C_n + M \longrightarrow C_{n-m} + C_m + M,
\end{equation}
secondly,  the oxidation of the chains through reaction of atomic oxygen with end-cap carbons to form CO
\begin{equation}
\label{c4}
O + C_n \longrightarrow C_{n-1} + CO,
\end{equation}
thirdly, the collision with Compton electrons
\begin{equation}
\label{c5}
C_n + e^-_{Compton} \longrightarrow C_{n-m} + C_m + e^-_{Compton}, 
\end{equation}
and fourthly, the destruction by He$^+$
\begin{equation}
\label{c6}
C_n + He^+ \longrightarrow C_{n-1} + C^+ + He. 
\end{equation}

The rates for reactions \ref{c3} are those used for $C_2$ in CD09, while the 
rate for reactions \ref{c4} is that of \citet{clay99,clay01}, and finally, the rate for reactions \ref{c5} is derived following the formalism explained in CD09.  

\subsubsection{Silicon Carbide}

Silicon carbide grains are ubiquitous in various evolved circumstellar environments \citep{mol03}. In AGB stars, the SiC 11.3 \micron\ band is observed either in emission or absorption for low and high mass loss rates, respectively \citep{sp97}, but there is no detection of this band in SN ejecta to date. Presolar SiC grains bearing the isotopic signatures pertaining to SNe are however identified in meteorites \citep{ber87}. In SN ejecta, the formation of SiC grains implies the mixing of Si- and C-rich non-adjacent layers, a situation not found in the dust formation layers of carbon stars where SiO and C-bearing chemical species are simultaneously present in large quantities \citep{wi98}. In the laboratory, SiC whiskers and nanorods are produced by different methods from vaporization of SiC bulk material, to the surface reaction of gas-phase SiO on carbon films or nanoclusters, or the gas phase formation from chlorosilanes, carbon monoxide, and methane \citep{zhou94}. In the present study, we consider the formation of SiC and (SiC)$_2$ according to the following reactions:

\begin{equation}
\label{sic1}
Si + CO \longrightarrow SiC +O,
\end{equation}
\begin{equation}
\label{sic2}
Si + C_2 \longrightarrow SiC +C,
\end{equation}
\begin{equation}
\label{sic3}
C + SiO \longrightarrow SiC +O,
\end{equation}
and 
\begin{equation}
\label{sic4}
C + Si \longrightarrow SiC + h\nu,
\end{equation}
followed by the nucleation reaction
\begin{equation}
\label{sic5}
SiC + SiC \longrightarrow (SiC)_2. 
\end{equation}

The rate for reaction \ref{sic5} is assumed to proceed with the rate of reaction \ref{sio7}. The (SiC)$_2$ cluster has a favored rhombic structure \citep{ya06} similar to that of (SiO)$_2$ in Figure 1, but a linear structure is also possible. The destruction channels for SiC are reactions with He$^+$, the reverse processes of reactions \ref{sic1}-\ref{sic3}, and thermal fragmentation, while the latter is the only destruction channel considered for (SiC)$_2$.  

\subsection{From Clusters to Dust: Considering Fe and Mg depletion}

The condensation stage in dust synthesis involves the coagulation of clusters and the growth of seeds through additional adsorption and chemical reactions on the grain surface. Those processes are extremely complex and their description is out of the scope of this paper. However, we can rely on useful information coming from the synthesis of smokes in the laboratory to make prescriptions for the dust formed from the various types of clusters synthesized in SN ejecta. We will adopt the following approach for the derivation of the dust upper limits in the rest of this paper. We may encountered situations where large amounts of (SiO)$_5$ rings are synthesized in a medium rich in atomic Mg and/or Fe. These SiO rings will lead to the formation of metastable solid silicon monoxide which disproportionates into a mixed phase of pure Si nano-crystals embedded in amorphous silica clusters at temperatures larger than $\sim$ 900 K \citep{fu99,kap05}. According to \S 2.1, such a rich Mg- or Fe-SiO$_2$/Si mixture does not lead to the formation of ferromagnesiosilica, but to separated populations of metastable ferrosilica or magnesiosilica compounds \citep{rie99,rie02,kai03,kam05}. Furthermore, the formation of additional magnesium silicide Mg$_2$Si and periclase MgO clusters is observed in the Mg-rich zone of the smoke resulting from the combined evaporation of Mg and SiO solids whereas pure Si and forsterite Mg$_2$SiO$_4$ clusters are observed in the Si-rich zone \citep{kai03,kim09}. 

In view of the complexity of the products and because the exact composition of those metastable compounds is not truly characterized, we here consider two simple cases, once the nucleation of molecular clusters has been completed: (1) a case for which all pure Fe and/or Mg atoms and clusters are depleted in the dust condensation process leading to either Mg$_2$SiO$_4$ (forsterite) or Fe$_2$SiO$_4$ (fayalite) dust;  (2) a case for which no Fe/Mg depletion occurs. For the sake of simplicity, we ignore the possible formation of magnesium silicide Mg$_2$Si and pyroxene compounds like enstatite MgSiO$_3$ and ferrosilite FeSiO$_3$. The former case is somehow extreme assuming a depletion efficiency during the grain formation process of 1 for the metals Fe and/or Mg, whereas the latter may depict a situation more in tune with the ejecta of local SNe. Indeed, in the CCSN SN1987A, optical lines of [MgI] and [OI] were observed to decline with time according to the ejecta optical depth whereas the [SiI] line was observed to have a sharp decline owing to Si depletion at the onset of dust formation \citep{luc89,wood93}. \citet{meik93} also observed a sharp decline in the 1.2 \micron~and 1.645 \micron~[SiI] lines at the onset of dust formation compared to the 1.257 \micron~[FeII] line. Observations thus indicate that Mg and Fe are not as depleted as Si, and it should be borne in mind that reality probably lies within these two extreme cases for metal depletion. We then derive upper limits for Mg$_2$SiO$_4$ and Fe$_2$SiO$_4$ according to these two cases. To satisfy stochiometry and form forsterite and fayalite and because of the large dioxygen ejecta content at late times after explosion, we assume that the magnesiosilica and ferrosilica pre-seeds will be oxidized through O$_2$ adsorption and decomposition on the grain surfaces at the low temperatures encountered at day 1000 \citep{ov09}. This is a plausible mechanism for oxidation as O$_2$ forms at $t > 700$ days after the decrease of He$^+$ abundance as seen in CD09 and \S 4. The oxidation process will partly deplete the O$_2$ reservoir, resulting in smaller O$_2$ mass yields ejected at day 1000 when metal depletion is considered.  In case no metal depletion and formation of forsterite or fayalite are assumed, the major SiO-based solid in the ejecta is silica SiO$_2$. 

\section{THE EJECTA MODELS}

For the present study, we pay attention to two Pop.~III  progenitor masses, a 170~\Ms~PISN, and a 20~\Ms~CCSN. We consider two mixing conditions: (1) a fully mixed ejecta in which all elements have been microscopically-mixed in the helium core; (2) an unmixed, stratified ejecta in which each layer elemental composition of the helium core reflects the prior nucleosynthesis stages of the progenitor, except for the inner most mass zone whose composition is the result of explosive nucleosynthesis and radioactive decay. None of these scenarios is fully satisfactory in term of ejecta description, but a stratified ejecta made of fully microscopically-mixed zones of various chemical compositions seems more appropriate to describe the chemistry of the gas filaments and clumps resulting from the explosion. However, we consider fully microscopically-mixed ejecta as those are considered in most of the existing studies. Mixing is a key parameter to the ejecta chemistry and we will discuss its impact on dust formation in \S 5, in the light of 2D/3D explosion models \citep{fry02,kif03,kif06,ham09}. A complete description of the ejecta models  is given in CD09, in which full details on the physics and chemistry at play and the model parameters used in the present study can be found.

As a brief summary, the gas parameters like temperature and number density are derived from the study of \citet[hereafter NK03]{noz03} for both mixed and unmixed ejecta, and include the radioactive decay of \Ni, \Co,~and \Fe~which induces the creation of a population of Compton electrons in the gas through \grays~down-scattering. The impact of the ultra violet (UV) radiation field resulting from  this \grays~degradation on molecules and dust clusters  was assessed by CD09 who found that the destruction of molecules and dust precursors by this UV radiation field was not important. We thus ignore UV radiation for the rest of the present study. The chemical compositions for fully microscopically-mixed ejecta are those of \citet[hereafter UN02]{um02} when elemental mass yields for unmixed ejecta are taken from NK03. The chemical compositions of our present models and the zoning considered in the unmixed ejecta are given in Tables 4 and 5 of CD09. 

The gas parameters of fully mixed ejecta over the timespan of interest (100 to 1000 days) are summarized in Table 3 for the two progenitors under study. For unmixed ejecta, the temperature variation with time is that of Table 1 for each mass zone of the helium core whereas the initial gas number density at 100 days shows small changes due to the different initial molecular weight characterizing the chemical composition of each mass zone in the ejecta. 

\section{RESULTS}
We now report  on cluster abundances and masses for the 170~\Ms~and the 20~\Ms~fully microscopically-mixed ejecta in \S 4.1. We also present a critical assessment of the various existing studies for fully-mixed ejecta to highlight some erroneous hypothesis assumed in certain models and the major problems encountered when using a CNT approach. Results for unmixed ejecta are presented in \S4.2.

\subsection{Fully-Mixed Ejecta}

As mentioned before, fully microscopically-mixed ejecta do not represent realistic cases of post-explosion SN environments as both theoretical explosion models \citep{mul91,kif03,ham09} and observations of SN remnants \citep{dou01} show evidence for the formation of inhomogeneities in the form of clumps and filaments in the ejecta. Because most of the existing studies on dust formation in SNe consider fully microscopically-mixed ejecta, and in order to compare and test  the CNT and the kinetic approach, we present the results for such ejecta in Figure 2 for the 170~\Ms\ progenitor. The abundance variation with time for silica precursors and metal-bearing clusters are shown along with the abundances for the main atoms, ions, and secondary electrons. As already stressed in \S2.1.1, silicon monoxide clusters (SiO)$_{n}$ will decompose at $T > 900$ K to $1 \over {2}$(SiO$_2$)$_n$ and $1 \over {2}$(Si)$_n$ clusters and can thus be considered as silica precursors. Until $\sim$ 250 days, some atomic silicon is converted to SiO and SiS, building up a reservoir of SiO (see CD09 for more detail). For t $>$ 250 days, the simultaneous and quick conversion of SiO to (SiO)$_2$ and its further implication in the growth of larger clusters through the polymerization reaction \ref{sio7} lead to the quick depletion of SiO and Si. The shape of the SiO abundance distribution is somewhat different than that presented in Figure 3 of CD09, and reflects the fact that in the present paper, the synthesis of silica precursors up to (SiO)$_5$ is followed whereas the growth up to (SiO)$_3$ was studied in CD09.  Silica precursors of smaller size than (SiO)$_5$, including SiO, are converted to larger clusters but also replenished by the thermal fragmentation of (SiO)$_5$ reaction \ref{sio8}, thus keeping a bell-shaped abundance distribution until $\sim$ 700~days when Si is depleted.  Hence, via SiO formation, atomic Si is very efficiently and totally converted to (SiO)$_5$ at $t>$ 700~days. 

Metal clusters form at late times, as illustrated in Figure 2b. Atomic iron, magnesium and aluminum are first destroyed by ionization to form  Fe$^+$, Mg$^+$, and Al$^+$ ions before 200 days but recombine afterwards, and their abundance gradually reaches steady state, as opposed to atomic Si. The formation of pure magnesium and iron clusters is delayed to $t > 700$~days and is driven by the formation and destruction of (Mg)$_2$ and (Fe)$_2$ clusters, respectively (see reactions \ref{f1}, \ref{f2}, and \ref{f3}). Indeed, the thermal decomposition of (Mg)$_2$ and (Fe)$_2$ is very efficient for temperatures in excess of 1000 K as derived by \citet{gie03}, but the decomposition reaction rate decreases dramatically for $T<1000$~K, or $t>$700 days, leading to the build-up of pure magnesium and iron clusters as observed in Figure 2b.  

We also see that the 170 \Ms, fully-mixed ejecta does not synthesize metal oxide (FeO)$_4$ and (MgO)$_4$ clusters. The dominant formation process of FeO and MgO molecules is via reactions \ref{feo1} and \ref{mgo1}, as O$_2$ abundance sharply increases at $t >$ 800~days. However, iron and magnesium oxidation competes with the fast (Fe)$_{2,3,4}$ formation process reaction \ref{f1} and the formation of iron and magnesium oxides is therefore hampered. A similar competition between processes applies to the formation of metal sulphide clusters (FeS)$_4$ and  (MgS)$_4$ clusters. The outcomes are very low abundances for metal oxides and sulphides.  As to alumina precursor AlO, the destruction by He$^+$ precludes AlO to form at early times. However, He$^+$ abundances start decreasing at $ t \sim 500$~days  owing to the decrease with time of the ionization rate of He by Compton electrons (CD09). These lower He$^+$ abundances trigger the formation of O$_2$ and the subsequent formation of AlO through reaction \ref{alo1} at $t \ge 800$~days. 

The derived cluster masses ejected at day 1000 are summarized in Table 4. (SiO)$_5$ clusters are by far the most abundant with a total mass of $\sim$ 25.4 \Ms, and decompose in 17.3 \Ms\ of (SiO)$_5$ and 8.1 \Ms\ of (Si)$_5$ clusters, followed by small amounts of pure magnesium, (Mg)$_4$, and iron, (Fe)$_4$, clusters and aluminum oxide molecules, AlO. The total mass yield for clusters is $\sim$ 25 \Ms, corresponding to 15\% of the progenitor mass. We notice that no carbon chains and rings, and no silicon carbide (SiC)$_2$ clusters form in this fully-mixed ejecta as atomic carbon is rapidly integrated into the  synthesis of CO, and CS to a less extent (see CD09).  

Clusters abundances are shown in Figure 3 for the 20~\Ms~progenitor. Chemical processes similar to those at play in the ejecta of the 170 \Ms\ progenitor are responsible for the nucleation of clusters. However, the initial chemical abundance of the ejecta at 100 days post-explosion is helium and carbon-rich, but oxygen and silicon-poor compared to that of the 170 \Ms\ progenitor (see Table 2 from CD09 for detail). The high helium content impedes the early nucleation of silicon monoxide and pure metal clusters due to the large He$^+$ abundances in the ejecta. The formation of silicon monoxide and pure Si, Mg, and Fe clusters is thus postponed to $t > 750$ days when the He$^+$ abundance decreases. The lower gas density at these late times combined to lower initial mass fractions for Si and O result in smaller cluster masses synthesized in the ejecta, as illustrated in Table 4. Indeed, the total cluster mass equals $\sim$ 0.16 \Ms\ and corresponds to 0.8 \% of the progenitor mass only. Thus, fully-mixed ejecta of primordial CCSNe are less efficient at nucleating solid clusters than PISNe due to their initial He-rich/metal-poor chemical composition. 

Prescriptions for the dust composition for the two mass progenitors under study are presented and discussed in \S 4.2.1.  

\subsection{Testing Existing Studies}
The formation of dust at high redshift in SN fully-mixed ejecta has been first modeled for zero-metallicity CCSNe by the pioneering study of \citet[hereafter TF01]{tod01}. They described the condensation of dust grains using CNT and found that a 20 \Ms~progenitor produced $\sim 0.09$ \Ms\ of amorphous carbon dust. This formalism was further used by \citet[hereafter SFS04]{sch04} to describe the formation of solids in PISNe, and by \citet{bian07} to study the production of dust in local CCSNe and its survival to the passage of the reverse shock. Finally, \citet{val09} used TF01 dust results in a comparative study of dust main providers in the far and local universe. Another study on primordial SNe was carried out by NK03 and considered both fully-mixed and unmixed ejecta of CCSNe and PISNe. 

Firstly, as the ejecta composition, mixing, and physical parameters of NK03 are used in CD09 and the present paper, we compared our results with those of NK03. Secondly, we test the validity of TF01 results in the framework of the chemical kinetic approach as TF01 results are used as a benchmark in many subsequent studies. Thirdly, we test the study of SFS04 in a similar context.

\subsubsection{The fully-mixed models by Nozawa et al. (2003)}

A direct comparison between results for our fully-mixed ejecta presented in \S 4.1 and those of NK03 is possible as NK03 ejecta models were used in CD09 and in the present study. NK03 results for the fully-mixed ejecta of the 170 \Ms\ and  20 \Ms\ progenitors are listed in Table 5 and compared to the present results. Despite identical initial conditions for the ejecta chemical composition and similar ejecta gas parameters, the present upper limits for the dust yields are lower than values derived by NK03, with a greater discrepancy (a factor of 2 to 5) for the 20 \Ms\ progenitor.  We also see that the chemical composition of the solids formed is totally different. The present models predict the formation of silica and pure silicon dust as the prominant condensates, along with alumina and pure magnesium and iron grains to a less extent. When full depletion of Fe and Mg is considered, fayalite and forsterite do also form. On the other hand, NK03 predict the formation of silica, forsterite/entastite (MgSiO$_3$), magnetite (Fe$_3$O$_4$) and alumina. In their model, iron is trapped in magnetite whereas it is in the form of pure iron clusters in the present model. Also, NK03 do not form pure magnesium grains or pure silicon grains resulting from the disproportionation of silicon monoxide clusters at high temperatures. The only agreement we find regarding the dust composition is related to the lack of carbon and silicon carbide dust in the ejecta. NK03 assume that the whole atomic C is quickly transformed into CO, and we confirm from chemical kinetics that this is indeed the case for the fully microscopically mixed ejecta, whatever the initial mass of the progenitor. 

More differences arise in the condensation sequence derived by both studies. NK03 find for both cases that the condensation of alumina first proceeds, followed by forsterite/entastite, silica, and magnetite. For their 170 \Ms\ progenitor, Al$_2$O$_3$ condenses at $\sim$ 500 days, followed by Mg$_2$SiO$_4$ and MgSiO$_3$ at $\sim$ 520 days, silica at $\sim$ 550 days, and finally Fe$_3$O$_4$ at $\sim$ 550 days. These various post-explosion times correspond to epochs when the nucleation rates of the various solids are maximum. However, as mentioned before, these nucleation rates are expressed as a function of physical quantities inappropriate to the description of molecular-size clusters, and the relevance of those condensation sequences have to be seriously questioned. Inspection of Figures 2 and 3 show different formation times for our cluster precursors, driven by chemical kinetics. In particular, the high initial silicon yields characterizing the 170 \Ms\ progenitor foster a rapid formation of silicon monoxide clusters (hence silica and pure Si precursors) at high temperatures as soon as 400 days, comparable to what is observed in ceramic-forming flames \citep{zac94}. The pure metallic clusters (Mg)$_4$ and (Fe)$_4$ on the other hand are forming at $t > 700$~days for the reasons explained in \S 4.1. Finally, AlO forms at day 900. Those strong discrepancies in formation times reflect the fact that condensation from the gas phase is controlled by non-equilibrium chemically-controlled nucleation processes, each of which has its own temperature dependence. 

For their 20 \Ms\ progenitor, NK03 find that Al$_2$O$_3$ condenses first at $\sim$ 380 days, followed by Mg$_2$SiO$_4$ and MgSiO$_3$ at $\sim$ 400 days, silica at $\sim$ 420 days, and finally Fe$_3$O$_4$ at $\sim$ 430 days. This condensation sequence is similar to that for their PISN case though solid formation proceeds at slightly earlier times owing to the smaller ejecta temperatures. In the present study, nucleation times are totally different for the PISN and the CCSN cases, reflecting different chemistries at play in the ejecta. As already stressed in \S 4.1, the large He$^+$ content of the 20 \Ms\ ejecta totally hampers cluster formation before day 700. Once again, these strong deviations are due to the inappropriate use of CNT applied to kinetically-commanded environments. 

\subsubsection{The CCSN model by Todini \& Ferrara (2001)}

We run our chemical network including the formation of molecules and nucleation of clusters for the initial elemental yields of the 20~\Ms~progenitor given by Table 2 of TF01 and their specific profiles for the temperature and number density of this ejecta. The elemental composition used by TF01 is that derived from explosion models by \citet{woos95} for their zero metallicity (Z = 0) progenitors. From TF01 Table 2, we notice for their 20 \Ms\ (and 18 \Ms) progenitor: (1) anomalously very low yields for heavy elements like Si, Mg, S and Al; (2) a carbon-to-oxygen ratio (C/O) larger than 1 compared to values for slightly lower or larger progenitor masses. Such very low mass yields for refractory elements and a C/O $>$ 1 are not present in other explosion models like those of UN02 or the more recent calculations of \citet{heg08} for a similar progenitor mass. Moreover, in those last two studies and for the specific 20 \Ms\ progenitor mass, the elemental yields are found quite similar. Refractory elements like Si, Mg, or Al being all key players in the dust condensation processes, their paucity drastically changes the chemical nature of the dust that condenses. We also notice that the ejecta temperature profile used by TF01 is derived from comparison with the ejecta of SN1987A. The photospheric temperature is chosen equal to 5400~K and slowly decreases with time to reach 960~K at day 1000, as illustrated in Table 2. The ejecta temperature from 100 to 1000 days post-explosion derived by NK03 and used in the present study and in CD09 takes into account the deposition of energy by radioactive elements through the degradation of $\gamma$- and $X$-rays using a radiative transfer formalism coupled to the energy equation. As seen from Table 2, the resulting temperature variation with post-explosion time has a much sharper decrease with a higher initial temperature at day 100.  As to gas number density, TF01 mentioned a value of $n=1 \times 10^8$ cm$^{-3}$ for temperatures less than 6000 K but do not provide a specific number density profile for their 20 \Ms\ progenitor. We thus assume that the gas number density has the  constant value of $10^8$ cm$^{-3}$ over the timespan involved in our test study. 

The initial He/C-rich, Si/Mg-poor ejecta elemental composition and temperature profile used by TF01 imply that only carbon dust is expected to form if it {\it can} form. Indeed, they find that most of the dust formed for this specific 20~\Ms~progenitor is amorphous carbon with a AC dust mass yield of $\sim$ 0.09~\Ms. When using a similar gas elemental composition, temperature and density profiles for our 20~\Ms~fully-mixed ejecta, our results show that the only molecules which are synthesized in the ejecta are C$_2$ and CO, as illustrated in Figure 4. Their abundances with respect to the total gas number density are very low and comprised between $1 \times 10^{-13}$  and $1 \times 10^{-11}$ from day 100 to day 1000 due to the overwhelming presence of He$^+$. Indeed, the relatively high ejecta temperatures at day 300 and later fosters the ionization of He and impede He$^+$ recombination, hence sustaining a strong destruction pathway to molecular synthesis over the whole timespan under study. With the specific high helium yield in TF01 initial elemental composition, it is thus not surprising that molecule formation is hampered in this ejecta. As a consequence, no carbon chains, starting with the synthesis of C$_2$, are formed and released at day 1000 owing to their destruction by He$^+$, as seen in Figure 4. Our results are in total contradiction with TF01 carbon dust mass yield of $\sim$ 0.09~\Ms. This deviation is due to the fact that their study considers the direct conversion of the initial carbon content decreased by its depletion in forming CO at steady state, without considering the ejecta full chemistry and the overwhelming destruction of small carbon clusters by He$^+$. CD09 show that a non-steady state chemistry commands the evolution of the ejecta gas phase until 1000 days after explosion, and the present results illustrate again the importance of considering the ejecta chemistry and all chemical processes responsible for dust nucleation in any treatment of dust formation in SNe. 

In summary, we suggest that the high carbon dust yields derived by TF01 for fully-mixed ejecta of zero-metallicity 20 \Ms\ CCSNe should be used with caution for they have been derived for questionable nucleosynthesis initial conditions and using the CNT approach. 

\subsubsection{The PISN model by Schneider et al. (2004)}

Dust synthesis in Pop. III~massive SNe is studied by SFS04. The CNT approach is used to describe the condensation of solids and SiC grains are not included among the condensates considered. The initial chemical compositions resulting from explosion nucleosynthesis for stellar progenitors with mass ranging from 140 to 260 \Ms\ are those of \citet[hereafter HW02]{heg02}. Firstly, we notice that very low temperature profiles are obtained for the various massive progenitors considered, e.g., the 195 \Ms\ progenitor ejecta has temperatures decreasing from $\sim $ 5400 K to $\sim 50$ K for $100 \le  t \le 1000$ days, a somewhat similar T variation than that found in CCSN ejecta. Secondly, inspection of SFS04 Figure 3 shows that the ejecta temperatures decrease with increasing progenitor masses. Low temperature profiles are questionable for PISNe, which are characterized by much larger explosion energies and  \Ni\ mass synthesized in the explosion nucleosynthesis than CCSNe  (HW02, UN02). As a matter of fact, solving for the radiative transfer equation taking into account the energy deposition by radioactive elements, NK03 find higher ejecta temperatures for PISNe than for CCSNe. Secondly, more massive PISNe are characterized by larger central temperatures, explosion energies, and synthesized \Ni\ mass than their less massive counterparts \citep[HW02, UN02]{fry01}, normally resulting in higher ejecta temperatures due to the conversion of part of the explosion energy in thermal energy and the deposition of energy through \grays\  production from radioactive decay. As the condensation of solids in TF01 and SFS04 formalism is assumed to start when the nucleation rates reach their maximum values, the low ejecta temperatures permit the condensation of grains as early as 150 days post-explosion where the gas densities are high enough to provide large final dust yields. 

Results from SFS04 and the present modeling are summarized in Table 6. SFS04 find that the amount of CO formed in the ejecta under steady state is equal to $\sim 6$ \Ms\ and $0.01$ \Ms\ for the 170 and 260 \Ms\ PISNe, respectively. Their dust mixture is dominated by silica, forsterite and amorphous carbon. They derive a total dust yield of $\sim 39$ \Ms\ and $\sim 70$ \Ms\ for the 170 \Ms\ and 260 \Ms\ progenitors, respectively. According to Table 6, the dust synthesized by a 260 \Ms\ PISN also includes 8 \Ms\ of magnetite Fe$_3$O$_4$, due to the large initial iron mass yield in the model of HW02. 
We apply our chemical kinetic approach to the SFS04 170 \Ms\ and 260 \Ms\ PISN ejecta models with similar initial compositions. Firstly, we find that a large molecular phase is synthesized in the ejecta for both models even when O$_2$ depletion is considered in the derivation of the dust yield upper limits. For the 170 \Ms\ progenitor, the molecular phase ranges from $\sim$ 34 \Ms\ and 51 \Ms\ depending on the metal depletion, while it ranges from 33  \Ms\ to 36 \Ms\  for the 260 \Ms\ CCSN. For both progenitors, molecules include CO, O$_2$, and SO and very massive PISNe are less efficient at forming chemical species in their ejecta than less massive PISNe, owing to their hotter gas and their more important radioactivity at play, as already shown by CD09. The present dust mass upper limits are different from the dust mass yields derived by SFS04 for the two massive progenitors, although somehow similar in term of order of magnitude. However, strong disparities exist in both the chemical composition of the dust and the epoch of its formation. For example, SFS04 form amorphous carbon as the atomic carbon not trapped in CO turns into AC grains. They argue that some CO is destroyed by the \grays\ produced by \Co\ radioactive decay and produces this pool of free atomic carbon, with no quantitative assessment of this process in their ejecta. CD09 show that this CO destruction pathway is of minor importance compared to alternative routes like neutral-neutral processes or destruction by He$^+$. When kinetics is considered, the entire carbon initial content is quickly included into CO, resulting in much larger CO masses than those derived from steady-state,  and the formation of small carbon chains is suppressed. No carbon dust can thus form in the present fully-mixed ejecta of PISN models as already seen in \S 4.1. 

As to other solids, we find that all the (SiO)$_5$ rings may be converted into silica, fayalite, forsterite, and pure silicon, or silica and pure silicon, dependent of the level of depletion of atomic Mg and Fe. Two other precursors are present in various amounts dependent of progenitor mass, AlO and (Fe)$_4$, leading to the formation of alumina and pure iron grains. The large initial Fe elemental yield for the 260 \Ms\ progenitor leads to very large amounts of pure iron dust at $t=1000$~days when Fe depletion is not considered, as seen in Table 6. SFS04 find that silica, forsterite, amorphous carbon and magnetite are the dominant dust products for both progenitor masses. Their dust chemical composition is then very different from that derived in this study. Also important is the discrepancy in the condensation sequence. SFS04 find a condensation time window of 150 to 250 days, dependent of their progenitor masses. Nucleation epochs in the present study are controlled by chemical kinetics, reproducing dust synthesis in the laboratory. For example, the very low ejecta temperatures considered by SFS04 imply that the (SiO)$_5$ clusters form at 110 days post-explosion (or at temperatures close to those found in flames) whereas (Fe)$_4$ are synthesized at $t =250$~days and AlO forms at much later times (500 days) and lower temperatures.

In summary, the SFS04 study uses very low ejecta temperatures for PISNe and the CNT approach to describe dust condensation. While the amount of dust synthesized is comparable to the upper limits we derive for similar progenitor masses, the present study points to a totally different composition for the condensed grains, in particular, the formation of pure iron grains and the lack of amorphous carbon and iron oxides like magnetite. Once again, caution should be taken when using CNT model outputs into star formation and chemical evolution studies in the early universe. 

\subsection{Unmixed Ejecta}

We present in this section the results for the PISN 170 \Ms\ and the CCSN 20 \Ms\ unmixed ejecta. The stratification of the ejecta results from: (1) the nucleosynthesis of the explosion itself for the innermost mass layers; (2) successive nucleosynthetic phases occuring during the evolution of the stellar progenitor. Such a stratification pattern implies that each zone has its proper chemical composition and therefore its specific chemistry and chemical products.  

\subsubsection{170 \Ms\ progenitor}

Cluster abundances versus time for the 170 \Ms\ progenitor are presented in Figures 5 and 6 for Zones 1 and 2, respectively, while the dust mass yields ejected at 1000 days assuming Mg depletion or no depletion, as explained in \S 2.4, are summarized in Table 7. While Zone 1 is characterized by a unique Si/Fe-rich chemistry, Zone 2 contributes the most to the cluster budget from all zones. We see from Figure 5 that the first clusters to form at $\sim 400$ days are iron sulphide clusters, followed by pure silicon clusters at $\sim 700$ days and pure iron clusters at $\sim 800$ days. The FeS molecule is quickly formed from reaction \ref{fes1} when its destruction occurs via thermal fragmentation and the reverse process of reaction \ref{fes1}. FeS formation leads to the depletion of molecular S$_2$ at $t > 300$~days, as seen in CD09. The polymerization of FeS clusters occurs via reaction \ref{fes3} at $t > 400$~days as thermal fragmentation remains important at early times due to the gas high temperatures. Once this process is switched off, the growth of the small cubic cluster (FeS)$_4$ proceeds, with no other chemical destruction channels taking place.  For pure silicon and iron clusters, formation of di-silicon and di-iron chains are assumed to proceed with a similar formation mechanism characterized by the three-body reaction \ref{f1}. The formation of (Si)$_4$ occurs at slightly earlier times than that of (Fe)$_4$ owing to the large amounts of available atomic silicon not locked in SiS. The high Si initial composition of Zone 1 spawns the large abundances of pure Si clusters, followed by iron sulphide and pure iron clusters. As to ions, the dominant species in Zone 1 is Fe$^+$ which forms from the charge exchange reaction between Si$^+$ and Fe, Si$^+$ being formed from the \grays\ ionization of atomic Si.  Fe$^+$ is further destroyed by its recombination to Fe, which controls the electron population in the zone. 

In Zone 2, characterized by a high Si/O initial content, the conversion of atomic Si into SiO occurs for $t < 250$ days as seen in Figure 6, followed by the quick polymerization of SiO ring clusters at $t > 350$ days (equivalent to a gas temperature of  $\sim$ 3700~K). The further growth of (SiO)$_5$ keeps depleting the SiO content of the ejecta zone, resulting in large (SiO)$_5$ abundances. We stress here that the SiO distribution obtained in the present study and shown in Figure 6 is slightly different from that presented in CD09 for the same mass zone. This is due to the fact that we treat SiO ring polymerization up to 5 (SiO) units in this study whereas (SiO)$_3$ was the end-product of cluster formation in CD09. The discrepancy results from the replenishment in SiO molecules during the destruction of (SiO)$_{3,4,5}$ rings, leading to a higher SiO abundance distribution with time, although the SiO abundance trend remains the same. As to other condensates, AlO forms at $T > $ 550~days from reaction \ref{alo1}, once O$_2$ has stopped to be efficiently destroyed by thermal fractionation. Finally, pure Mg clusters are synthesized at $t > 800$~days when their thermal fractionation has ceased. 

In table 7, we list the results on the upper limits for dust masses formed at 1000 days post-explosion for Zones 1 - 5 and for the two cases of total Mg depletion and no depletion. In this regard, the latter case gives almost directly the mass yields of the clusters synthesized at 1000 days.  As already stated, the dominant dust formation zones are Zones 2 and 3, followed by Zone 1. From Table 7, we see that there exist two mass classes of condensates. The first class gathers the most abundant solids, all Si-bearing compounds, silica, forsterite, and pure silicon. Indeed, the prevalent condensate in both depletion cases is silica (SiO$_2$), forming essentially in Zones 2 and 3. If Mg depletion is considered, forsterite (Mg$_2$SiO$_4$) also forms in these zones. In both depletion cases, pure silicon clusters are too abundant owing to the disproportionation of silicon monoxide clusters, as explained in \S 2.1.1. This first class of solids represents $\sim$ 99.5 \% of the dust content. The second class of condensates gathers the minor compounds, alumina, iron sulphide, and pure magnesium and iron grains. All of those form separately from the first class and their mass yields are independent of the depletion case. This class accounts for $\sim$ 0.5 \% of the total dust budget. It is noteworthy that neither carbon grains nor periclase (MgO) and SiC precursors form in significant amounts in the unmixed ejecta of the 170 \Ms\ progenitor. Despite the presence of carbon in Zones 4 and 5, these zones are characterized by a C/O ratio less than unity, a very low atomic Si content, and by large amounts of helium in Zone 5. Although the formation of carbon chains is triggered by the rapid thermal conversion of CO into C$_2$ at high temperatures and not by slow radiative association processes, as seen in CD09, those chains never form in significant quantities due to either their overwhelming oxidation by atomic O which hinders their growth in Zone 4, or the presence of He$^+$ in Zone 5. As to SiC precursors, they do not form owing to the small Si initial content of Zone 5 and the fact that the synthesis of gas-phase SiC competes with the faster formation channels for C-chains, and suffers severe destruction by He$^+$. 

Upper limits of the mass yields for each solid versus depletion case are shown in Figure 7 for the present 170 \Ms\ progenitor along with results from NK03 for a similar unmixed ejecta. According to Table 7, the total amount of dust formed for the 170 \Ms\ progenitor ranges between  5.7 and 7.1 \Ms. This is well below the $\sim$ 25 \Ms\ of dust synthesized in NK03 170 \Ms\ ejecta. As seen from Figure 7, our dust chemical composition also disagrees with that of NK03 who find that $\sim$ 75 \% of the grains are made of pure silicon, forsterite and silica grains, when the last 25 \%  are pure iron and iron sulphide grains, periclase, and amorphous carbon, the latter accounting for $\sim$ 1\% of the total dust mass.  

\subsubsection{20 \Ms\ progenitor}

Cluster abundances for Zones~1 and 2 of the 20 \Ms\ progenitor are shown in Figures 8 and 9, respectively while upper limits for dust masses formed at 1000 days post-explosion for Zones~1 - 4 and for the two cases of total Mg depletion and no depletion are presented in Table 8. The chemical processes at play in the formation of clusters are similar to those already highlighted in \S 4.3.1 for Zones 1 and 2 of the 170 \Ms\ ejecta.  In Figure 8 and for Zone 1, the prevalent clusters are, in order of decreasing abundances, (FeS)$_4$, (Si)$_4$, (SiO)$_3$, and (Fe)$_4$. A major departure arises from the large initial mass of oxygen compared to Zone~1 of the 170 \Ms\ ejecta. In this Si-rich zone, the large O mass combined to the low gas temperature fosters the early nucleation of (SiO)$_3$ clusters at t $\sim$ 250 days, while SiO clusters are absent from Zone 1 of the 170 \Ms\ ejecta. Moreover, Zone~1 contains eight times more Fe than Zone~1 of the 170 \Ms\ ejecta. All these discrepancies spawn the larger efficiency at making dust in Zone 1 for the CCSN, as shown in Table 8, where dust mass yields at 1000 days are presented. However, the formation of Fe-rich silicate like fayalite (Fe$_2$SiO$_4$) is hampered in Zone 1 due to the fact that all the initial oxygen is trapped into the formation of SiO, leaving no oxygen for further oxidation by O$_2$ or atomic O. Thus, the large amounts of (SiO)$_3$ result in silica formation. When Zone~2 forms essentially silica and/or forsterite, we see from Table 8 that Zone 3 and Zone~4 do not contribute to the cluster budget. Zone~3 is Si-poor with a C/O ratio less than unity, implying that CO is the prevalent species to form in this zone (CD09). Zone 3 is thus not conducive to cluster synthesis. Zone 4 is characterized by a large C/O ratio (29.5) and a high He content. These conditions preclude the synthesis of C-chains and rings, and the subsequent AC grain condensation because of the destruction of small clusters by He$^+$, as explained in \S 4.4.2. For SiC precursors, conclusions similar to those drawn for the 170 \Ms\ case apply, and no silicon carbide dust forms in Zones~3 and 4. 

Upper limits of the mass yields for each solid versus depletion case are shown in Figure 10 along with the results from NK03 for a similar unmixed ejecta. According to Table 8, the total amount of dust formed for the 20 \Ms\ progenitor ranges between  0.103 and 0.154 \Ms. This is well below the $\sim$ 0.58 \Ms\ of dust synthesized in NK03 170 \Ms\ ejecta. As seen from Figure 10, our dust chemical composition disagrees with that of with NK03 who find that $\sim$ 64 \% of the dust mass is in the form of pure silicon, forsterite and silica grains, 16 \% and 5 \% are iron sulphide and pure iron grains, respectively, and 10 \% is in the form of amorphous carbon. 

\subsubsection{Depletion efficiencies}

Elements initially in atomic form are first depleted in molecules and molecular clusters as a result of nucleation chemistry. To assess this trapping independent of the depletion cases defined in \S 2.5, we define the efficiency $x$ as 
\begin{equation}
\label{d1}
x = N_{Cluster}(y) / N_{Total}(y),
\end{equation}
where $y$ is the element, $N_{Cluster}(y) $ is the total number of $y$ atoms locked up in one type of cluster, and $N_{Total}(y)$ is the initial total number of $y$ atoms in the ejecta. The efficiency $x$ measures how the chemistry locks up elements in the nucleation processes of molecular clusters. It is not linked to the full depletion of Mg and Fe during the condensation phase of dust seeds. Depletion efficiencies are shown as a function of zoning in Table 9 for the two progenitor masses.

For both progenitor masses, the depletion of Si atoms into silicon oxide precursors leading to silica and pure silicon clusters is efficient in silicate forming zones and starts at early times when Si is converted into SiO and small SiO clusters (see Figures 6 and 9). The depletion is then total at day 1000 as seen in Table 9 and  \S 4.3.1 and \S 4.3.2. This is in agreement with the steeper fading of the 1.65 \micron\ [SiI] line observed in SN1987A compared to the 4571 \AA\ [MgI] and the 6300 \AA\ [OI] lines observed by \citet{luc89}. In the Fe-rich, O-poor Zone 1, the conversion of Si into pure silicon clusters proceeds with a low efficiency of $\sim$ 1\% to 3\% due to the prevalent conversion of Si in SiS molecules with an efficiency ranging from 26 to 28\%. 

Aluminum also suffers total depletion into molecular AlO. The formation of alumina clusters has not been modeled here but AlO should react with itself and O$_2$ to build up small Al$_2$O$_3$ clusters \citep{ca03}. A total depletion of Al into alumina clusters is thus to be expected.

Carbon is massively depleted into CO in the zones where it is present. In the C-rich Zones 4 and 5 of the 20 and 170 \Ms\ ejecta, respectively, its depletion depends on the presence of He$^+$, as fully explained in \S 4.4.2. In Zone 5 of the 170 \Ms\ PISN, which has a C/O ratio of 0.66, most of the carbon is tied up in CO once the He$^+$ abundance decreases for $t > 600$ days (see Figure 7 of CD09). Conversely, in Zone 4 of the 20 \Ms\ CCSN characterized by a C/O ratio of 29.5, C remains in atomic form due to the overabundance of He, hence He$^+$. However, atomic carbon is almost entirely depleted in the carbon ring C$_{10}$ if He is not microscopically mixed to the gas, as seen in \S 4.4.2. 

Atomic oxygen is depleted in molecular O$_2$ with efficiency values ranging from 12\%  to 61\% in all zones where O is present, and into CO ($x \equiv  29 - 43$\%) where atomic carbon is abundant. The formation of silica clusters locks up atomic oxygen to efficiency values spanning $\sim$ 3\% to 10\% only. Thus part of the decline in the oxygen lines is due to molecular formation rather than dust condensation.  

Iron is depleted in the formation of FeS clusters in Zone 1 of both ejecta. Its depletion efficiency in FeS reaches the high level of 44\% for the 20 \Ms\ progenitor ejecta. We also see that Fe is never efficiently transformed into Fe clusters. To assess if this low efficiency in nucleating iron small clusters is due to the presence of sulphur and the formation of FeS, we ran the chemistry for the S-free Zone 1 of the 20 \Ms\ progenitor. We find that the depletion of iron into small Fe clusters stays very low ($\sim 0.3$\%) and that most of the iron remains in atomic form. As to Mg, it is seldom depleted by the nucleation of small clusters except in Zones~3 and 2 for the 170 and 20 \Ms\ ejecta, respectively, with depletion efficiencies of 1 and 2\%. However, the coagulation into large silica clusters may trapped atomic Mg or Fe into the cluster lattices to form forsterite and fayalite, respectively, a situation described by our 'total depletion' case defined in \S 2.1. However, this depletion processes are beyond the aims of the present paper and will be presented in a forthcoming publication. 

\subsection{Critical Parameters under Study}

We now explore the impact of two critical parameters in the cluster nucleation process, i.e., the nucleation rate of reactions \ref{sio7} for silica clusters and the impact of He diffusion on carbon dust formation in the outer zones of the ejecta. 

\subsubsection{Pressure Dependence of Nucleation Reactions }

A critical parameter to the formation of silicates and metal oxides is the rate at which small clusters react one to another to nucleate small dust seeds that can later coagulate and condense into lager dust grains. In previous chapters, we assume that reactions \ref{sio7}, \ref{feo4}, \ref{mgo4}, \ref{mgs3}, and \ref{fes3} all proceed with similar rates listed in Table 2, as the only information available on nucleation rates is for small silicon oxide clusters (ZT93). To test the impact of varying the nucleation rates of small clusters of silicon oxide, metal oxides and sulphides, we consider a case study describing the nucleation of small silicon oxide clusters in Zone 2 of the 170 \Ms\ progenitor unmixed ejecta. The rates for reactions such as reactions \ref{sio7} are listed in Table 2 and are taken from the ZT93 study for the 10$^{-2}$ atm pressure case. Using this series of rates as benchmarks, we now study the impact of increasing and decreasing by a factor 100 the benchmark rates. The increase by 100 corresponds to the 1 atm pressure case of ZT93 and we calculate the abundances  normalized to total gas number density of clusters as a function of post-explosion time. 

Abundances for the (SiO)$_5$ clusters and gas-phase SiO are presented in Figure 11 for the different rates of reactions \ref{sio7}. As the nucleation rate decreases (e.g., in the 10$^{-4}$ case for which all rates have been decreased by a factor 10$^{-2}$ compared to our benchmark case), the formation of clusters is postponed to late times while the SiO abundance increases first and starts dropping once (SiO)$_n$ clusters form. However, the abundance of the (SiO)$_5$ cluster is unchanged at day 1000. Not shown in Figure 11, the O$_2$ and CO abundances follow similar trends, implying that the lower the rates, the larger the molecular phase and its mass in the ejecta while the final cluster mass yields at day 1000 are similar for all cases. As seen in \S 4.3.3, elemental Si and SiO are totally depleted in the process of silica and silicate dust formation. Therefore, observations of Si and SiO optical and IR lines in local young SNe are good indicators of the timing and the efficiency of the nucleation of silicate dust in SN ejecta, once excitation effects are taken into account in the analysis of line declines as a function of time.  

\subsubsection{How Can Carbon Dust Form?}

No carbon rings form in the unmixed ejecta of the 170 and 20 \Ms\ progenitors owing to the large amounts of He and hence He$^+$ in C-rich Zones~5 and 4, respectively. However, evidence for the presence of carbon grains in CCSNe ejecta exists from various sources of investigation. Graphite grains bearing the isotopic anomalies of CCSNe are found in meteorites \citep{zin06}. Furthermore, models aiming at explaining the IR excess emission of several low-z CCSNe and SN remnants show that amorphous carbon is a necessary component of the dust mixture required to reproduce IR data \citep{sug06,erco07,rho08}. In order to assess the impact of He/He$^+$ on the formation of carbon chains and rings, we re-investigate the chemistry of  Zone 4 of our 20 \Ms\ progenitor, which is characterized by a C/O ratio of 29.47. Keeping the initial metal mass constant, we gradually increase the amount of He present in the zone up to its value assumed in the previous section (0.884 \Ms\ according to Table 3 of CD09). We consider the following increase factors: 0\% ($\equiv$ no He), 1\%, 20\%, 40\%, and finally 100\% of the He value of Zone 4. For increasing He masses, the initial total number of elements in the gas also increases while the partial number of metals is kept constant. Doing so, we aim at reproducing the chemical composition of a C-rich clump in the ejecta partially embedded in a helium mushroom cap, a configuration commonly found in 2D/3D modeling of CCSN explosion \citep{mul91,kif03,ham09}. We then allow some diffusion of He inside the C-rich Zone 4. 

The abundance of the ring C$_{10}$  as a function of post-explosion time for our various He diffusion values are shown in Figure 12. We see that the larger the He content of the zone, the later the formation and the smaller the abundances of C$_{10}$ rings. Indeed, for very large He content, the amount of He$^+$ in the zone is large ($x({\rm He^+}) \sim 0.1\% \times x({\rm He})$, where $x({\rm He})$ and $x({\rm He^+})$ are the abundances of He and He$^+$, respectively). The destruction of small carbon chains like C$_2$ then proceeds until He$^+$ abundances diminishes at late times owing to a decrease in He ionization by Compton electrons and the recombination of He$^+$. The resulting chain and ring formation is thus delayed. When the zone is free of He, the rapid conversion of the atomic carbon not locked in CO into C$_2$  via the radiative association reaction \ref{c1} triggers the formation of larger chains, with instantaneous closure of the end-product ring C$_{10}$. For the He-free Zone 4, the final abundance of C$_{10}$ with respect to total gas number density at 1000 days is 0.55, equivalent to a AC dust mass yield of 0.0145 \Ms\, in which $\sim$ 95\% of the initial carbon is depleted. When He diffusion occurs, the conversion of free atomic carbon into C$_{10}$ is as effective as for the He-free case, with similar final mass yield and formation efficiency for the C$_{10}$ ring, except that the conversion takes place at later times until the He$^+$ abundance decreases. 

In summary, the build-up of carbon clusters in SN ejecta is significantly hampered by oxidation in zones where the C/O ratio is less than unity, and by the microscopic mixing of He$^+$ in the carbon-rich zones of the ejecta. More generally, He$^+$ has a detrimental impact on molecular formation in SN ejecta as first shown by \citet{lepp90} and CD09. In C-rich, He-free zones, carbon condenses with a high efficiency. Any formation of carbon dust in SNe must thus occur in C-rich inhomogeneities deprived of He$^+$ and is thus highly dependent of the mixing induced by the SN explosion. 

\section{SUMMARY AND DISCUSSION}

We have investigated the synthesis of molecular clusters precursors to dust in Pop.~III supernova ejecta and show that dust clusters form despite the harsh, post-explosion environments. Of special interest are the following points: 
\begin{itemize}

\item In general terms, we outline that a classical nucleation theory (CNT) formalism is not appropriate to the descrition of dust formation in SNe ejecta as it involves hypothesis  which do not hold for SN ejecta out of chemical equilibrium and concepts not prevailing for small molecular clusters. We highlight some specific problems encountered by previous studies of fully-mixed SN ejecta using CNT, specifically: (1) erroneous initial conditions for the \citet{tod01} 18 and 20 \Ms\ progenitor models leading to the formation of carbon dust in Pop.~III SNe; (2) very low ejecta temperatures used by \citet{sch04} for their PISN models. A chemical kinetic approach applied to the 20 \Ms\ TF01 model makes the synthesis of dust merely impossible because of the large He content of the ejecta. When applied to SFS04 models, the chemical kinetic approach points to a totally different chemical composition for the dust than that derived by SFS04. In both cases, chemical kinetics prevents the formation of carbon dust. 

\item Fully microscopically-mixed and unmixed models produce smaller dust masses ejected at day 1000 after explosion than those derived by studies using CNT and similar ejecta conditions such as the study by \citet{noz03}. Dust mass upper limits ranging from 25 to 33 \Ms\ and 0.16 to 0.33 \Ms\ are derived for the fully-mixed ejecta of 170 \Ms\ and 20 \Ms\ progenitors, respectively. These upper limits depend on the level of Fe/Mg depletion and lie between 16\% and 21 \% of the 170 \Ms\ progenitor mass and 0.8\% and 2\% of the 20 \Ms\ progenitor mass. NK03 find that 170 and 20 \Ms\ progenitors form dust mass yields representing 21\% and 4\% of the progenitor mass, respectively. Our upper limits are thus down by a factor of 2 to 5 for the 20 \Ms\ progenitor. Our dust composition is dominated by silica (and silicates when metal depletion is considered), pure silicon, alumina and pure iron. No iron oxides, amorphous carbon dust, or silicon carbides form. These dust compositions disagree with those derived by NK03. 

\item For the unmixed ejecta, we find dust upper limits dependent of the degree of metal depletion ranging from 5.7 to 7 \Ms\  for the 170 \Ms\ progenitor, and from 0.09 to 0.14 \Ms\  for the 20 \Ms\ progenitors. These limits are much smaller than the dust masses found by NK03 by factors of $\sim4$ and $\sim$ 5 for the 170 and 20 \Ms\ progenitors, respectively. The chemical composition of the dust is dominated by silica/silicates, pure silicon dust and iron sulphide grains. No magnesium and ferrous oxides, pure iron grains, silicon carbides, and amorphous carbon dust form, as opposed to the results from NK03. 

\item Negligible amounts of carbon dust form in unmixed ejecta, despite the large carbon mass initially present in the outer ejecta zone of the 20 \Ms\ progenitor. We show that the presence of He$^+$ hampers the formation via small carbon chains and rings even for high C/O ratios. Consequently, carbon dust formation in SNe must occur in inhomogeneities where He/He$^+$ is not microscopically mixed with the gas. In zones where the C/O ratio is less than unity, thermal fragmentation of CO triggers the formation of C$_2$ and longer carbon chains, as opposed to radiative association reactions, but the oxidation of chains prevent those from growing into rings. No silicon carbide precursors form in mixed and unmixed ejecta as their synthesis channels compete with those for carbon precursors. Furthermore, the extremely low silicon content of the outer carbon-rich layers precludes SiC nucleation in unmixed ejecta. 

\item The depletion of metals resulting from the formation of molecules and the nucleation of dust clusters prior to any depletion stemming from the dust condensation phase varies according to the elements and their local environment. In a O-rich gas, Si is depleted at 100\% in the formation of silicon monoxide clusters, when a S-rich/O-poor gas depletes Si to a 30\% level, mainly in gas phase SiS. Al is also totally depleted by the formation of gas-phase aluminum oxide. However, Fe is scarcely depleted except in the innermost mass zone of the CCSN ejecta where its conversion to FeS clusters accounts for an efficiency of 40\%. Mg does not suffer severe nucleation depletion and gets depleted into small pure Mg clusters to a level of $\sim$ 2 \%. 

\item The molecular content of the ejecta before 1000 days after explosion is highly dependent of the nucleation reaction rates at which small clusters grow. Smaller rates imply a larger molecular content for the ejecta. From a chemical point of view, optical and IR lines of [SiI] and SiO are thus direct tracers of the timing and efficiency of silicate condensation in local young SNe ejecta. 
\end{itemize}

In spite of the fact that the amounts of dust clusters forming in our unmixed PISN and CCSN ejecta§ are smaller than those derived by existing studies, the upper limits for dust are still high compared to what is observed in local CCSNe. Observations with Spitzer \citep{sug06,ko09} point to dust mass yields not exceeding 10$^{-2}$ \Ms\, i.e., a factor 10 down to the upper limit we derived for a primordial 20 \Ms\ CCSN. A possible reason for this discrepancy is the existence of cool dust escaping IR detection and its contribution to the final dust budget \citep{gom09}. Another possibility is that existing studies and the present work do not properly consider {\it real} SN ejecta for their dust synthesis models. SN ejecta and remnants are known to be highly inhomogeneous, with the presence of clumps, knots and filaments forming after the explosion as a result of Rayleigh-Taylor instabilities developing at the interfaces of the nucleosynthesis zones of the progenitor. On the one hand, fully microscopically-mixed ejecta are highly unrealistic since the mixing going inward to the deep core layers as a result of Rayleigh-Taylor instabilities is macroscopic. On the other hand, unmixed ejecta do not describe the mixing with (non)-adjacent layers. Such a microscopic mixing is necessary to explain the formation of silicon carbide and silicon nitride pre-solar inclusions bearing SN isotopic signatures found in meteorites \citep{ott03}. Therefore, we see that mixing in knots and clumps is a key parameter in the determination of the chemical composition of the dust and in its formation efficiency. This is best exemplified with amorphous carbon, as we see that carbon chains, rings and grains can only grow in primordial CCSNe if the formation locus is carbon-rich and deprived of He$^+$. More generally, any He$^+$ microscopic mixing impedes the formation of molecules and molecular clusters, as seen in CD09. The recent 3D model by \citet{ham09} shows the formation of O-rich blobs encompassing up to 10$^{-2}$ \Ms\  a few days after explosion. Depending on the blob initial chemical composition resulting from this mixing, a certain type of dust will form during the blob expansion. We see that any atomic Si will be totally depleted in a O-rich inhomogeneity due to the formation process of silica/silicate clusters, and the final amount of condensates will essentially depends on the initial Si/O ratio characterizing the inhomogeneity.  If part of the Si resides in O-free blobs, its efficiency at forming pure Si clusters is rather low ($\sim$ 2\% according to Table 7), and silicon will then stays in atomic form or may be depleted in SiS to a level of $\sim$ 27 \% if sulphur is present. Thus, mixing itself by the way of the chemical processes involved in the formation of molecules and dust clusters in inhomogeneities contributes towards the limitation of the total amount of dust synthesized in PISN and CCSN ejecta. 

Pop.~III stars exploding as PISNe are definitely the first dust makers in the early universe and the first dust enrichment of the primordial gas should be silicon/silica/silicate-rich to a level of 99 \% of the total dust mass ejected. No carbon grains are formed as atomic carbon is trapped in molecular CO in the outer layers of the PISN ejecta. Therefore, carbon dust at high redshift must come from different stellar sources than PISNe and its formation may be delayed until the lower-mass Pop.II.5/II stars have evolved to their AGB stage or as Supergiants, depending on their initial masses \citep{dwek07,val09}. PISN dust will be further reprocessed by the reverse shock several thousands years after the explosion. \Citet{noz07} find for an initial 170 \Ms\ progenitor a grain survival rate ranging from $\sim$ 0.02\% to 48\% of the initial dust mass depending on the pre-shock gas number density and the initial explosion energy. Thus between $\sim$ 0.001 to 3 \Ms\ of Si-based dust will be injected into the dense shell of the radiative phase of a 170 \Ms\ PISN explosion. This dust will further provide cooling for the gas fragmentation and Pop.~II.5 star formation. As an extension of the present work, dust mass yields, composition, and grain size distributions will be derived using a Monte-Carlo formalism for dust condensation in PISN ejecta. The impact on cooling of this Si-based dust and its related molecular phase as well as its contribution to Pop.~II.5 star formation will be further explored.

\acknowledgments We thank the anonymous referee for comments that have contributed to the improvement of the manuscript. IC acknowledges support from the Swiss National Science Foundation through the Maria Heim-V{\"o}gtlin subsidies PMPD2-114347 and PMPDP2-1241159. 

\clearpage



\clearpage

\begin{deluxetable}{lcll}
\tabletypesize{\scriptsize}
\tablecaption{Molecular clusters and their related dust products considered in this study. }
 \tablewidth{0pt}
 \tablehead{
\multicolumn{1}{c}{ Parent molecule } &\multicolumn{1}{c}{ Clusters } & \multicolumn{1}{c}{Dust } &\multicolumn{1}{c}{Symbol }\\
}
\startdata
SiO &(SiO)$_{2}$ to (SiO)$_{5}$ & Silica   & SiO$_2$ \\
  & & Forsterite & Mg$_2$SiO$_4$\tablenotemark{a} \\
  & &  Fayalite& Fe$_2$SiO$_4$\tablenotemark{a} \\
MgO & (MgO)$_{2}$ to (MgO)$_{4}$&  Magnesium oxide or Periclase& MgO  \\
FeO & (FeO)$_{2}$ to (FeO)$_{4}$ & Iron oxide or Troilite & FeO   \\
AlO & -- & Alumina& Al$_2$O$_3$  \\
FeS & (FeS)$_{2}$ to (FeS)$_{4}$ & Solid iron sulphide & FeS  \\
MgS & (MgS)$_{2}$ to (MgS)$_{4}$& Solid magnesium sulphide  & MgS  \\
Fe &(Fe)$_{2}$ to (Fe)$_{4}$& Solid iron& Fe  \\
Mg & (Mg)$_{2}$ to (Mg)$_{4}$& Solid magnesium& Mg  \\
Si & (Si)$_{2}$ to (Si)$_{4}$&Solid silicon & Si \\
C & C$_{2}$ to C$_{10}$ &  Amorphous carbon& AC \\
SiC & (SiC)$_2$ & Silicon carbide&SiC \\
 \enddata
\tablenotetext{a}{Formed in the full metal (Mg and/or Fe) depletion case. }
\end{deluxetable}

\clearpage

\begin{deluxetable}{llclrrrr}
\tabletypesize{\scriptsize}
\tablecaption{Reaction rates for the chemical processes involved in the nucleation of molecular clusters \tablenotemark{a}. Other reactions included in the total network are listed in Cherchneff \& Dwek (2009).}
 \tablewidth{0pt}
 \tablehead{
\colhead{ } &\multicolumn{3}{c}{Chemical processes}  & \colhead{A${ij}$ } & \colhead{$\nu$} & \colhead{E$_a$} & \colhead{Reference \tablenotemark{b}}
}
\startdata
\multicolumn{8}{c}{\bf Aluminum - Al}  \\
\tableline
A1 &Al + O$_2$ &$\longrightarrow$& AlO + O  & 1.61$\times 10^{-10}$& -1& 0 & NIST\\
A2 & Al + CO$_2$& $\longrightarrow$ &AlO + CO & 2.89$\times 10^{-10}$& 0& 3221.0 & Garland et al. 1992\\
A3 &Al + H$_2$O &$\longrightarrow$ &AlO + H$_2$ & 1.59$\times 10^{-10}$& 0& 2868.6 & Mc Lean et al. 1993\\
A4 &Al + O + M& $\longrightarrow$& AlO + M & 8.27$\times 10^{-31}$& 0& 0 & NIST\\
A5 &AlO + O &$\longrightarrow$ &Al + O2 & 1.93$\times 10^{-11}$& -0.5& 1199.9 & Garland et al. 1992\\
A6 &AlO+ He$^+$&$\longrightarrow$ &Al + O$^+$ + He & 8.60$\times 10^{-10}$& 0& 0 & NIST \\
A7 &AlO + He$^+$& $\longrightarrow$& Al$^+$O +He& 8.60$\times 10^{-10}$& 0& 0& NIST \\
A8&AlO + M &$\longrightarrow$ &Al + O + M & 4.40$\times 10^{-10}$& 0& 98600.0& Same as S2\\
A9&AlO + e$^-_{Compton}$ &$\longrightarrow$ &Al + O +  e$^-_{Compton}$  & 5.81$\times 10^{-6}$& 0& 3464.1& See CD09 \tablenotemark{c}\\
A10&AlO + e$^-_{Compton}$ &$\longrightarrow$ &Al$^+$ + O +  e$^-_{Compton}$  & 2.95$\times 10^{-6}$& 0& 3464.1& See CD09 \tablenotemark{c}\\
A11&AlO + e$^-_{Compton}$ &$\longrightarrow$ &Al + O$^+$ +  e$^-_{Compton}$  & 9.48$\times 10^{-7}$& 0& 3464.1& See CD09 \tablenotemark{c}\\
\tableline
\multicolumn{8}{c}{\bf Silicon - Si}  \\
\tableline
S1&SiO + SiO &$\longrightarrow$ &Si$_2$O$_2$ & 4.61$\times 10^{-17}$& 0& -2821.4& Zachariah \& Tsang 1993\\
S2&Si$_2$O$_2$ + M&$\longrightarrow$ &SiO + SiO + M& 4.40$\times 10^{-10}$& 0& 98600.0& Reaction TF15 in CD09\\
S3&SiO + Si$_2$O$_2$+ M &$\longrightarrow$& Si$_3$O$_3$ & 2.24$\times 10^{-15}$& 0& -2878.9&Zachariah \& Tsang 1993 \\
S4&Si$_3$O$_3$ + M&$\longrightarrow$ &SiO + Si$_2$O$_2$ + M& 4.40$\times 10^{-10}$& 0& 98600.0& Zachariah \& Tsang 1993\\
S5 &SiO + Si$_3$O$_3$+ M &$\longrightarrow$& Si$_4$O$_4$ & 1.53$\times 10^{-14}$& 0& -2386.8 & Zachariah \& Tsang 1993\\
S6&Si$_2$O$_2$ + Si$_2$O$_2$ &$\longrightarrow$& Si$_4$O$_4$ & 1.53$\times 10^{-14}$& 0& -2386.8 & Zachariah \& Tsang 1993\\
S7&Si$_2$O$_2$ + Si$_2$O$_2$ &$\longrightarrow$& Si$_3$O$_3$ + SiO& 1.53$\times 10^{-14}$& 0& -2386.8 & Zachariah \& Tsang 1993\\
S8&Si$_4$O$_4$ + M &$\longrightarrow$& Si$_3$O$_3$ + SiO & 4.40$\times 10^{-10}$& 0& 98600.0& Same as S2\\
S9&Si$_4$O$_4$ + M &$\longrightarrow$& Si$_2$O$_2$ + Si$_2$O$_2$ & 4.40$\times 10^{-10}$& 0& 98600.0& Same as S2\\
S10 &SiO + Si$_4$O$_4$+ M &$\longrightarrow$& Si$_5$O$_5$ & 1.53$\times 10^{-14}$& 0& -2386.8 & Same as S5\\
S11 &Si$_2$O$_2$ + Si$_3$O$_3$+ M &$\longrightarrow$& Si$_5$O$_5$ & 1.53$\times 10^{-14}$& 0& -2386.8 & Same as S5\\
S12 &Si$_2$O$_2$ + Si$_3$O$_3$+ M &$\longrightarrow$& Si$_4$O$_4$ +SiO& 1.53$\times 10^{-14}$& 0& -2386.8 & Same as S5\\
S13&Si$_5$O$_5$ + M &$\longrightarrow$& Si$_2$O$_2$ + Si$_3$O$_3$ & 4.40$\times 10^{-10}$& 0& 98600.0& Same as S2\\
S14&Si$_5$O$_5$ + M &$\longrightarrow$& SiO + Si$_4$O$_4$ & 4.40$\times 10^{-10}$& 0& 98600.0& Same as S2\\
S15&Si + Si + M &$\longrightarrow$ &Si$_2$ +M & 2.76$\times 10^{-29}$& 0& -2821.4& Same as F23\\
S16&Si$_2$ + M &$\longrightarrow$ &Si + Si +M & 7.14$\times 10^{-5}$& 0& 17800& Same as F24\\
S17&Si + Si$_2$ + M &$\longrightarrow$ &Si$_3$ +M & 2.76$\times 10^{-29}$& 0& -2821.4& Same as F25\\
S18&Si$_3$ + M &$\longrightarrow$ &Si + Si$_2$ +M & 1.66$\times 10^{-5}$& 0& 19200& Same as F26\\
S19&Si + Si$_3$ + M &$\longrightarrow$ &Si$_4$ +M & 2.76$\times 10^{-29}$& 0& -2821.4& Same as F27\\
S20&Si$_4$ + M &$\longrightarrow$ &Si + Si$_3$ +M & 8.30$\times 10^{-7}$& 0& 21600& Same as F28\\
S21&Si$_2$ + Si$_2$ + M &$\longrightarrow$ &Si$_3$ +Si & 8.30$\times 10^{-10}$& 0& 0& Same as F29\\
S22&Si$_2$ + Si$_2$ + M &$\longrightarrow$ &Si$_4$ & 8.30$\times 10^{-10}$& 0& 0& Same as F30\\
S23&Si + CO &$\longrightarrow$ &SiC + O  & 5.43$\times 10^{-7}$& -1.50& 57200& Same as NN57 in CD09\\
S24&O+ SiC &$\longrightarrow$ &Si + CO  & 5.99$\times 10^{-10}$& 0.0& 1420& Same as C38\\
S25& C + SiO &$\longrightarrow$ &SiC + O  & 5.43$\times 10^{-7}$& -1.50& 57200& Same as S23\\
S26&O+ SiC &$\longrightarrow$ &SiO + C   & 5.99$\times 10^{-10}$& 0.0& 1420& Same as C38\\
S27&Si + C$_2$ &$\longrightarrow$ &SiC + C &  5.99$\times 10^{-10}$& 0.0& 1420& Same as C38\\
S28&SiC + SiC &$\longrightarrow$ &Si$_2$C$_2$ & 4.61$\times 10^{-17}$& 0& -2821.4& Same as S1\\
S29&SiC + SiC + M &$\longrightarrow$ &Si$_2$C$_2$ + M& 2.48$\times 10^{-21}$& -3.0& 0& Same as C37\\
S30&Si$_2$C$_2$ + M &$\longrightarrow$ &SiC + SiC+M& 4.40$\times 10^{-10}$& 0& 98600.0& Same as S2\\

\tableline
\multicolumn{8}{c}{\bf Iron - Fe}  \\
\tableline
F1 &Fe+ O$_2$ &$\longrightarrow$ &FeO+ O & 2.09$\times 10^{-10}$& 0& 10199.6& Akhmadov et al. 1988\\
F2 &Fe + CO$_2$ &$\longrightarrow$ &FeO + CO& 5.38$\times 10^{-10}$& 0& 15033.1& NIST \\
F3 &FeO + O &$\longrightarrow$ &Fe + O$_2$ & 4.60$\times 10^{-10}$& 0& 360& Self \& Plane 2003\\
F4&FeO + FeO &$\longrightarrow$ &Fe$_2$O$_2$ & 4.61$\times 10^{-17}$& 0& -2821.4& Same as S1\\
F5&Fe$_2$O$_2$ + M&$\longrightarrow$ &FeO + FeO + M& 4.40$\times 10^{-10}$& 0& 98600.0& Same as S2\\
F6 &FeO + Fe$_2$O$_2$ &$\longrightarrow$& Fe$_3$O$_3$ & 2.24$\times 10^{-15}$& 0& -2878.9 & Same as S3\\
F7&Fe$_3$O$_3$ + M&$\longrightarrow$ &FeO + Fe$_2$O$_2$ + M& 4.40$\times 10^{-10}$& 0& 98600.0& Same as S4\\
F8 &FeO + Fe$_3$O$_3$ &$\longrightarrow$& Fe$_4$O$_4$ & 1.20$\times 10^{-32}$& 0& 2160.0 & Same as S5\\
F9&Fe$_4$O$_4$ + M&$\longrightarrow$ &FeO + Fe$_3$O$_3$ + M& 4.40$\times 10^{-10}$& 0& 98600.0& Same as S8\\
F10&Fe$_4$O$_4$ + M&$\longrightarrow$ &Fe$_2$O$_2$ + Fe$_2$O$_2$ + M& 4.40$\times 10^{-10}$& 0& 98600.0& Same as S9\\
F11 &Fe+ SO &$\longrightarrow$ &FeS+ O & 2.09$\times 10^{-10}$& 0& 10199.6& Same as F1\\
F12 &Fe+ S$_2$ &$\longrightarrow$ &FeS+S & 2.09$\times 10^{-10}$& 0& 10199.6& Same as F1\\
F13&FeS + O &$\longrightarrow$ &Fe + SO$_2$ & 1.00$\times 10^{-10}$& 0& 0& E\\
F14&FeS + S &$\longrightarrow$ &Fe + S$_2$ & 1.00$\times 10^{-10}$& 0& 0& E\\
F15&FeS + FeS &$\longrightarrow$ &Fe$_2$S$_2$ & 4.61$\times 10^{-17}$& 0& -2821.4& Same as F5\\
F16&Fe$_2$S$_2$ + M&$\longrightarrow$ &FeS + FeS + M& 4.40$\times 10^{-10}$& 0& 98600.0& Same as F6\\
F17&FeS + Fe$_2$S$_2$ &$\longrightarrow$& Fe$_3$S$_3$ & 2.24$\times 10^{-15}$& 0& -2878.9 & Same as F7\\
F18&Fe$_3$S$_3$ + M&$\longrightarrow$ &FeS + Fe$_2$S$_2$ + M& 4.40$\times 10^{-10}$& 0& 98600.0& Same as F8\\
F19 &FeS + Fe$_3$S$_3$+ M &$\longrightarrow$& Fe$_4$S$_4$ & 1.20$\times 10^{-32}$& 0& 2160.0 &Same as F9\\
F20&Fe$_4$S$_4$ + M&$\longrightarrow$ &FeS + Fe$_3$S$_3$ + M& 4.40$\times 10^{-10}$& 0& 98600.0& Same as F10\\
F21&Fe$_4$S$_4$ + M&$\longrightarrow$ &Fe$_2$S$_2$ + Fe$_2$S$_2$ + M& 4.40$\times 10^{-10}$& 0& 98600.0& Same as F11\\
F22&Fe + Fe + M &$\longrightarrow$ &Fe$_2$ +M & 2.76$\times 10^{-29}$& 0& -2821.4& Giesen et al. 2003\\
F23&Fe$_2$ + M &$\longrightarrow$ &Fe + Fe +M & 7.14$\times 10^{-5}$& 0& 17800& Giesen et al. 2003\\
F24&Fe + Fe$_2$ + M &$\longrightarrow$ &Fe$_3$ +M & 2.76$\times 10^{-29}$& 0& -2821.4& Giesen et al. 2003\\
F25&Fe$_3$ + M &$\longrightarrow$ &Fe + Fe$_2$ +M & 1.66$\times 10^{-5}$& 0& 19200& Giesen et al. 2003\\
F26&Fe + Fe$_3$ + M &$\longrightarrow$ &Fe$_4$ +M & 2.76$\times 10^{-29}$& 0& -2821.4& Giesen et al. 2003\\
F27&Fe$_4$ + M &$\longrightarrow$ &Fe + Fe$_3$ +M & 8.30$\times 10^{-7}$& 0& 21600& Giesen et al. 2003\\
F28&Fe$_2$ + Fe$_2$ + M &$\longrightarrow$ &Fe$_3$ +Fe & 8.30$\times 10^{-10}$& 0& 0& Giesen et al. 2003\\
F29&Fe$_2$ + Fe$_2$ + M &$\longrightarrow$ &Fe$_4$ & 8.30$\times 10^{-10}$& 0& 0& Giesen et al. 2003\\
\tableline
\multicolumn{8}{c}{\bf Magnesium - Mg}  \\
\tableline
M1 &Mg+ O$_2$ &$\longrightarrow$ &MgO+ O & 1.18$\times 10^{-10}$& 0& 5769.1& Kashireninov et al. 1982\\
M2 &Mg + CO$_2$ &$\longrightarrow$ &MgO + CO& 5.38$\times 10^{-10}$& 0& 15033.1& Same as F2\\
M3 &MgO + O &$\longrightarrow$ &Mg + O$_2$ & 1.00$\times 10^{-11}$& 0& 0& E\\
M4&MgO + MgO &$\longrightarrow$ &Mg$_2$O$_2$ & 4.61$\times 10^{-17}$& 0& -2821.4& Same as S1\\
M5&Mg$_2$O$_2$ + M&$\longrightarrow$ &MgO + MgO + M& 4.40$\times 10^{-10}$& 0& 98600.0& Same as S2\\
M6&MgO + Mg$_2$O$_2$ &$\longrightarrow$& Mg$_3$O$_3$ & 2.24$\times 10^{-15}$& 0& -2878.9 & Same as S3\\
M7&Mg$_3$O$_3$ + M&$\longrightarrow$ &MgO + Mg$_2$O$_2$ + M& 4.40$\times 10^{-10}$& 0& 98600.0& Same as M5\\
M8&MgO + Mg$_3$O$_3$+ M &$\longrightarrow$& Mg$_4$O$_4$ & 1.20$\times 10^{-32}$& 0& 2160.0 & Same as S5\\
M9&Mg$_4$O$_4$ + M&$\longrightarrow$ &MgO + Mg$_3$O$_3$ + M& 4.40$\times 10^{-10}$& 0& 98600.0& Same as S8\\
M10&Mg$_4$O$_4$ + M&$\longrightarrow$ &Mg$_2$O$_2$ + Mg$_2$O$_2$ + M& 4.40$\times 10^{-10}$& 0& 98600.0& Same as S9\\
M11 &Mg+ SO &$\longrightarrow$ &MgS+ O & 2.09$\times 10^{-10}$& 0& 10199.6& Same as F11\\
M12 &Mg+ S$_2$ &$\longrightarrow$ &MgS+S & 2.09$\times 10^{-10}$& 0& 10199.6& Same as F12\\
M13&MgS + O &$\longrightarrow$ &Mg + SO$_2$ & 1.00$\times 10^{-10}$& 0& 0& E\\
M14&MgS + S &$\longrightarrow$ &Mg + S$_2$ & 1.00$\times 10^{-10}$& 0& 0& E\\
M15&MgS + MgS &$\longrightarrow$ &Mg$_2$S$_2$ & 4.61$\times 10^{-17}$& 0& -2821.4& Same as F15\\
M16&Mg$_2$S$_2$ + M&$\longrightarrow$ &MgS + MgS + M& 4.40$\times 10^{-10}$& 0& 98600.0& Same as F16\\
M17&MgS + Mg$_2$S$_2$ &$\longrightarrow$& Mg$_3$S$_3$ & 2.24$\times 10^{-15}$& 0& -2878.9 & Same as F17\\
M18&Mg$_3$S$_3$ + M&$\longrightarrow$ &MgS + Mg$_2$S$_2$ + M& 4.40$\times 10^{-10}$& 0& 98600.0& Same as F18\\
M19 &MgS + Mg$_3$S$_3$+ M &$\longrightarrow$& Mg$_4$S$_4$ & 1.20$\times 10^{-32}$& 0& 2160.0 &Same as F19\\
M20&Mg$_4$S$_4$ + M&$\longrightarrow$ &MgS + Mg$_3$S$_3$ + M& 4.40$\times 10^{-10}$& 0& 98600.0& Same as F20\\
M21&Mg$_4$S$_4$ + M&$\longrightarrow$ &Mg$_2$S$_2$ + Mg$_2$S$_2$ + M& 4.40$\times 10^{-10}$& 0& 98600.0& Same as F21\\
M22&Mg + Mg + M &$\longrightarrow$ &Mg$_2$ +M & 2.76$\times 10^{-29}$& 0& -2821.4& Same as F22\\
M23&Mg$_2$ + M &$\longrightarrow$ &Mg + Mg +M & 7.14$\times 10^{-5}$& 0& 17800& Same as F23\\
M24&Mg + Mg$_2$ + M &$\longrightarrow$ &Mg$_3$ +M & 2.76$\times 10^{-29}$& 0& -2821.4& Same as F24\\
M25&Mg$_3$ + M &$\longrightarrow$ &Mg + Mg$_2$ +M & 1.66$\times 10^{-5}$& 0& 19200& Same as F25\\
M26&Mg + Mg$_3$ + M &$\longrightarrow$ &Mg$_4$ +M & 2.76$\times 10^{-29}$& 0& -2821.4& Same as F26\\
M27&Mg$_4$ + M &$\longrightarrow$ &Mg + Mg$_3$ +M & 8.30$\times 10^{-7}$& 0& 21600& Same as F27\\
M28&Mg$_2$ + Mg$_2$ + M &$\longrightarrow$ &Mg$_3$ +Mg & 8.30$\times 10^{-10}$& 0& 0& Same as F28\\
M29&Mg$_2$ + Mg$_2$ + M &$\longrightarrow$ &Mg$_4$ & 8.30$\times 10^{-10}$& 0& 0& Same as F29\\
\tableline
\multicolumn{8}{c}{\bf Carbon - C}  \\
\tableline
C1 &C+ C &$\longrightarrow$ &C$_2$+ h$\nu$ & 4.36$\times 10^{-18}$& 0.35& 161.3& Andreazza \& Singh 1997\\
C2 &C + C$_2$ &$\longrightarrow$ &C$_3$ + h$\nu$& 1.00$\times 10^{-17}$& 0& 0& Clayton et al. 1999\\
C3 &C + C$_3$ &$\longrightarrow$ &C$_4$ + h$\nu$& 1.00$\times 10^{-10}$& 0& 0&  Clayton et al. 1999\\
C4 &C + C$_4$ &$\longrightarrow$ &C$_5$ + h$\nu$& 1.00$\times 10^{-13}$& 0& 0&  Clayton et al. 1999\\
C5 &C + C$_4$ &$\longrightarrow$ &C$_2$ + C$_3$& 1.00$\times 10^{-10}$& 0& 0&  Clayton et al. 1999\\
C6 &C + C$_5$ &$\longrightarrow$ &C$_6$ + h$\nu$& 1.00$\times 10^{-10}$& 0& 0&  Clayton et al. 1999\\
C7 &C + C$_6$ &$\longrightarrow$ &C$_7$ + h$\nu$& 1.00$\times 10^{-13}$& 0& 0&  Clayton et al. 1999\\
C8 &C + C$_6$ &$\longrightarrow$ &C$_2$ + C$_5$ & 1.00$\times 10^{-10}$& 0& 0&  Clayton et al. 1999\\
C9 &C + C$_6$ &$\longrightarrow$ &C$_3$ + C$_4$ & 1.00$\times 10^{-10}$& 0& 0&  Clayton et al. 1999\\
C10 &C + C$_7$ &$\longrightarrow$ &C$_8$ + h$\nu$& 1.00$\times 10^{-10}$& 0& 0&  Clayton et al. 1999\\
C11 &C + C$_8$ &$\longrightarrow$ &C$_9$ + h$\nu$& 1.00$\times 10^{-13}$& 0& 0&  Clayton et al. 1999\\
C12 &C + C$_8$ &$\longrightarrow$ &C$_2$ + C$_7$& 1.00$\times 10^{-10}$& 0& 0&  Clayton et al. 1999\\
C13 &C + C$_8$ &$\longrightarrow$ &C$_3$ + C$_6$& 1.00$\times 10^{-10}$& 0& 0&  Clayton et al. 1999\\
C14 &C + C$_8$ &$\longrightarrow$ &C$_4$ + C$_5$& 1.00$\times 10^{-10}$& 0& 0&  Clayton et al. 1999\\
C15 &C + C$_9$ &$\longrightarrow$ &C$_{10}$ + h$\nu$& 1.00$\times 10^{-10}$& 0& 0&  Clayton et al. 1999\\
C16 &C$_2$ + C$_2$ &$\longrightarrow$ &C$_4$ + h$\nu$& 1.00$\times 10^{-10}$& 0& 0&  Same as C2\\
C17&C$_2$ + C$_2$ &$\longrightarrow$ &C$_3$ + C& 5.38$\times 10^{-10}$& 0& 0& Kruse \& Roth 1997\\
C18 &C$_2$ + C$_3$ &$\longrightarrow$ &C$_5$ + h$\nu$& 1.00$\times 10^{-10}$& 0& 0&  Same as C3\\
C19 &C$_2$ + C$_4$ &$\longrightarrow$ &C$_6$ + h$\nu$& 1.00$\times 10^{-10}$& 0& 0&  Same as C18\\
C20&C$_2$ + C$_4$ &$\longrightarrow$ &C$_5$ + C& 5.38$\times 10^{-10}$& 0& 0&  Same as C17\\
C21 &C$_2$ + C$_5$ &$\longrightarrow$ &C$_7$ + h$\nu$& 1.00$\times 10^{-10}$& 0& 0&  Same as C18\\
C22 &C$_2$ + C$_6$ &$\longrightarrow$ &C$_8$ + h$\nu$& 1.00$\times 10^{-10}$& 0& 0&  Same as C18\\
C23&C$_2$ + C$_6$ &$\longrightarrow$ &C$_7$ + C& 5.38$\times 10^{-10}$& 0& 0&  Same as C17\\
C24 &C$_2$ + C$_7$ &$\longrightarrow$ &C$_9$ + h$\nu$& 1.00$\times 10^{-10}$& 0& 0&  Same as C18\\
C25&C$_2$ + C$_8$ &$\longrightarrow$ &C$_{10}$ + h$\nu$& 1.00$\times 10^{-10}$& 0& 0&  Same as C18\\
C26&C$_2$ + C$_8$ &$\longrightarrow$ &C$_9$ + C& 5.38$\times 10^{-10}$& 0& 0&  Same as C17\\
C27 &C$_3$ + C$_3$ &$\longrightarrow$ &C$_6$ + h$\nu$& 1.00$\times 10^{-10}$& 0& 0&  Curl \& Haddon 1993\\
C28 &C$_3$ + C$_4$ &$\longrightarrow$ &C$_7$ + h$\nu$& 1.00$\times 10^{-10}$& 0& 0&  Same as C27\\
C29 &C$_3$ + C$_5$ &$\longrightarrow$ &C$_8$ + h$\nu$& 1.00$\times 10^{-10}$& 0& 0&  Same as C27\\
C30 &C$_3$ + C$_6$ &$\longrightarrow$ &C$_9$ + h$\nu$& 1.00$\times 10^{-10}$& 0& 0&  Same as C27\\
C31 &C$_3$ + C$_7$ &$\longrightarrow$ &C$_{10}$ + h$\nu$& 1.00$\times 10^{-10}$& 0& 0&  Same as C27\\
C32 &C$_4$ + C$_4$ &$\longrightarrow$ &C$_{8}$ + h$\nu$& 1.00$\times 10^{-10}$& 0& 0&  Same as C27\\
C33 &C$_4$ + C$_5$ &$\longrightarrow$ &C$_{9}$ + h$\nu$& 1.00$\times 10^{-10}$& 0& 0&  Same as C27\\
C34 &C$_4$ + C$_6$ &$\longrightarrow$ &C$_{10}$ + h$\nu$& 1.00$\times 10^{-10}$& 0& 0&  Same as C27\\
C35 &C$_5$ + C$_5$ &$\longrightarrow$ &C$_{10}$ + h$\nu$& 1.00$\times 10^{-10}$& 0& 0&  Same as C27\\
C36 &C+ C+ M&$\longrightarrow$ &C$_{2}$ + M& 5.46$\times 10^{-31}$& 1.60& 0&  NIST\\
C37 &C$_2$ + C$_2$ +M&$\longrightarrow$ &C$_{4}$ + M& 2.48$\times 10^{-21}$& -3.00& 0&  Kruse \& Roth 1997\\
C38 &O + C$_2$ &$\longrightarrow$ &C + CO& 5.99$\times 10^{-10}$& 0& 1420&  NIST\\
C39 &O + C$_3$ &$\longrightarrow$ &C$_2$ + CO& 1.00$\times 10^{-11}$& 0.30& 1130&  Woon \& Herbst 1996\\
C40 &O + C$_4$ &$\longrightarrow$ &C$_3$ + CO& 3.00$\times 10^{-10}$& 0.00& 1420&  Terzieva \& Herbst 1998\\
C41 &O + C$_5$ &$\longrightarrow$ &C$_4$ + CO& 1.00$\times 10^{-11}$& -0.30& 1130&  Same as C39\\
C42 &O + C$_6$ &$\longrightarrow$ &C$_5$ + CO& 3.00$\times 10^{-10}$&  0.00& 1420&  Same as C40\\
C43 &O + C$_7$ &$\longrightarrow$ &C$_6$ + CO& 1.00$\times 10^{-11}$& -0.30& 1130&  Same as C39\\
C44 &O + C$_8$ &$\longrightarrow$ &C$_7$ + CO& 3.00$\times 10^{-10}$&  0.00& 1420&  Same as C40\\
C45 &O + C$_9$ &$\longrightarrow$ &C$_8$ + CO& 1.00$\times 10^{-11}$& -0.30& 1130&  Same as C39\\
C46 &C$_3$ + M&$\longrightarrow$ &C + C$_2$+M& 6.64$\times 10^{-8}$& 0& 75406&Kruse \& Roth 1997\\
C47 &C$_4$ + M&$\longrightarrow$ &C$_2$ + C$_2$+M& 2.49$\times 10^{-8}$& 0& 69753&Reaction TF18 in CD09s\\
C48 &C$_5$ + M&$\longrightarrow$ &C$_2$ + C$_3$+M& 6.64$\times 10^{-8}$& 0& 75406&Same as C46\\
C49 &C$_6$ + M&$\longrightarrow$ &C$_2$ + C$_4$+M& 2.49$\times 10^{-8}$& 0& 69753&Same as C47\\
C50 &C$_7$ + M&$\longrightarrow$ &C$_2$ + C$_5$+M& 6.64$\times 10^{-8}$& 0& 75406&Same as C46\\
C51 &C$_8$ + M&$\longrightarrow$ &C$_2$ + C$_6$+M& 2.49$\times 10^{-8}$& 0& 69753&Same as C47\\
C52 &C$_9$ + M&$\longrightarrow$ &C$_2$ + C$_7$+M& 6.64$\times 10^{-8}$& 0& 75406&Same as C46\\
C53 &C$_{10}$ + M&$\longrightarrow$ &C$_2$ + C$_8$+M& 2.49$\times 10^{-9}$& 0& 69753&E \\
C54 &He$^+$ + C$_3$ &$\longrightarrow$ &C$^+$ + C$_2$ + He& 1.60$\times 10^{-09}$& 0& 0& Reaction IM15 in CD09\\
C55 &He$^+$ + C$_4$ &$\longrightarrow$ &C$^+$ + C$_3$ + He& 1.60$\times 10^{-09}$& 0& 0& Same as C54\\
C56 &He$^+$ + C$_5$ &$\longrightarrow$ &C$^+$ + C$_4$ + He& 1.60$\times 10^{-09}$& 0& 0& Same as C54\\
C57 &He$^+$ + C$_6$ &$\longrightarrow$ &C$^+$ + C$_5$ + He& 1.60$\times 10^{-09}$& 0& 0& Same as C54\\
C58 &He$^+$ + C$_7$ &$\longrightarrow$ &C$^+$ + C$_6$ + He& 1.60$\times 10^{-09}$& 0& 0& Same as C54\\
C59 &He$^+$ + C$_8$ &$\longrightarrow$ &C$^+$ + C$_7$ + He& 1.60$\times 10^{-09}$& 0& 0& Same as C54\\
C60 &He$^+$ + C$_9$ &$\longrightarrow$ &C$^+$ + C$_8$ + He& 1.60$\times 10^{-09}$& 0& 0& Same as C54\\
\enddata
\tablenotetext{a}{The rates are given in the Arrhenius form $k_{ij}(T) = A_{ij} \times ({T \over 300})^{\nu}  \times \exp (-E_{a} / T)$, where $T$ is the gas temperature, $A_{ij}$ the Arrhenius coefficient in s$^{-1}$ molecule$^{-1}$, cm$^3$ or cm$^6$ s$^{-1}$ molecule$^{-1}$ for a unimolecular, bimolecular or termolecular processes respectively, $\nu$ has no unit and reflects the temperature dependance of the reaction, and  $E_{a}$ is the activation energy barrier in K$^{-1}$.}
\tablenotetext{b}{NIST is the NIST chemical kinetics database. UDFA06 is by Woodall et al. (2007). Other references are listed in the bibliography. 'E' means 'estimated'. Reaction C53 has been estimated according to reaction C47 decreased by a factor of 10 to account for the lower reactivity of the C$_{10}$ ring compared to other chains.}
\tablenotetext{c}{The destruction rates by Compton electrons are calculated for the 170 \Ms\ progenitor. For other progenitor masses, see Table 6 of CD09.}
\end{deluxetable}
\begin{deluxetable}{ccccccccc}
\rotate
\tabletypesize{\scriptsize}
\tablewidth{0pt}
\tablecaption{Gas temperature and number density profiles as a function of time for the 20~\Ms~CCSN  and the 170~\Ms~PISN used in this study. Similar profiles for the 20 \Ms~CCSN of Todini \& Ferrrara (2001) and the 170 \Ms~PISN of Schneider et al. (2004) are also given for comparison. }     
 \tablehead{
  & \multicolumn{2}{c}{20 M$_{\odot}$} & \multicolumn{2}{c}{170 M$_{\odot}$}& \multicolumn{2}{c}{ Todini \& Ferrara} &\multicolumn{2}{c}{Schneider et al.} \\
\colhead{Time (Days)} & \colhead { T (K)} & \colhead {n$_{gas}$ (cm$^{-3}$)}& \colhead { T (K)} & \colhead {n$_{gas}$ (cm$^{-3}$)}& \colhead { T (K)}& \colhead {n$_{gas}$ (cm$^{-3}$)}& \colhead { T (K)} & \colhead {n$_{gas}$ (cm$^{-3}$)}}
\startdata  
 100 & 18000 & $1.4\times 10^{12}$ & 21000 & $3.9\times 10^{11} $ &5400& $1.0\times 10^{8}$& 5400& $1.0\times 10^{12}$ \\
 200 &  5200 & $1.8\times 11^{11} $&  11800 & $4.9\times 10^{10} $ &3200& $1.0\times 10^{8}$&3200& $1.2\times 10^{11}$\\
 300 & 2500 & $5.2\times 10^{10}$ &  6340 & $1.4\times 10^{10}$ &2400& $1.0\times 10^{8}$&2400& $3.7\times 10^{10}$\\    
 400 & 1500 & $2.2\times 10^{10}$ &  3330 & $6.0\times 10^{9}$ &1900& $1.0\times 10^{8}$&1900 &$1.6\times 10^{12}$\\   
 500 & 1000 & $1.2\times 10^{10} $& 1880 & $3.1\times 10^{9}$ &1600& $1.0\times 10^{8}$&1600& $8.0\times 10^{9}$\\ 
 600 & 740 & $6.7\times 10^{9} $& 1190 & $1.8\times 10^{9}$ &1400& $1.0\times 10^{8}$&1400& $4.6\times 10^{9}$\\ 
 700 & 560 & $3.8\times 10^{8}$ & 857 & $1.1\times 10^{9}$ &1300& $1.0\times 10^{8}$&1300& $2.9\times 10^{9}$\\ 
 800 & 440 & $2.8\times 10^{9}$ &  636 & $7.5\times 10^{8}$&1100& $1.0\times 10^{8}$&1100 & $1.9\times 10^{9}$\\ 
 900 & 360 & $2.0\times 10^{9} $&  479 & $5.2\times 10^{8}$ &1000& $1.0\times 10^{8}$&1000 & $1.4\times 10^{9}$\\ 
 1000 & 300 & $1.4\times 10^{9}$ & 470 & $3.9\times 10^{8}$ &960& $1.0\times 10^{8}$&960 & $1.0\times 10^{9}$   
\enddata
\end{deluxetable}

\begin{deluxetable}{lcc}
\tabletypesize{\scriptsize}
\tablecaption{Mass yields of clusters (in \Ms) ejected at day 1000 for the 170 and 20 \Ms\ fully-mixed ejecta. \tablenotemark{a} }
 \tablewidth{0pt}
 \tablehead{
\colhead{ Clusters } & \colhead{170 M$_{\odot}$ } & \colhead{20 M$_{\odot}$\tablenotemark{b}  } \\
}
\startdata
(SiO)$_5$ & 25.39 & 0.154 \\
(MgO)$_4$& 0 & 0  \\
(FeO)$_4$ & 0 & 0  \\
AlO & 2.86 $\times 10^{-3}$ & 7.42 $\times 10^{-4}$  \\
(FeS)$_4$ & 0 & 0 \\
(MgS)$_4$ & 0  & 0 \\
(Fe)$_4$ & 4.40$ \times 10^{-3} $ & 4.32 $\times 10^{-4}$  \\
(Mg)$_4$& 3.18$ \times 10^{-3}$ & 1.56$ \times 10^{-3}$  \\
(Si)$_4$ & 0 &1.59 $\times 10^{-4}$  \\
C$_{10}$& 0 &  0  \\
(SiC)$_2$ & 0 & 0 \\
\hline

Total Cluster Mass&{\bf 25.4}  & {\bf 0.157} \\
Efficiency & {\bf 30.9  \%} & {\bf 2.7  \%}\\
 \enddata
\tablenotetext{a}{The efficiency is defined as the ratio of the cluster mass to the He core mass given in CD09.} 
\tablenotetext{b}{A mass cut of 2.4 \Ms~is assumed as in Nozawa et al. (2003).}
\end{deluxetable}
\clearpage

\begin{deluxetable}{lcrcrcr}
\tabletypesize{\scriptsize}
\tablewidth{0pt}
\tablecaption{Dust mass yields (in \Ms) ejected at day 1000 for the 170 \Ms\ and the 20 \Ms\ fully-mixed ejecta. The dust mass yields from Nozawa et al. (2003) for the same mass progenitors are also listed. Values quoted for the present study are upper limits to dust yields. Two cases of metal depletion are considered: a case where Mg/Fe atoms are totally depleted by dust formation and a case where no metal depletion occurs (see text). The \% columns represent the mass fraction of the total dust mass for each ejecta. }
 \tablehead{
 \multicolumn{1}{c}{  }& \multicolumn{4}{c}{This Study} & \multicolumn{2}{c}{NK03}  \\
\\
\colhead{Dust}&\colhead{Mg/Fe depleted} & \colhead {\% of Dust} &\colhead{No depletion}& \colhead {\% of Dust} &\colhead{Dust Mass} & \colhead {\% of Dust}}
\startdata  
{\bf 170 M$_{\odot}$}& & & & & & \\ 
 Silica (SiO$_2$) & 12.9  &38.7 \% &17.3&68.1 \%&  20.0 & 56.1 \% \\
Fayalite (Fe$_2$SiO$_4$) &6.6& 19.8 \%& 0& -- &0 & -- \\
 Forsterite (Mg$_2$SiO$_4$ ) & 5.7 &17.1 \% & 0& -- & 11.3&31.7 \% \\
 Pure silicon (Si) & 8.1&24.3 \% &8.1&31.8 \% &0& -- \\
Alumina (Al$_2$O$_3$) &0.003 &0.009 \%& 0.003 &0.012 \%& 0.05&0.14 \% \\
 Magnetite (Fe$_3$O$_4$)    & 0 & -- & 0& --  &4.3 &12.0 \% \\
 Pure iron (Fe) &0.004&0.012 \% & 0.004&0.016 \% & 0 & -- \\ 
Carbon (AC) & 0 & -- & 0 & -- & 0& -- \\
Silicon carbide (SiC) & 0 & -- & 0 & -- & 0& -- \\
\hline
 Total Dust & {\bf 33.3} & 100 \% & {\bf 25.4}& 100 \% & {\bf 35.6} & 100 \% \\
 \hline
{\bf 20 M$_{\odot}$} & & & & & & \\  
  Silica (SiO$_2$) & 0  & --  &0.105 &67.5 \%&  0.42 & 57.5\% \\
  Fayalite (Fe$_2$SiO$_4$) &0.125& 37.4 \%& 0& -- &0 & -- \\
  Forsterite (Mg$_2$SiO$_4$ ) & 0.160 &47.7 \% & 0& -- & 0.23&31.5 \% \\
  Pure silicon (Si)& 0.049&14.7 \% &0.049&31.5 \% &0& -- \\
  Alumina (Al$_2$O$_3$) &8.8 $\times 10^{-4}$  &0.26 \% & 8.8 $\times 10^{-4}$ &0.56 \% & 9.0 $\times 10^{-4}$& 0.12 \% \\
  Magnetite (Fe$_3$O$_4$)    & 0 & -- & 0& --  &0.08 &10.9 \% \\
  Pure iron (Fe) &0&--& 4.32 $\times 10^{-4}$&0.28 \% & 0 & -- \\
  Pure Magnesium (Mg)& 0&--&0.0014&0.88 \%&0&--\\
  Carbon (AC) & 0 & -- & 0 & -- & 0& -- \\
  Silicon carbide (SiC) & 0 & -- & 0 & -- & 0& -- \\
  \hline
 Total Dust & {\bf 0.33} & 100 \% & {\bf 0.16}& 100 \% & {\bf 0.73} & 100 \% 
 \enddata
\end{deluxetable}
\clearpage

\begin{deluxetable}{lcccccccc}
\rotate
\tabletypesize{\scriptsize}
\tablewidth{0pt}
\tablecaption{Molecular and dust yields (in M$_{\odot}$) ejected at 1000 days for our comparative study of Schneider at al. (2004, SFS04) models. Values quoted for the present study are upper limits to dust yields. Cluster yields are listed under 'Ejecta". For each progenitor, two cases of metal depletion have been considered (see text) : 1) Mg and Fe are totally depleted by dust formation, and 2) no metal depletion is considered.}      

 \tablehead{
 \multicolumn{1}{c}{  }& \multicolumn{4}{c}{{\bf 170 M$_{\odot}$}} & \multicolumn{4}{c}{{\bf 260 M$_{\odot}$}} \\
\\
\colhead{Molecules/Clusters/Dust}&\colhead{Ejecta} & \colhead {Mg/Fe depleted} &\colhead{No depletion} & \colhead {SFS04}&\colhead{Ejecta} & \colhead {Mg/Fe depleted} &\colhead{No depletion} & \colhead {SFS04}}
\startdata  
CO & 10.5 & 10.5 & 10.5 & 6 & 8.5&8.5&8.5& 0.01 \\
SO & 9.7&9.7&9.7& -- & 17.8 &17.8 &17.8& --  \\
O$_2$ & 30.1 & 13.3 & 30.1 &-- &9.6 & 6.3 & 9.6 & --\\
\hline
Total (Molecules)& {\bf 50.5} & {\bf 33.5} & {\bf 50.5} &{\bf 6}&  {\bf 35.9} &  {\bf 32.6} & {\bf 35.9} & {\bf 0.01}\\
\hline
(SiO)$_5$ & 27.6& & &  & 39.6 & & &  \\
(MgO)$_4$& -- &  &   & & -- & & & \\
(FeO)$_4$ & -- &   &   & & -- & & &  \\
AlO & 2.82 $\times 10^{-2}$ &  &   &  & 2.22 $\times 10^{-2}$ & & &  \\
(FeS)$_4$ & -- &  & &  & -- & & &  \\
(MgS)$_4$ & -- &  &   &  & -- & & &  \\
(Fe)$_4$ & 7.28$ \times 10^{-4} $ &   &  & & 34.3 & & &  \\
(Mg)$_4$ & 1.78$ \times 10^{-1} $ &   &  & & 2.5 & & &  \\
(Si)$_4$ & -- &  &   &  & -- & & & \\
C$_{10}$& -- &   &   &  & -- & & &  \\
\hline 
Total (Clusters)&{\bf27.8} & & & &{\bf 74.8}& & & \\
\hline
 Silica (SiO$_2$) &  & 12.1  &18.8 &  26 &   &14.4& 25.9 & 46.0  \\
Fayalite (Fe$_2$SiO$_4$) & & 4.7& --& -- &   &12.6 & --& -- \\
 Forsterite (Mg$_2$SiO$_4$ )&  & 12.3  & --& 11.0& & 6.0 & -- &11.0 \\
 Pure silicon (Si) & & 8.8&8.8&-- &  & 12.1 & 12.1 & -- \\
Alumina (Al$_2$O$_3$) &  &0.03 & 0.03 & 0.03&   &0.02 & 0.02 & 0.03\\
 Magnetite (Fe$_3$O$_4$)  &   & -- & -- &0.06 &   & -- & --& 8.7\\
 Pure iron (Fe) &   &0& 7.28$ \times 10^{-4} $ & -- &   & 34.3 &34.3&-- \\
 Pure iron (Mg) &   &0& $1.8 \times 10^{-1} $ & -- &   & 2.5 &2.5&-- \\
 Carbon (AC)&   & -- &-- & 1.6&   & --& -- & 3.3 \\
 \hline
 Total (Dust) & & {\bf 37.9} & {\bf 27.8} & {\bf 38.7}& & {\bf 81.9}& {\bf 74.8}&{\bf 69.1}
\enddata
\end{deluxetable}
\clearpage

\begin{deluxetable}{lccccccccccc}
\tabletypesize{\scriptsize}
\rotate
\tablewidth{0pt}
\tablecaption{Upper limits to dust mass yields (in \Ms) ejected at day 1000 for the 170 \Ms\ unmixed ejecta. Two cases of metal depletion are considered: 1) a case where Mg is totally depleted by dust formation (D), and 2) a case where no metal depletion occurs (ND) - see text for more detail.\tablenotemark{a} }
 \tablehead{
 \multicolumn{1}{c}{  }& \multicolumn{1}{c}{Zone 1}& \multicolumn{2}{c}{Zone 2} & \multicolumn{2}{c}{Zone 3} & \multicolumn{2}{c}{Zone 4}&\multicolumn{2}{c}{Zone 5}&\multicolumn{2}{c}{Zone 1-5} \\
  \multicolumn{1}{c}{Zone mass}& \multicolumn{1}{c}{(20 \Ms)}& \multicolumn{2}{c}{(15 \Ms)} & \multicolumn{2}{c}{(15 \Ms)} & \multicolumn{2}{c}{(23 \Ms)}&\multicolumn{2}{c}{(4.3 \Ms)}&\multicolumn{2}{c}{(82.3 \Ms)} \\
\multicolumn{1}{c}{Major elements\tablenotemark{b}}&   \multicolumn{1}{c}{Si/S/Fe}& \multicolumn{2}{c}{O/Si/S/Mg}& \multicolumn{2}{c}{O/Mg/Si}& \multicolumn{2}{c}{O/C/Mg} & \multicolumn{2}{c}{O/C}&\multicolumn{2}{c}{  } 
 \\
 \hline
\colhead{  }& \colhead {ND} &\colhead{D} & \colhead {ND} &\colhead{D} & \colhead {ND} &\colhead{D} &\colhead {ND}&\colhead{D} & \colhead {ND}& \colhead{D} & \colhead {ND}}
\startdata  
Silica  &--&2.577&2.976 &0&0.661& 0 & 1.8$\times 10^{-3}$&0&3.2$\times 10^{-6}$&2.577&3.638\\
Forsterite  &--&0.927&0 & 1.542&0&4.3$\times 10^{-3}$&0&7.5$\times 10^{-6}$&0&2.474&0 \\
Alumina &--&9.2$\times 10^{-4}$ &9.2$\times 10^{-4}$ & 0.028 &0.028 &8.0$\times 10^{-4}$&8.0$\times 10^{-4}$&0&0 &0.0296&0.0297\\
Periclase &--&2.8$\times 10^{-9}$&2.8$\times 10^{-9}$& 2.5$\times 10^{-6}$&2.5$\times 10^{-6}$&0&0&0&0&2.5$\times 10^{-6}$&2.5$\times 10^{-6}$\\
Pure iron&6.7$\times 10^{-5}$& -- & --&--&--&--&--& --&--&6.7$\times 10^{-5}$&6.7$\times 10^{-5}$\\ 
Pure silicon  &0.265 &1.388&1.388&0.308&0.308&8.6$\times 10^{-4}$&8.6$\times 10^{-4}$&1.5$\times 10^{-6}$&1.5$\times 10^{-6}$& 1.963&1.963\\ 
Pure magnesium  & -- & 3.1$\times 10^{-4}$& 3.1$\times 10^{-4}$&7.4$\times 10^{-3}$&7.4$\times 10^{-3}$&2.5$\times 10^{-4}$&2.5$\times 10^{-4}$&3.3$\times 10^{-8}$&3.3$\times 10^{-8}$&2.5$\times 10^{-4}$&0.00802\\ 
Iron sulphide &0.011&--&--&--&--&--&--&--&--&0.011&0.011\\ 
Carbon & -- &-- & --&--&--&1.5$\times 10^{-14}$&1.5$\times 10^{-14}$&1.1$\times 10^{-13}$&1.1$\times 10^{-13}$&1.2$\times 10^{-13}$&1.2$\times 10^{-13}$\\ 
Silicon carbide& -- &-- & --&--&--&--&--&--&--&--&--\\ 
\hline

 Total Dust  & 0.277&{4.894} & {4.365}&{1.885}&{1.004}&{ 0.0062}&{ 0.0038}&{9.0$\times 10^{-6}$}&{4.7$\times 10^{-6}$}&{\bf 7.06}&{\bf 5.65}\\ 
 Efficiency &{1.38 \%}&32.63 \%& 25.1 \%&{12.57 \%}&6.69 \%&0.027 \%&0.016 \%&0 \%&0 \%&{\bf 8.58 \%}&{\bf 6.86 \%}\\ 
 \enddata
 \tablenotetext{a}{The efficiency is defined as the ratio of the molecular mass to the zone mass.} 
 \tablenotetext{b}{In order of decreasing mass yield (see Table 3 of CD09).} 
\end{deluxetable}

\clearpage

\begin{deluxetable}{lccccccccc}
\tabletypesize{\scriptsize}
\rotate
\tablewidth{0pt}
\tablecaption{Upper limits to dust mass yields (in \Ms) ejected at day 1000 for the 20 \Ms\ unmixed ejecta. Two cases of metal depletion are considered: 1) a case where Mg is totally depleted by dust formation (D), and 2) a case where no metal depletion occurs (ND) - see text for more detail.\tablenotemark{a} }
 \tablehead{
 \multicolumn{1}{c}{  }& \multicolumn{1}{c}{Zone 1}& \multicolumn{2}{c}{Zone 2} & \multicolumn{2}{c}{Zone 3} & \multicolumn{2}{c}{Zone 4}&\multicolumn{2}{c}{Zone 1-4} \\
  \multicolumn{1}{c}{Zone mass}& \multicolumn{1}{c}{(0.6 \Ms)\tablenotemark{b} }& \multicolumn{2}{c}{(0.6 \Ms)} & \multicolumn{2}{c}{(1.35 \Ms)} & \multicolumn{2}{c}{(0.9 \Ms)}&\multicolumn{2}{c}{(3.45 \Ms)}\\
\multicolumn{1}{c}{Major elements\tablenotemark{c}}& \multicolumn{1}{c}{Si/S/Fe}& \multicolumn{2}{c}{O/Mg/Si}& \multicolumn{2}{c}{O/C/Mg}& \multicolumn{2}{c}{He/C/O} & \multicolumn{2}{c}{  } 
 \\
\hline
\colhead{  }& \colhead {ND} &\colhead{D} & \colhead {ND} &\colhead{D} & \colhead {ND} &\colhead{D} &\colhead {ND}& \colhead{D} & \colhead {ND}}
\startdata  
Silica  &1.4$\times 10^{-3}$&0&0.038&0&2.8$\times 10^{-5}$& 0&0&0&0.039\\
Forsterite  &--&0.089&0 & 6.5$\times 10^{-5}$&0&0&0&0.089&0 \\
Alumina &--&7.9$\times 10^{-5}$ &7.9$\times 10^{-5}$ & 7.4$\times 10^{-6}$ &7.4$\times 10^{-6}$&0&0&8.6$\times 10^{-5}$&8.6$\times 10^{-5}$\\
Periclase &--&4.9$\times 10^{-4}$&4.9$\times 10^{-4}$& 0&0&0&0&4.9$\times 10^{-4}$&4.9$\times 10^{-4}$ \\
Pure iron&4.6$\times 10^{-5}$& -- & --&--&--&--&--& 4.6$\times 10^{-5}$&4.6$\times 10^{-5}$\\ 
Pure silicon  &0.012 &0.018&0.018&1.3$\times 10^{-5}$&1.3$\times 10^{-5}$&0&0& 0.030&0.030\\ 
Pure magnesium  &-- & 3.9$\times 10^{-4}$& 3.9$\times 10^{-4}$&2.1$\times 10^{-6}$&2.1$\times 10^{-6}$&0&0&3.9$\times 10^{-4}$&3.9$\times 10^{-4}$\\ 
Iron sulphide &0.033&--&--&--&--&--& --&0.033&0.033\\ 
Carbon &-- & --& -- &1.8$\times 10^{-15}$&1.8$\times 10^{-15}$&1.3$\times 10^{-13}$&1.3$\times 10^{-13}$&1.3$\times 10^{-13}$&1.3$\times 10^{-13}$\\ 
Silicon carbide& -- &-- & --&--&--&--&--&--&--\\ 
\hline
 Total Dust  & 0.046&0.108& 0.057&8.8$\times 10^{-5}$&5.1$\times 10^{-5}$&0&0&{\bf 0.154}&{\bf 0.103}\\ 
 Efficiency &7.67 \% & 18.0 \% & 9.5 \%& 0.0065 \%&0.0038 \%&0 \%&0\%&{\bf 4.47 \%}&{\bf 2.98 \%}\\ 
 \enddata
 \tablenotetext{a}{The efficiency is defined as the ratio of the molecular mass to the zone mass.}
 \tablenotetext{b}{A mass cut of 2.4 \Ms\ is assumed for Zone 1 as in NK03}
 \tablenotetext{c}{In order of decreasing mass yield (see Table 3 of CD09).} 
\end{deluxetable}

\clearpage

\begin{deluxetable}{llccccccccc}
\tabletypesize{\scriptsize}
\tablewidth{0pt}
\tablecaption{Depletion efficiencies of metals in molecular clusters and molecules (underlined) for the 170 and 20 \Ms\ unmixed ejecta. \tablenotemark{a} }
\tablehead{
\multicolumn{2}{c}{ }& \multicolumn{5}{c}{170 M$_{\odot}$}&\multicolumn{4}{c}{20 M$_{\odot}$} \\
\colhead{Element} & \colhead{ Clusters } & \colhead{Zone 1} & \colhead{Zone 2 } &\colhead{Zone 3} &\colhead{Zone 4}& \colhead{Zone 5} &\colhead{Zone 1} & \colhead{Zone 2 }& \colhead{Zone 3} &\colhead{Zone 4}}
\startdata  
Si & (SiO$_2$)$_5$ & -- & 50.0 \% & 50.0 \% & 46.7 \% & 37.2 \% &0.12 \% &49.5 \% &48.2 \%& --\\
  & (Si)$_{4,5}$ & 1.8 \% &50.0 \% & 50.0 \% & 46.7 \% & 37.2 \%& 3.1 \% & 49.5 \% & 48.2 \% & --  \\
  & {\underline{SiS}} & {\underline {26.5} \%} & -- & -- & --& -- & {\underline{ 28.3} \%} & -- & -- & --\\
 O & (SiO$_2$)$_5$  & -- & 9.6 \%& 2.7 \% & -- & -- & -- & 3.9 \% & -- & --  \\
   & {\underline {O$_2$}} & -- & {\underline { 38.0} \%} & {\underline {46.4} \%} & {\underline {12.3} \%}  & -- & --  & {\underline {60.9} \%} & {\underline {27.6} \%} & --  \\
      & {\underline {CO}} & -- & --  & -- & {\underline {29.1} \%} &{\underline {43.0} \%}& --  & --  & {\underline {32.7} \%} & --  \\
      & {\underline {SO}} & -- & {\underline {1.2} \%} & {\underline {0.3} \%} & {\underline {0.1} \%} &--&  {\underline {0.7} \%}   & --  & -- & --  \\
 S &(FeS)$_4$ & 0.09 \% & -- & --& --& --& 8.7 \% &-- &--&-- \\
  & {\underline {SiS}} & {\underline {98.9} \%} & -- & -- & --& -- & {\underline {91.4} \%} & -- & -- & --\\
  & {\underline {SO}} & -- & {\underline {100} \%} & {\underline {99.8} \%} & --& -- & --& {\underline {99.2} \%}  & -- & --\\
Al &AlO & -- & 97.0 \% & 98.3\% & 92.1\% & -- & -- & 99.1 \% & 96.8 \% & -- \\
Fe &(FeS)$_4$ & 1.8 \% & -- & --& --& --& 44.0 \% &-- &--&-- \\
  &(Fe)$_4$ & 0.02 \% & -- & --& --& -- & 0.1 \% & -- & -- & --\\
Mg & (Mg)$_4$& -- & 0.1 \% & 2.3 \% & 0.1 \% & -- &--& 0.91 \% &0.04 \% & -- \\
  & (MgO)$_4$& -- & --  & -- & -- & -- & --&-- &0.73 \% & -- \\
C & C$_{10}$& -- & -- & --& --& 0 - 21.6 \% \tablenotemark{b} & -- & --& --& 0 - 95.3 \% \tablenotemark{b}  \\
  & {\underline {CO}} & --  & --  & {\underline {100.0} \%} & {\underline {100.0} \%} & {\underline {77.3} \%} & -- & {\underline {99.8} \%} & {\underline {99.7} \%} & {\underline {0 - 0.14} \%} \tablenotemark{b} \\
\enddata
\tablenotetext{a}{The depletion efficiency is defined by Eq. (41) in the text}
\tablenotetext{b}{The efficiencies are for a He-rich and He-free Zone.} 
\end{deluxetable}
\clearpage


\begin{figure}
\label{}
\epsscale{1.}
\plotone{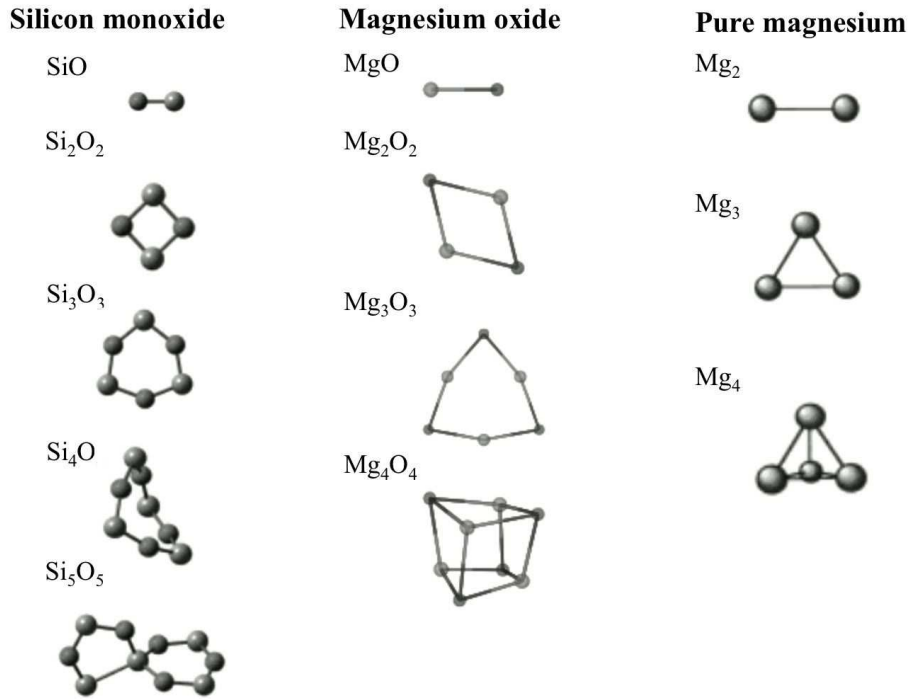}
\caption{The structures of some small clusters are illustrated for silicone monoxide (adapted from Lu et al. 2006), magnesium monoxide (adapted from K{\"o}hler et al. 1997), and pure magnesium (adapted from Jellinek \& Acioli 2002). Structure similar to (MgO)$_n$ applies to iron oxide (FeO)$_n$ and magnesium sulphide (MgS)$_n$ and iron sulphides (FeS)$_n$. Pure silicon and iron clusters have structures similar to pure magnesium (Mg)$_n$ clusters. }
\end{figure}


\clearpage
\begin{figure}
\epsscale{1.15}
\plottwo{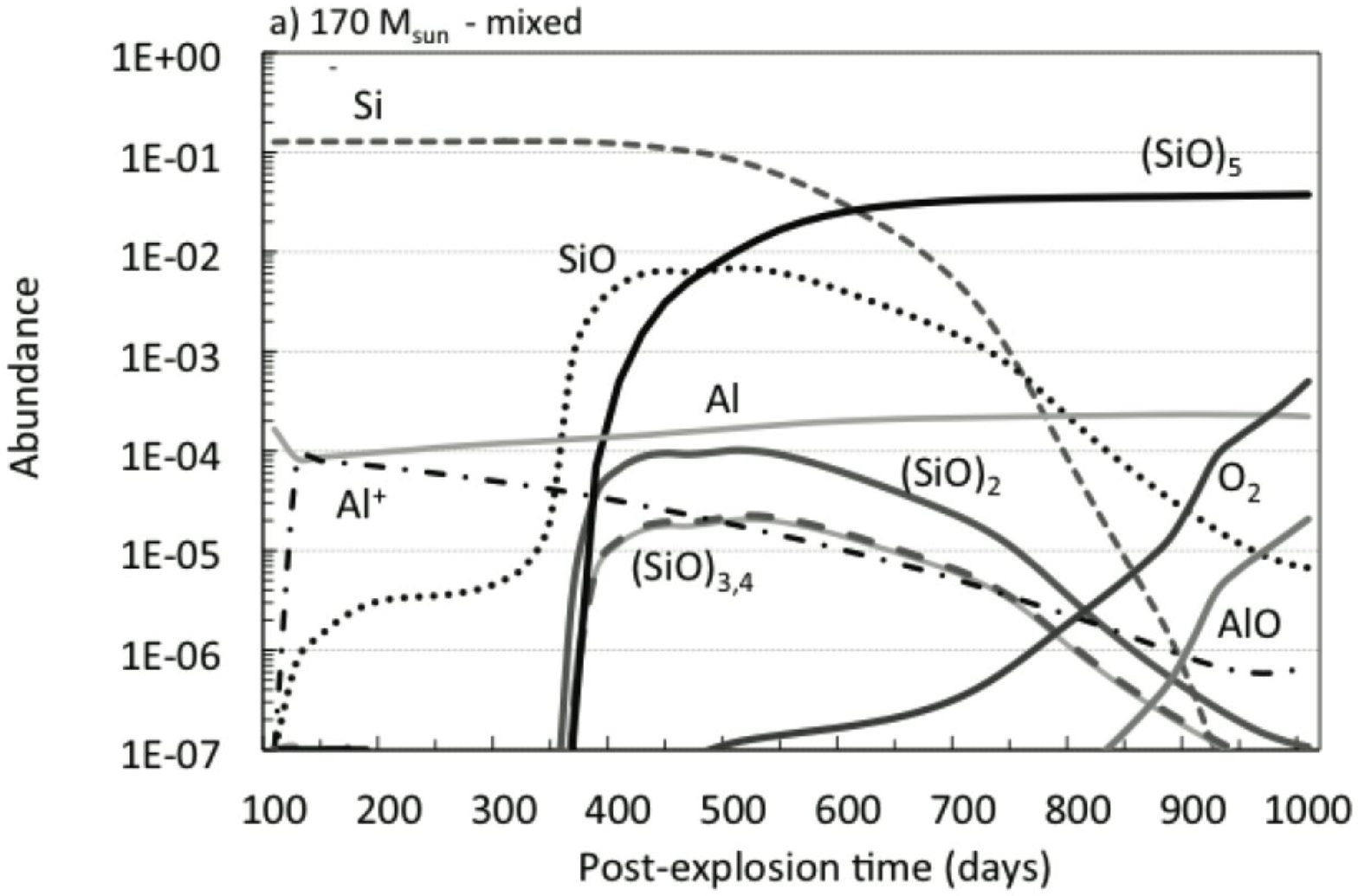}{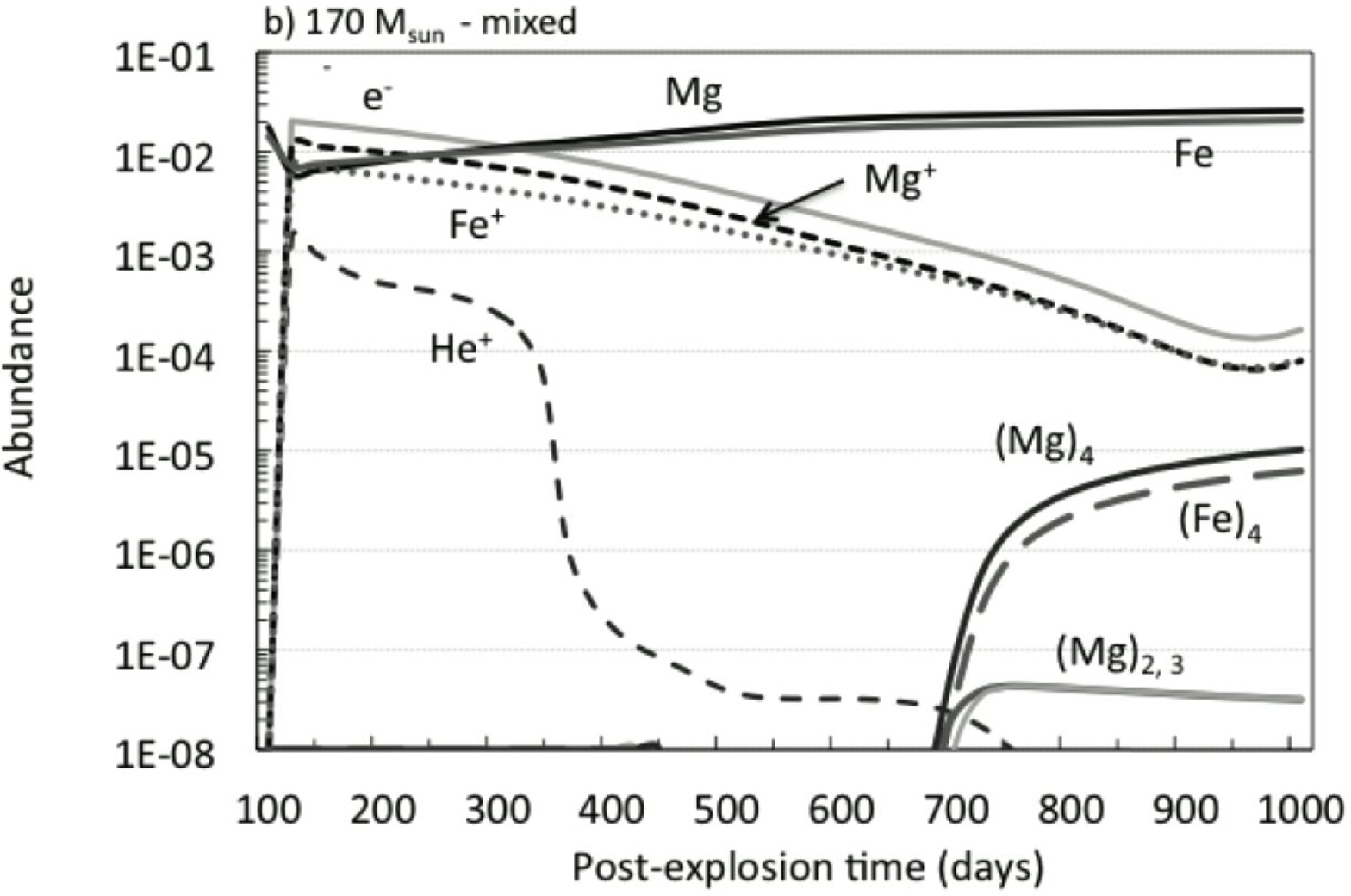}
\caption{The evolution of the cluster abundances normalized to total gas number density for the 170~\Ms~fully-mixed ejecta: a) Silicon oxide clusters and aluminum monoxide; b) Iron clusters, most important ions, and electrons. The dominant precursors are silicon monoxide, pure magnesium and iron clusters, and molecular aluminum monoxide. Carbon chains and rings do not form in fully-mixed ejecta. The abundance of secondary electrons is commanded by the ionization of the metals Mg and Fe. The He$^+$ abundance is high from 100 to 400 days impeding the efficient synthesis of molecules during this timespan (see CD09). The O$_2$ abundance is also shown.  }
\end{figure}


\begin{figure}
\epsscale{1.15}
\plottwo{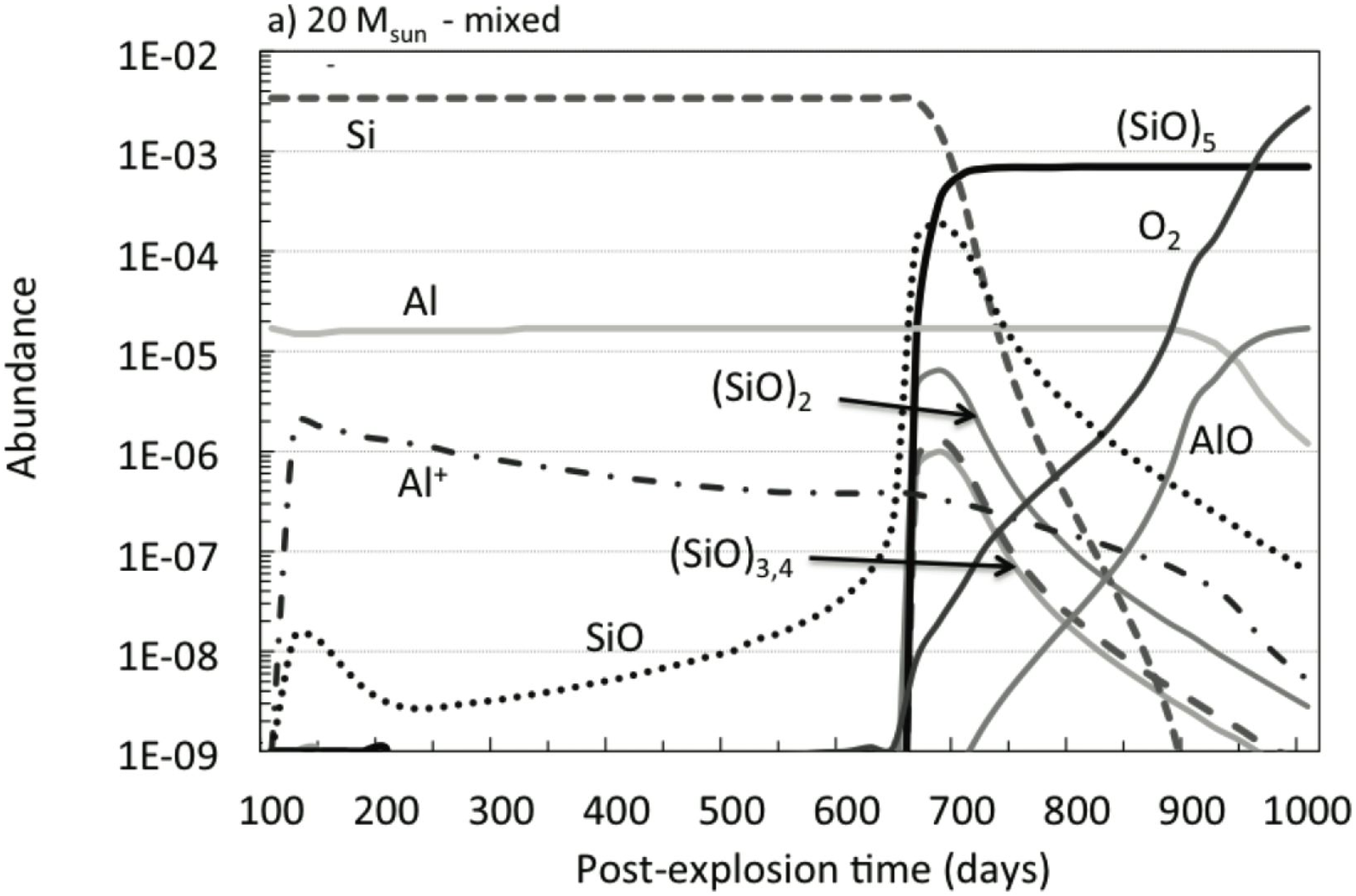}{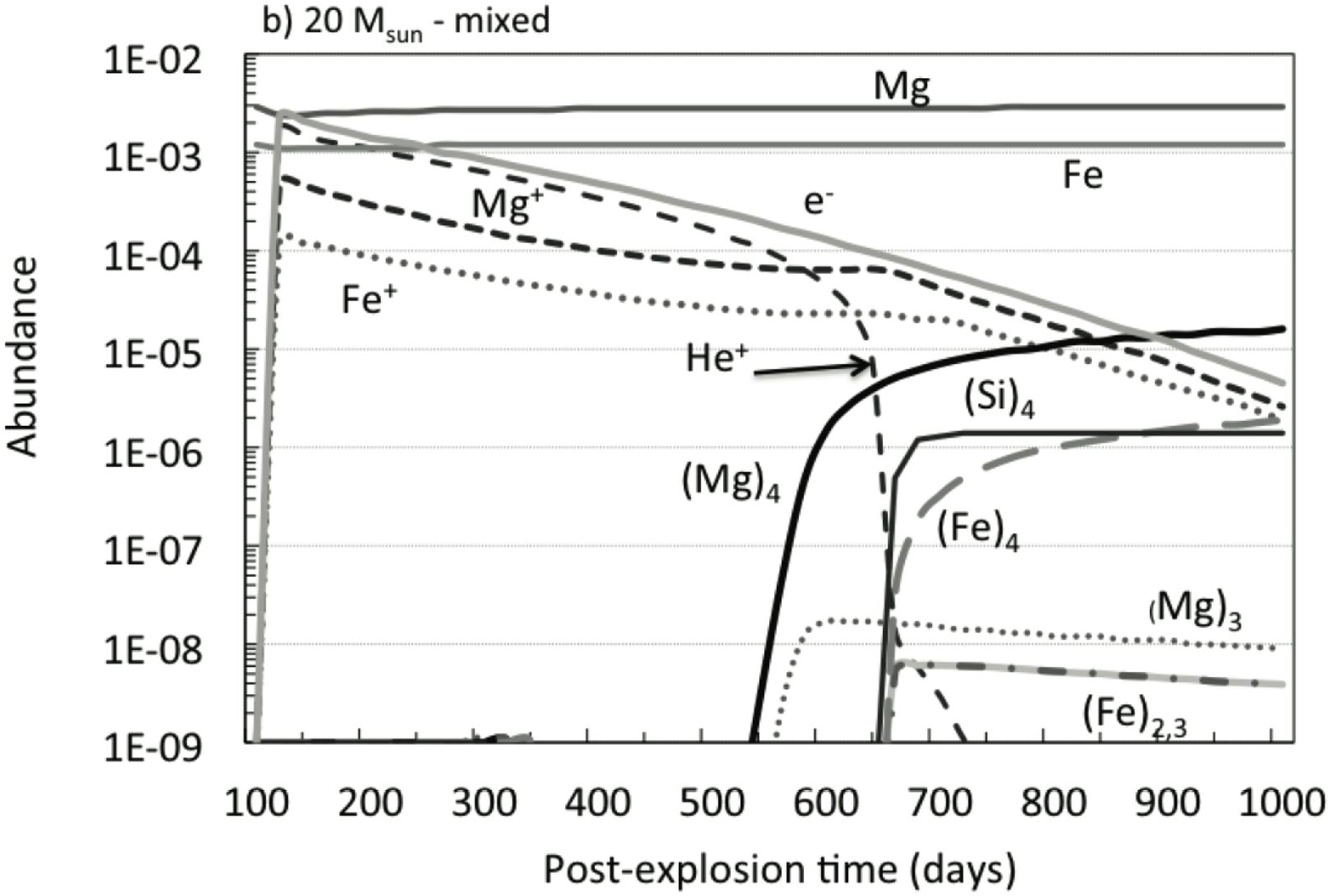}
\caption{The evolution of the cluster abundances normalized to total gas number density for the 20~\Ms~fully-mixed ejecta: a) Silicon oxide clusters and aluminum monoxide; b) Iron clusters, most important ions, and electrons. Carbon chains/rings do not form in significant amounts. The He$^+$ abundance is high from 100 to 700 days impeding the efficient synthesis of molecules during this timespan (see CD09). The abundance of secondary electrons is commanded by the ionization of He and  metals like Mg and Fe. The O$_2$ abundance is also shown. }
\end{figure}


\clearpage
\begin{figure}
\epsscale{0.8}
\plotone{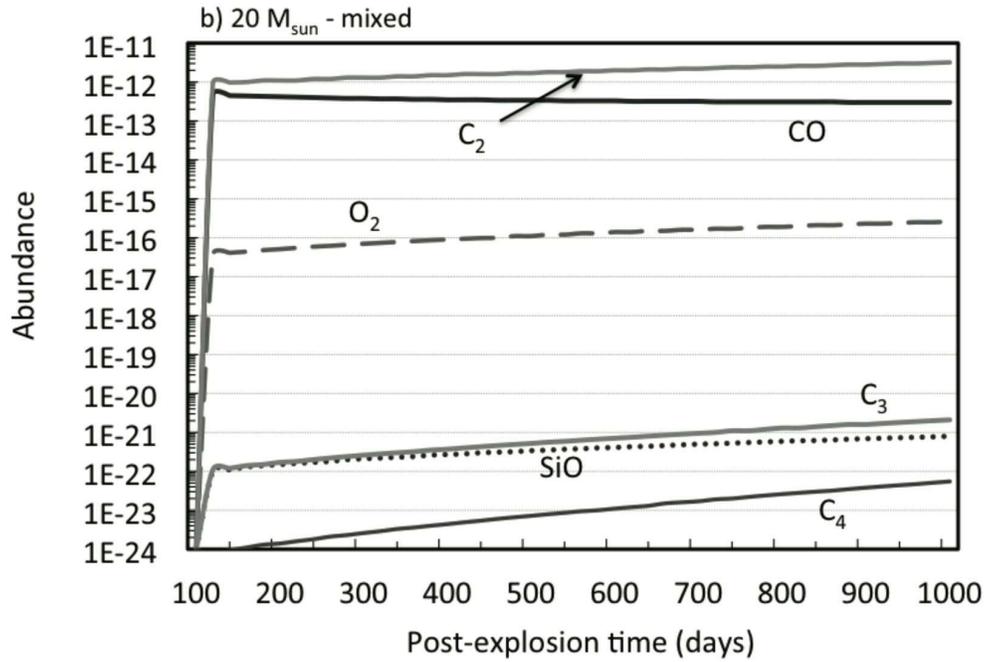}
\caption{Carbon clusters abundances normalized to total gas number density as a function of post-explosion time for the 20~\Ms~fully-mixed ejecta model of Todini \& Ferrara (2001, TF01). Carbon chains abundances (with respect to total gas density) are extremely small due to the large amounts of He$^+$ in the ejecta. These results totally contradict the findings by TF01 who predict the formation of amorphous carbon grains in the ejecta of primordial SNe with 20~\Ms~progenitors -  see \S4.2.2. }
\end{figure}


\clearpage
\begin{figure}
\epsscale{1.15}
\plottwo{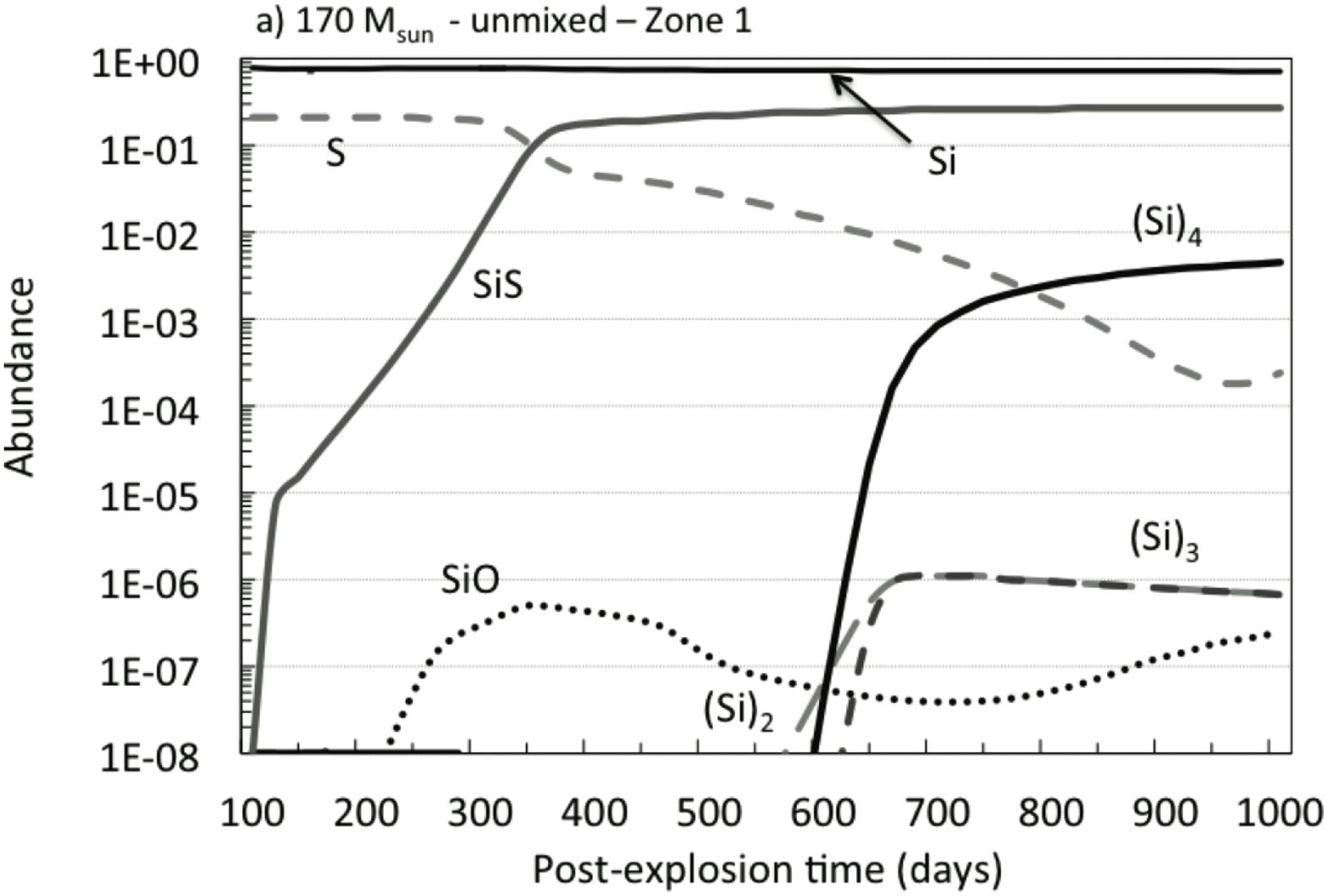}{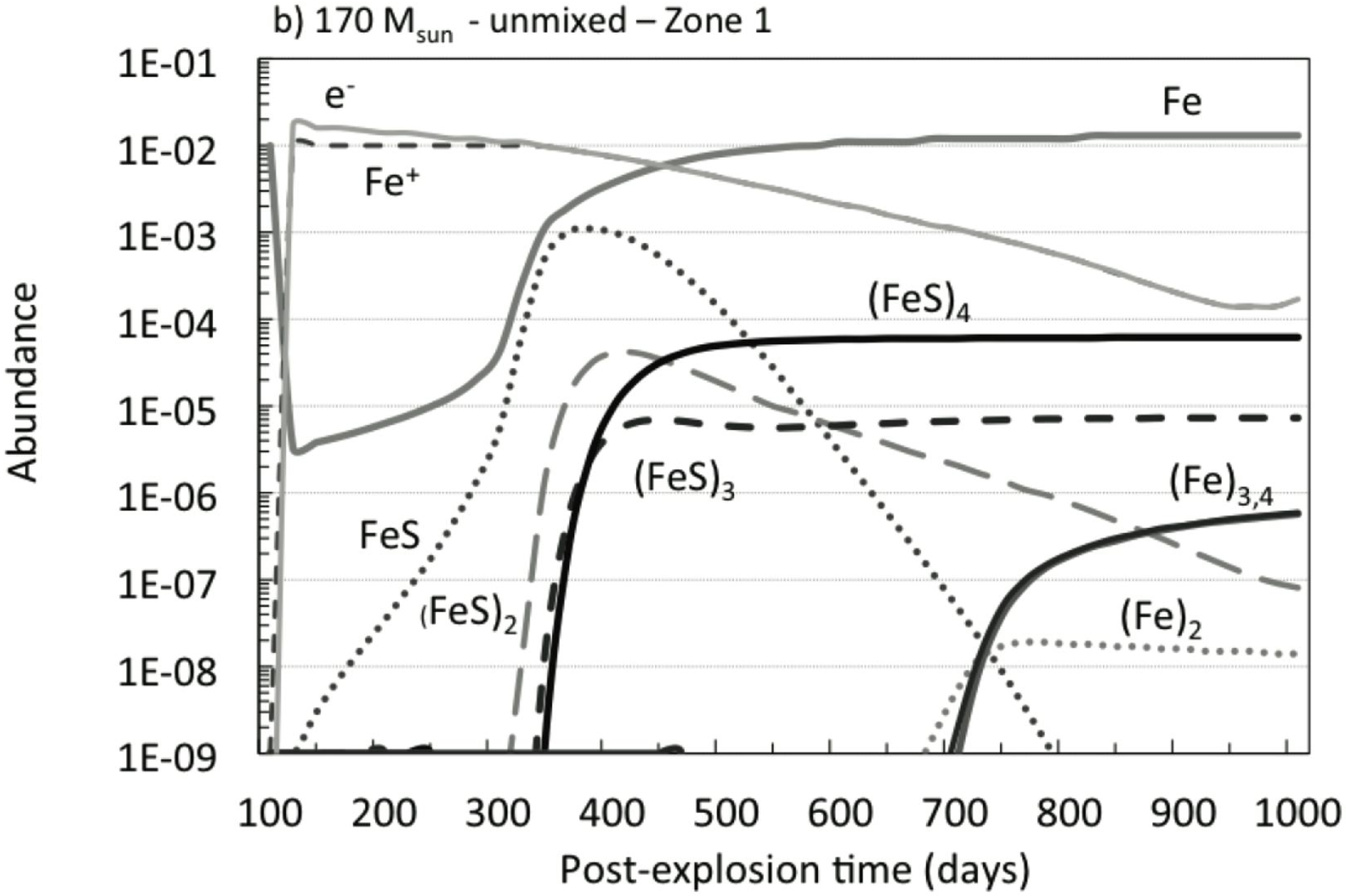}
\caption{The evolution of the cluster abundances normalized to total gas number density for the 170~\Ms~unmixed ejecta - Zone 1: a) Silicon clusters - molecular SiS and SiO are also shown; b) Iron-bearing clusters, most important ions, and electrons. The dominant precursors are pure silicon and iron sulphide clusters.The abundance of secondary electrons is commanded by the ionization of atomic Fe.  }
\end{figure}


\begin{figure}
\epsscale{1.15}
\plottwo{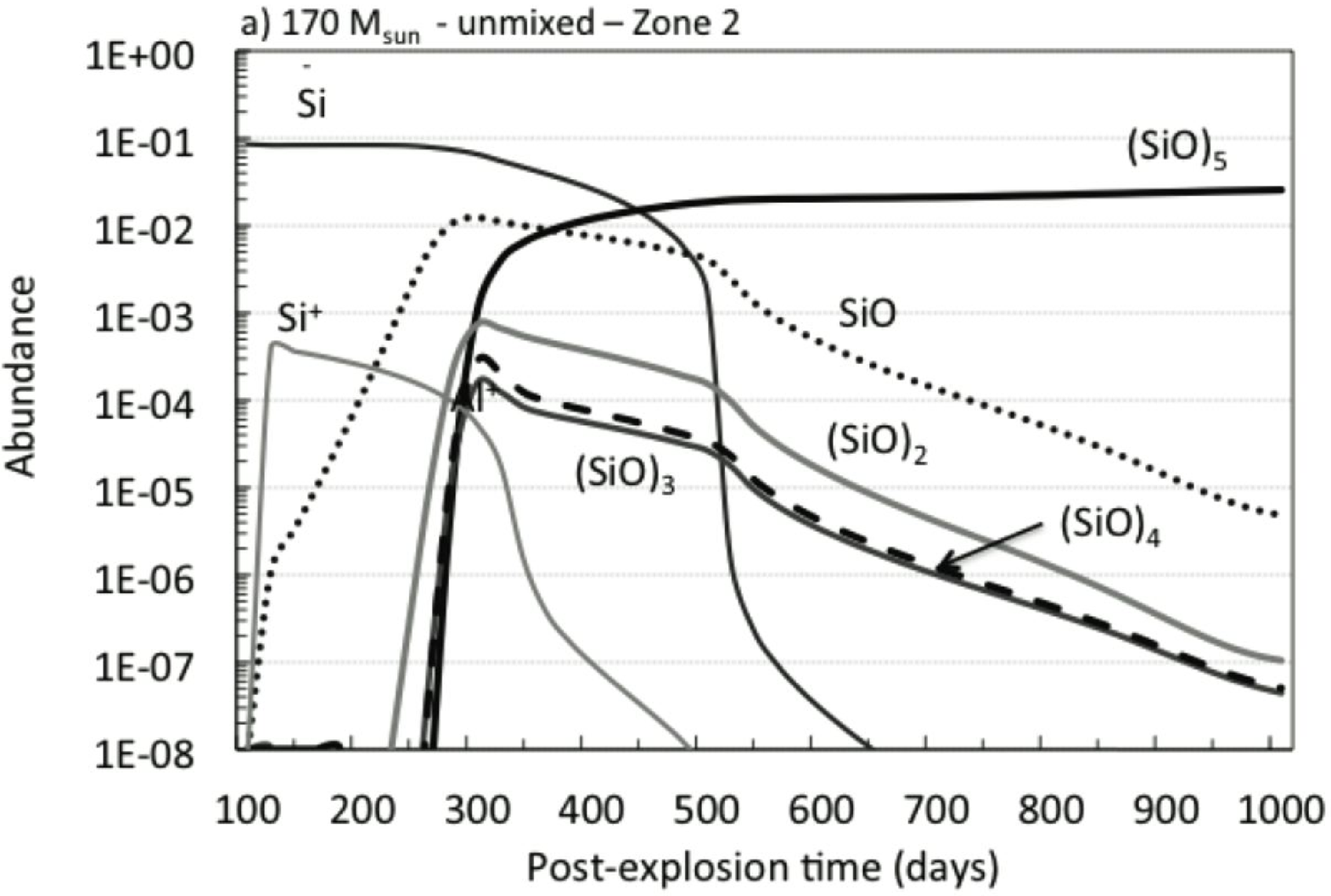}{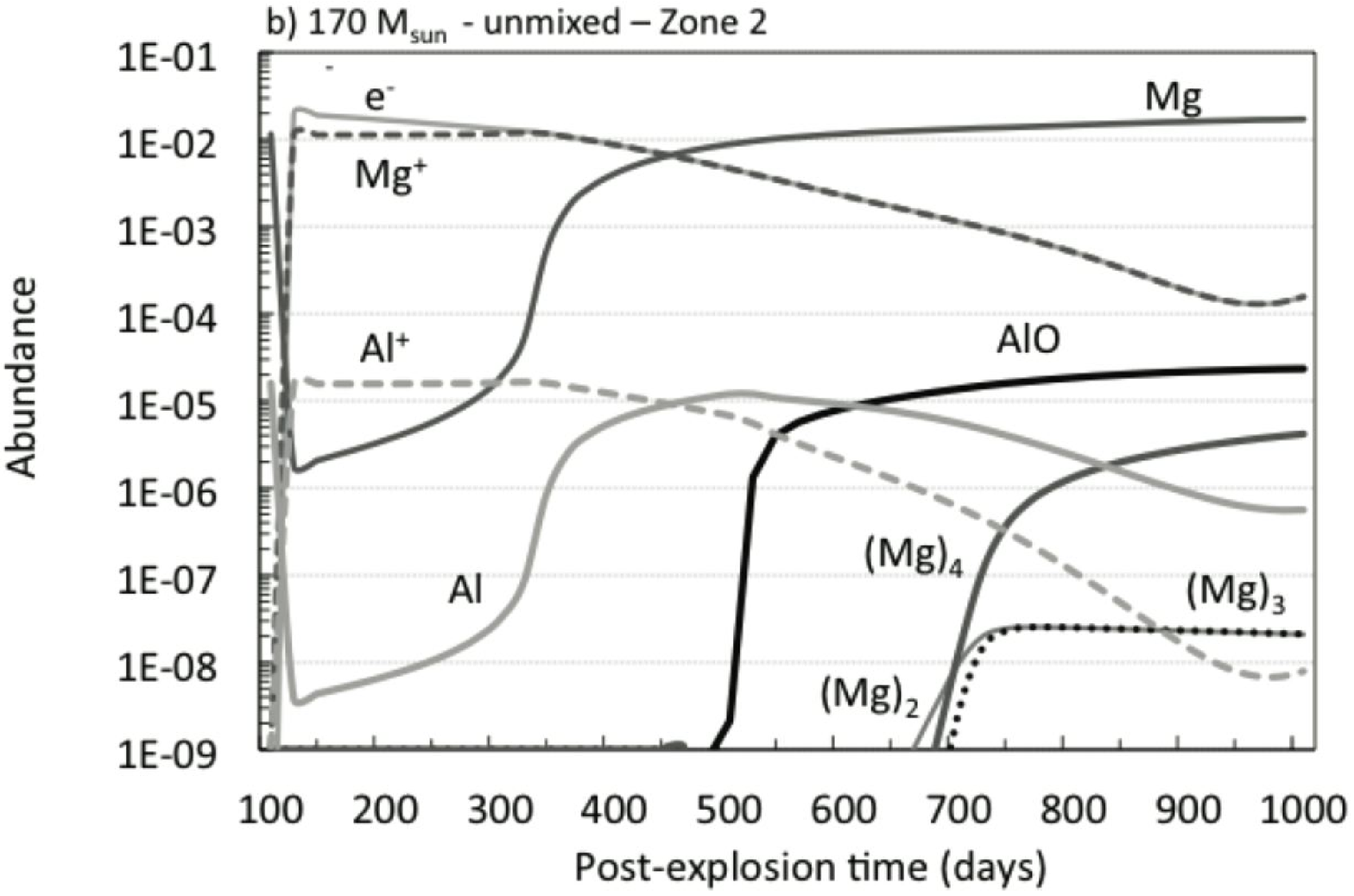}
\caption{The evolution of the cluster abundances normalized to total gas number density for the 170~\Ms~unmixed ejecta - Zone 2: a) Silicon monoxide clusters; b) Magnesium clusters and aluminum monoxide, most important ions, and electrons. The dominant precursors are silicon monoxide clusters and aluminum monoxide. The abundance of secondary electrons is commanded by the ionization of atomic Mg.  }
\end{figure}


\clearpage
\begin{figure}
\epsscale{0.8}
\plotone{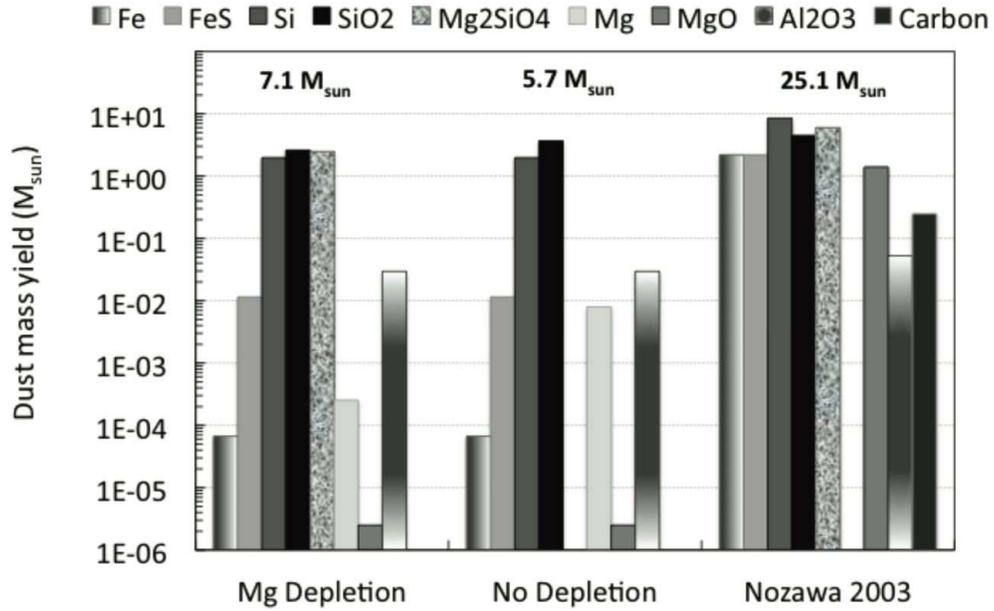}
\caption{Dust chemical composition and mass yields (in units of  \Ms) at t = 1000 days versus depletion case for the 170 \Ms\ unmixed ejecta. Results from Nozawa et al. (2003) are shown for comparison. The total upper limit for dust mass and the total dust mass from NK03 are also indicated. Zone 1 synthesizes pure Si, FeS, and Fe clusters. Silica and/or silicates (forsterite), pure silicon clusters are formed in Zone 2 and Zone 3. No dust is formed in Zone 4. Very small amounts of silica and/or silicates (forsterite), and pure silicon clusters form in Zone 5, when AC and SiC grains do not. }
\end{figure}


\clearpage
\begin{figure}
\epsscale{1.15}
\plottwo{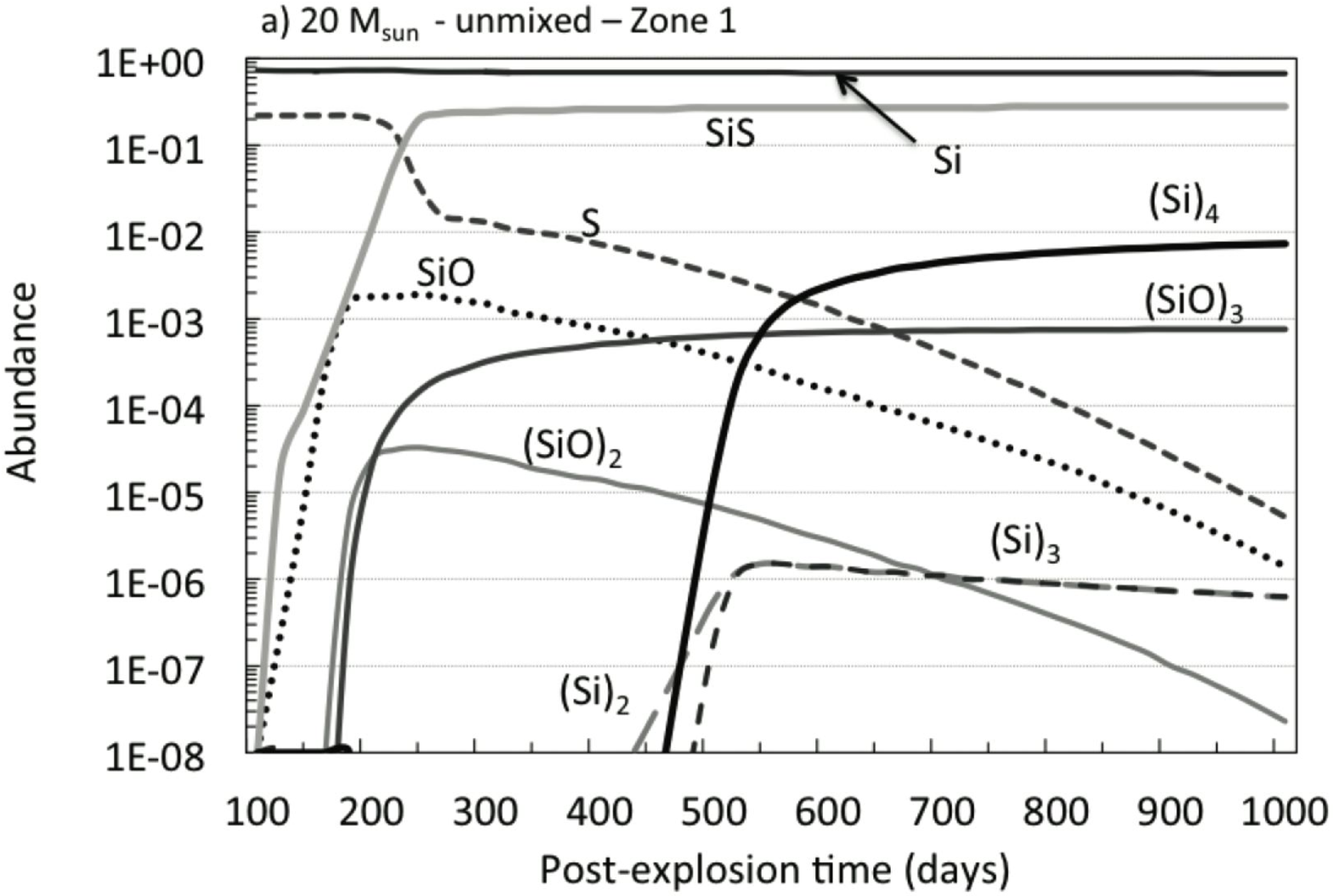}{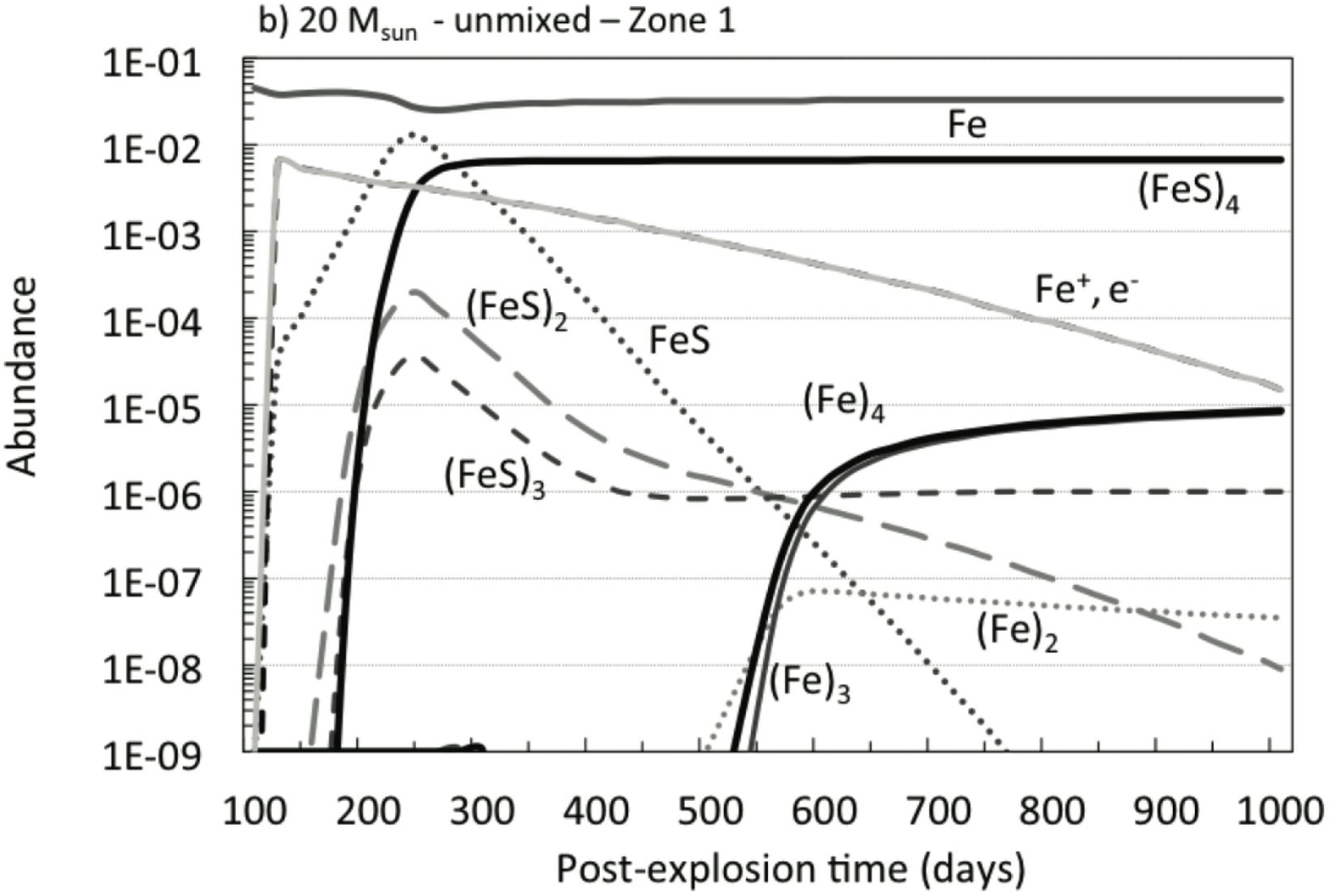}
\caption{The evolution of the cluster abundances normalized to total gas number density for the 20~\Ms~unmixed ejecta - Zone 1: a) Silicon-bearing clusters - Si/S-bearing species; b) Iron-bearing clusters, most dominant atoms and ions, and electrons. The dominant precursors are pure silicon, iron sulphide, silicon monoxide and pure iron clusters. The abundance of secondary electrons is commanded by the ionization of atomic Fe. }
\end{figure}


\begin{figure}
\epsscale{1.15}
\plottwo{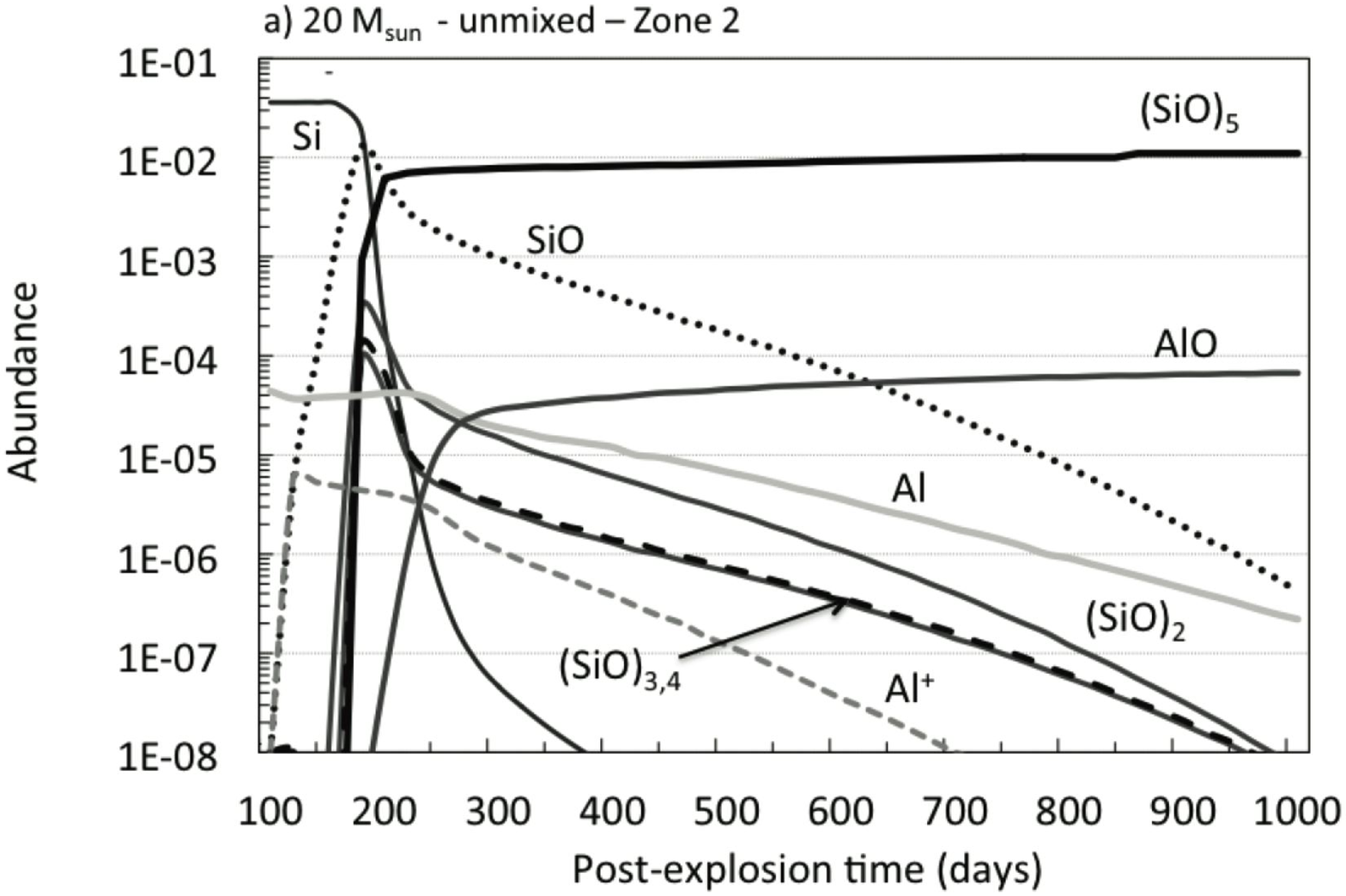}{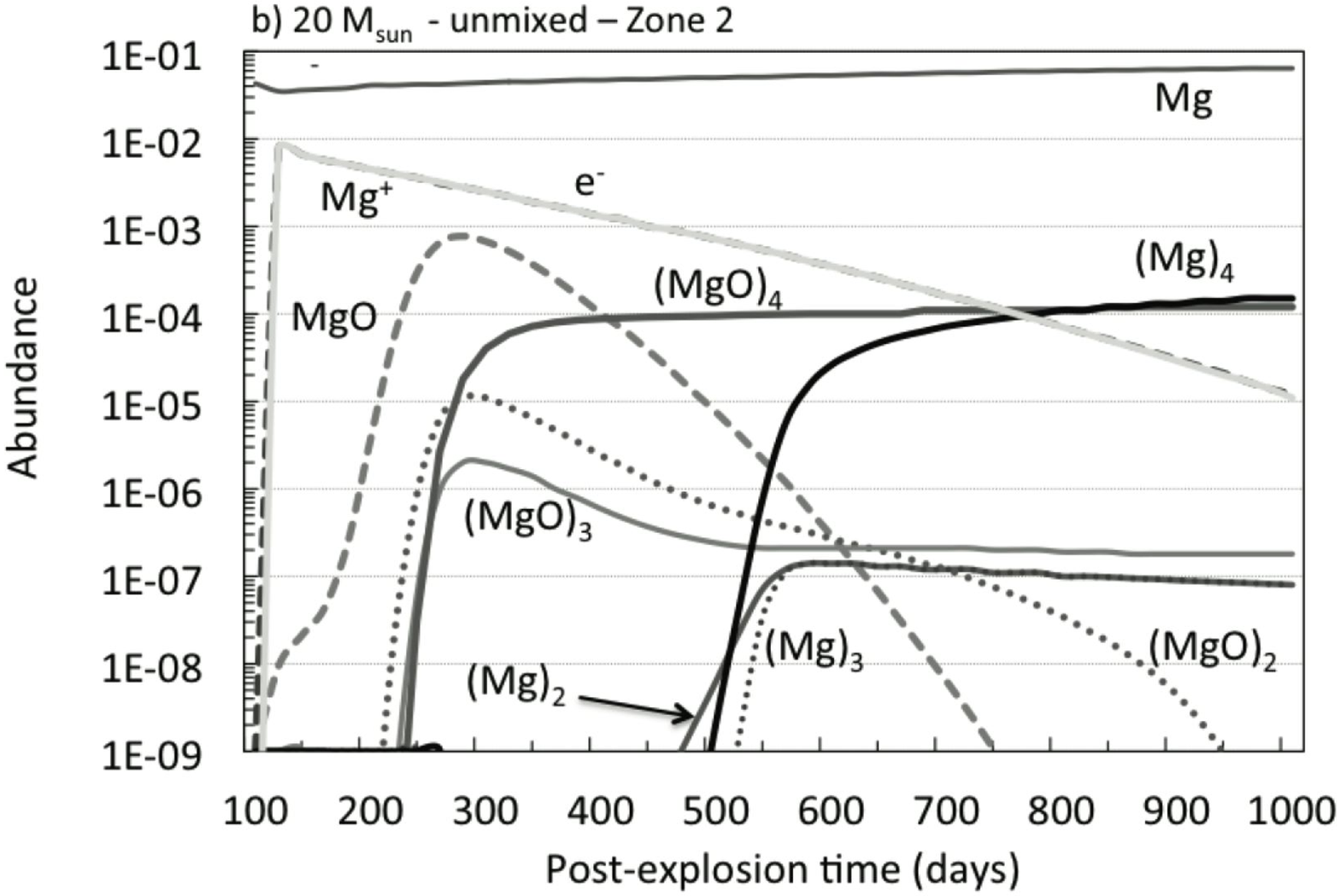}
\caption{The evolution of the cluster abundances normalized to total gas number density for the 20~\Ms~unmixed ejecta - Zone 2: a) Silicon monoxide clusters, aluminum-bearing species; b) Magnesium-bearing clusters, atoms, ions and electrons. The dominant precursors are silicon oxide and pure magnesium clusters, and aluminum monoxide. The abundance of secondary electrons is commanded by the ionization of atomic Mg.}
\end{figure}


\clearpage
\begin{figure}
\epsscale{0.8}
\plotone{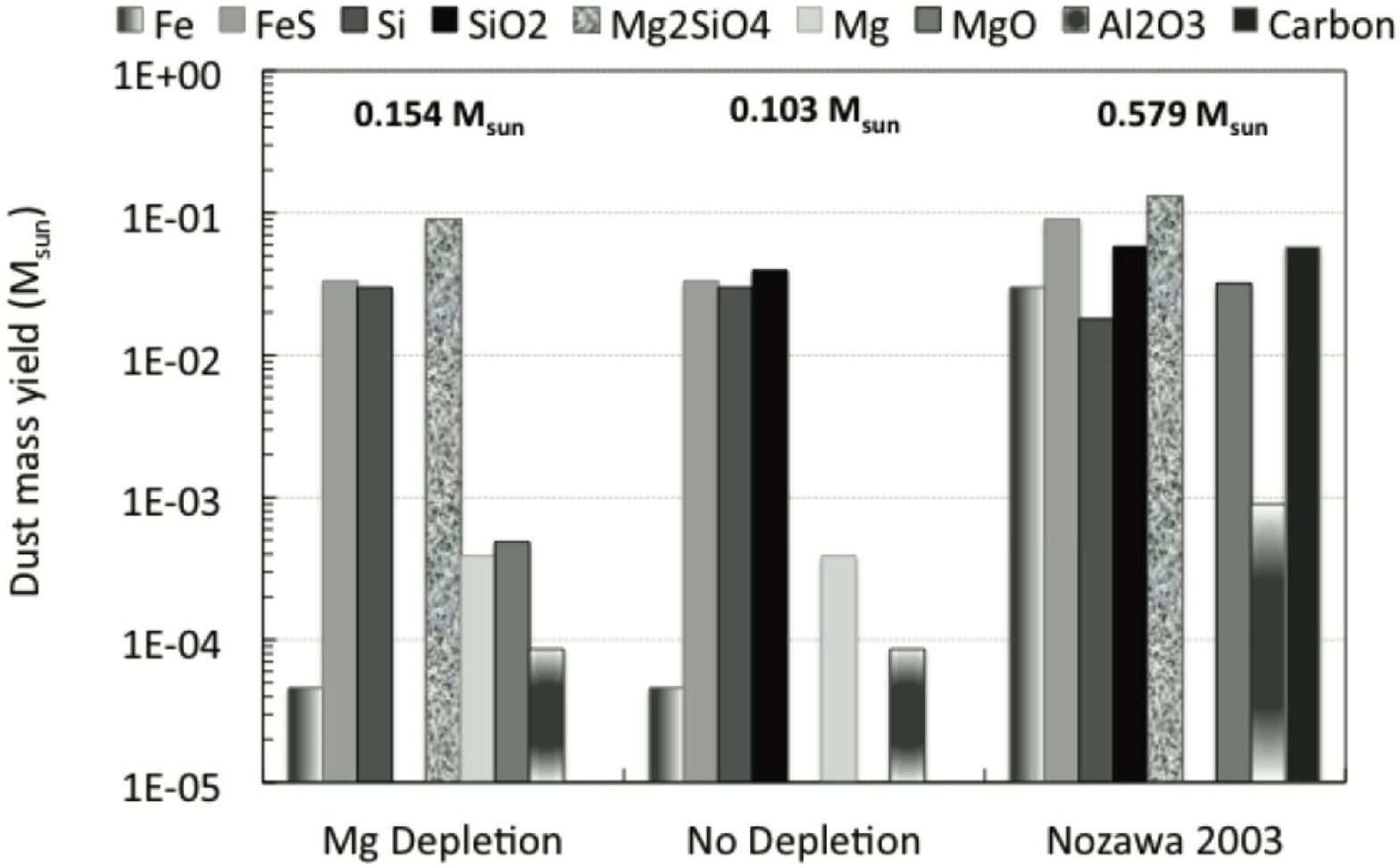}
\caption{Dust chemical composition and mass yields (in units of  \Ms) at t = 1000 days versus depletion case for the 20 \Ms\ unmixed ejecta. Results from Nozawa et al. (2003) are shown for comparison. The total upper limit for dust mass and the total dust mass from NK03 are also indicated. Zone 1 synthesizes iron sulphide, pure Si clusters, and Fe clusters. Silica or forsterite, pure silicon clusters and alumina are formed in Zone 2 and Zone 3. Extremely small amounts of carbon chains are formed in Zone 4.  }
\end{figure}

\begin{figure}
\epsscale{0.8}
\plotone{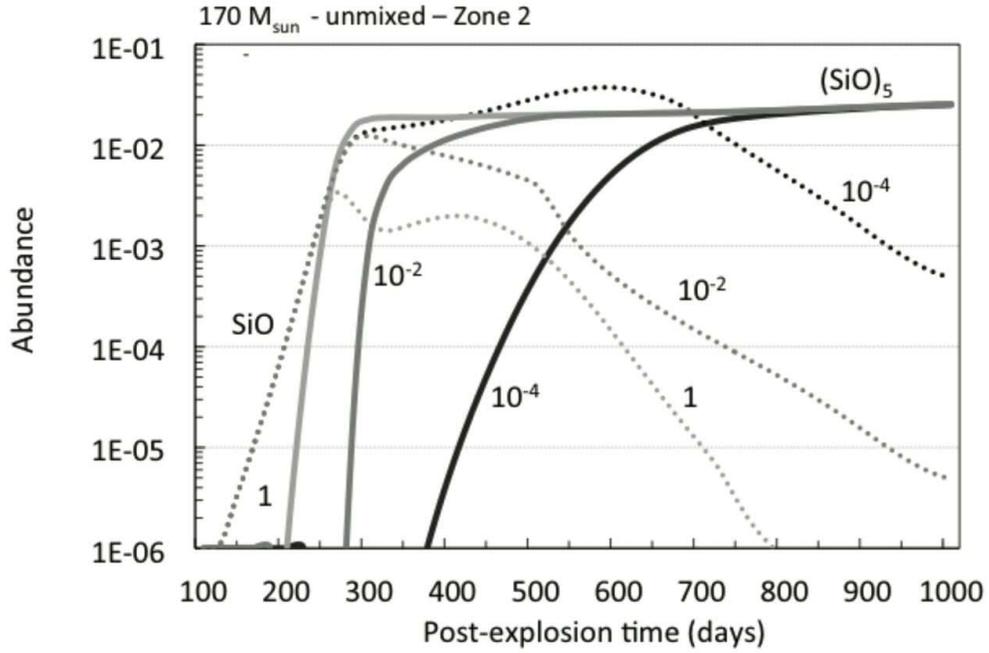}
\caption{Abundances normalized to total gas number density of (SiO)$_5$ clusters (solid lines) and gas-phase SiO (dotted lines) as a function of time in Zone 2 of the 170~\Ms~unmixed ejecta. Curves are labelled by the decrease factor applied to reaction \ref {sio7} rates according to similar decrease in total gas pressures: 1 corresponds to the rate at 1 atm, while results presented in \S 4.2 correspond to a decrease of 10$^{-2}$. For decreasing nucleation rates, the formation of (SiO)$_5$ clusters is delayed to late times though the final mass yield for the clusters is similar for all cases. As expected, the SiO abundance is larger and its decreases with time is less pronounced for the smaller reaction \ref {sio7} rate. }
\end{figure}


\begin{figure}
\epsscale{0.8}
\plotone{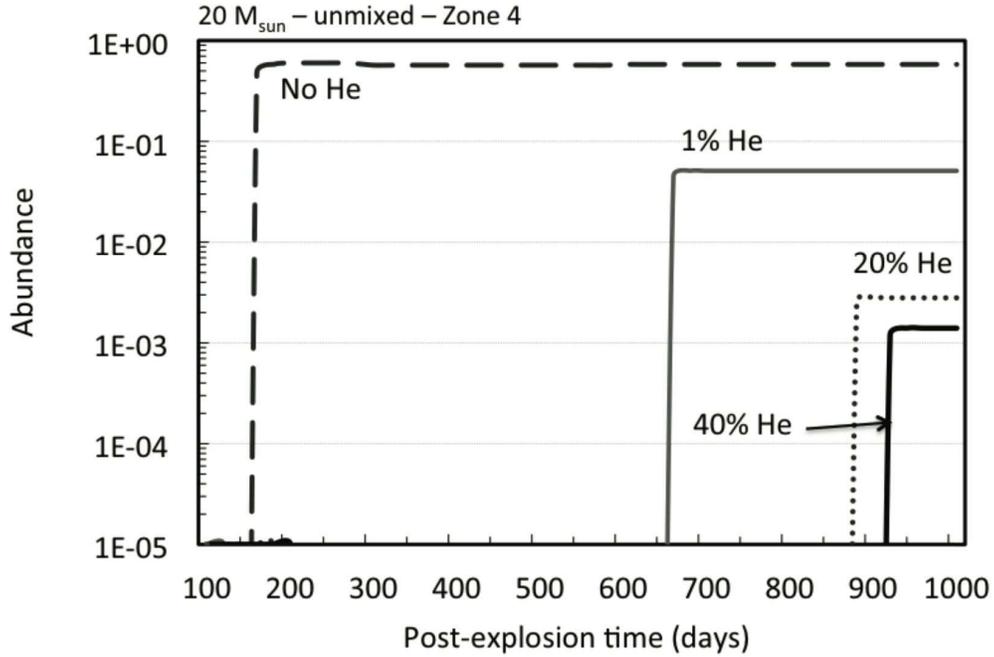}
\caption{Abundances for C$_{10}$ rings normalized to total gas number density for various He initial composition in Zone 4 of the 20~\Ms~unmixed ejecta. Curves are labelled by the percentage of the initial He mass present. The 100 \% labelled curve corresponds to the results for Zone 4 discussed in \S 4.3.2.  The larger the He content, the lower the abundance of the end-product ring C$_{10}$ due to the destruction of C-chains by He$^+$, and the larger the delay in formation time, which depends on the He$^+$ abundance in the ejecta. }
\end{figure}


\end{document}